\documentclass[12pt,twoside, a4paper]{article}
\def\pd{\partial}
\def\mc{\mathcal}

\usepackage[dvips]{graphicx}
\usepackage{amssymb}
\usepackage[bbgreekl]{mathbbol}
\usepackage{amssymb,amsmath}
\usepackage{graphicx}
\usepackage{dsfont}
\usepackage{caption}
\usepackage{subcaption}
\usepackage{mathtools}
\usepackage{verbatim}
\usepackage{graphicx}
\usepackage{multirow}
\usepackage[outline]{contour}
\usepackage{xcolor,colortbl}
\input{epsf.sty}  
\begin{document}
\begin{center}
\Large{\textbf{New supersymmetric $AdS_5$ black strings from 5D $N=4$ gauged supergravity}}
\end{center}
\vspace{1 cm}
\begin{center}
\large{\textbf{Parinya Karndumri}}
\end{center}
\begin{center}
String Theory and Supergravity Group, Department
of Physics, Faculty of Science, Chulalongkorn University, 254 Phayathai Road, Pathumwan, Bangkok 10330, Thailand \\
E-mail: parinya.ka@hotmail.com \vspace{1 cm}
\end{center}
\begin{abstract}
We find a large class of new supersymmetric $AdS_5$ black strings from five-dimensional $N=4$ gauged supergravity coupled to five vector multiplets with $SO(2)_D\times SO(3)\times SO(3)$ gauge group. These solutions have near horizon geometries of the form $AdS_3\times \Sigma^2$ for $\Sigma^2$ being a two-sphere ($S^2$) or a hyperbolic space ($H^2$). There are four supersymmetric $AdS_5$ vacua with $N=4$ and $N=2$ supersymmetries. By performing topological twists along $\Sigma^2$ with $SO(2)\times SO(2)_{\textrm{diag}}$ and $SO(2)_{\textrm{diag}}$ gauge fields, we find a number of $AdS_3\times \Sigma^2$ fixed points describing near horizon geometries of black strings in asymptotically $AdS_5$ spaces. Most of the solutions take the form of $AdS_3\times H^2$ with only one being $AdS_3\times S^2$ preserving $SO(2)_{\textrm{diag}}$ symmetry. We also give the corresponding black string solutions interpolating between asymptotically locally $AdS_5$ vacua and the near horizon $AdS_3\times \Sigma^2$ geometries. There are a number of solutions flowing from one, two or three $AdS_5$ vacua to an $AdS_3\times \Sigma^2$ fixed point. These solutions can also be considered as holographic RG flows across dimensions from $N=2$ and $N=1$ SCFTs in four dimensions to two-dimensional SCFTs with $N=(2,0)$ or $N=(0,2)$ supersymmetry.    
\end{abstract}
\newpage
\section{Introduction}
Supersymmetric black hole solutions in asymptotically $AdS$ spaces have attracted much attention recently since the groundbreaking work on microstate counting of $AdS_4$ black holes from supersymmetric localization in \cite{BH_entropy_Zaffaroni1}, see also \cite{BH_entropy_Zaffaroni1_2,BH_entropy_Zaffaroni2,BH_entropy_Zaffaroni2_1,BH_entropy_Zaffaroni3,
BH_entropy_IIA1,BH_entropy_IIA2,Bobev_AdS4_BH,BH_microstate_Cabo,Zaffaroni_lecture}. This technique has also been extended to supersymmetric $AdS_5$ black strings in \cite{microstate_black_string1,microstate_black_string2,microstate_black_string3,microstate_black_string4}. Due to these results, constructing new supersymmetric $AdS_5$ black string solutions to further test the proposed holographic relation is interesting. Along this line, gauged supergravities in five dimensions provide a very useful tool in which $AdS_5$ black strings become domain walls interpolating between asymptotically $AdS_5$ spaces and near horizon geometries of the form $AdS_3\times \Sigma^2$ for $\Sigma^2$ being a two-sphere ($S^2$) or a two-dimensional hyperbolic space ($H^2$).     
\\
\indent On the other hand, supersymmetric $AdS_5$ black string solutions also describe holographic RG flows from four-dimensional superconformal field theories (SCFTs) in the UV, dual to $AdS_5$ vacua, to two-dimensional SCFTs in the IR dual to near horizon geometries \cite{Maldacena_Nunez_nogo}. In this point of view, the SCFTs in four dimensions undergo twisted compactifications on $\Sigma^2$ to conformal field theories in two dimensions. Since the pioneering work \cite{Maldacena_Nunez_nogo}, a number of similar solutions have been found \cite{black_string_Klemm1,black_string_Klemm2,black_string_Klemm3,BB,Wraped_D3,3D_CFT_from_LS_point,BBC,black_string_Klemm4,flow_acrossD_bobev,black_string_Klemm5,flow_across_Betti,5D_N4_flow,5D_N4_flow1}. Some of the solutions can be uplifted to string/M-theory, and the SCFTs dual to these solutions have also been identified. 
\\
\indent In this paper, we will look for new supersymmetric $AdS_5$ black string solutions from five-dimensional $N=4$ gauged supergravity with $SO(2)_D\times SO(3)\times SO(3)$ gauge group. This gauged supergravity can be constructed from $N=4$ supergravity coupled to five vector multiplets and has been studied recently in \cite{5D_flowII} in which a number of $N=2$ and $N=4$ supersymmetric $AdS_5$ vacua and holographic RG flows interpolating among them have been found. Similar solutions within the framework of $N=4$ gauged supergravity have also appeared in \cite{5D_N4_flow,5D_N4_flow1}, see \cite{flow_across_Gauntlett1,flow_across_Gauntlett2,BB,Wraped_D3,Wraped_M5,6D_twist,2DSCFT_from_N2_7D} for $AdS$ black string solutions in higher dimensions. 
\\
\indent Unlike the solutions given in \cite{5D_N4_flow,5D_N4_flow1}, due to a richer structure of $AdS_5$ vacua in the gauged supergravity considered in the present paper, there is a large number of new and more interesting $AdS_5$ black string solutions. In particular, there are solutions connecting one, two or three $AdS_5$ critical points to $AdS_3\times \Sigma^2$ fixed points. A truncation of this $SO(2)_D\times SO(3)\times SO(3)$ gauged supergravity to two or three vector multiplets and $SO(2)_D\times SO(3)$ gauge group can be embedded in eleven dimensions as shown in \cite{Malek_AdS5_N4_embed}. Accordingly, both the $AdS_5$ vacua and $AdS_5$ black strings within this truncation could be uplifted to higher dimensions leading to new $AdS_5$ and $AdS_3$ solutions of string/M-theory. However, apart from the proof of the consistency of the truncation using the powerful framework of exceptional field theories in \cite{Malek_AdS5_N4_embed}, no complete truncation ansatze in this case have appeared to date.
\\
\indent The paper is organized as follows. In section \ref{N4_SUGRA},
we review the construction of five-dimensional $N=4$ gauged supergravity with $SO(2)_D\times SO(3)\times SO(3)$ gauge group in the embedding tensor formalism. In sections \ref{SO2_SO2diag_twist} and \ref{SO2diag_twist}, we give $AdS_3\times \Sigma^2$ geometries with $SO(2)_{\textrm{diag}}\times SO(2)$ and $SO(2)_{\textrm{diag}}$ symmetries together with black string solutions interpolating between these geometries and supersymmetric $AdS_5$ vacua. We end the paper with some conclusions and comments in section \ref{conclusion}. 

\section{Five dimensional $N=4$ gauged supergravity with $SO(2)_D\times SO(3)\times SO(3)$ gauge group}\label{N4_SUGRA} 
In this section, we review $N=4$ gauged supergravity as constructed in \cite{N4_gauged_SUGRA,5D_N4_Dallagata} by using the embedding tensor formalism. The supergravity multiplet consists of the graviton
$e^{\hat{\mu}}_\mu$, four gravitini $\psi_{\mu i}$, six vectors $(A^0_\mu,A_\mu^m)$, four spin-$\frac{1}{2}$ fields $\chi_i$ and one real
scalar $\Sigma$ or the dilaton. Space-time and tangent space indices are denoted respectively by $\mu,\nu,\ldots =0,1,2,3,4$ and
$\hat{\mu},\hat{\nu},\ldots=0,1,2,3,4$. The fundamental representation of $SO(5)_R\sim USp(4)_R$
R-symmetry is described by $m,n=1,\ldots, 5$ for $SO(5)_R$ and $i,j=1,2,3,4$ for $USp(4)_R$. A vector multiplet contains a vector field $A_\mu$, four gaugini $\lambda_i$ and five scalars $\phi^m$. For $N=4$ supergravity coupled to $n$ vector multiplets, we will label these multiplets by indices $a,b=1,\ldots, n$ of the form $(A^a_\mu,\lambda^{a}_i,\phi^{ma})$. 
\\
\indent From the gravity and vector multiplets, there are $6+n$ vector fields denoted collectively by $A^{\mc{M}}_\mu=(A^0_\mu,A^m_\mu,A^a_\mu)$ and $5n+1$ scalars in the $SO(1,1)\times SO(5,n)/SO(5)\times SO(n)$ coset manifold. For later convenience, we have introduced an index $\mc{M}=(0,M)$ as in \cite{N4_gauged_SUGRA}. The $5n$ scalars parametrizing the $SO(5,n)/SO(5)\times SO(n)$ coset can be described by a coset representative $\mc{V}_M^{\phantom{M}A}$ transforming under the global $G=SO(5,n)$ and the local $H=SO(5)\times SO(n)$ by left and right multiplications, respectively. We use the global $SO(5,n)$ indices $M,N,\ldots=1,2,\ldots , 5+n$ while the local $H$ indices $A,B,\ldots$ can be split as $A=(m,a)$. The coset representative can then be written as
\begin{equation}
\mc{V}_M^{\phantom{M}A}=(\mc{V}_M^{\phantom{M}m},\mc{V}_M^{\phantom{M}a}).
\end{equation}
For later convenience, we also define a symmetric matrix 
\begin{equation}
M_{MN}=\mc{V}_M^{\phantom{M}m}\mc{V}_N^{\phantom{M}m}+\mc{V}_M^{\phantom{M}a}\mc{V}_N^{\phantom{M}a}
\end{equation}
which is manifestly $SO(5)\times SO(n)$ invariant. Finally, all fermionic fields are symplectic Majorana spinors subject to the condition
\begin{equation}
\xi_i=\Omega_{ij}C(\bar{\xi}^j)^T 
\end{equation}
with $C$ and $\Omega_{ij}$ being the charge conjugation matrix and $USp(4)$ symplectic matrix, respectively.
\\
\indent Gaugings of $N=4$ supergravity are characterized by the embedding tensor with components $\xi^{M}$, $\xi^{MN}=\xi^{[MN]}$ and $f_{MNP}=f_{[MNP]}$. These components determine the embedding of gauge groups in the global symmetry group $SO(1,1)\times SO(5,n)$. We are only interested in gaugings with $\xi^M=0$ which admit supersymmetric $AdS_5$ vacua as shown in \cite{AdS5_N4_Jan}. Accordingly, we will set $\xi^{M}=0$ which leads to considerable simplification in various expressions. In particular, the quadratic constraints on the embedding tensor simply reduce to
\begin{equation}
f_{R[MN}{f_{PQ]}}^R=0\qquad \textrm{and}\qquad {\xi_M}^Qf_{QNP}=0\, .\label{QC}
\end{equation}
Furthermore, for $\xi^{M}=0$, the gauge group is embedded solely in $SO(5,n)$ with the corresponding gauge generators in $SO(5,n)$ fundamental representation given by
\begin{equation}
{(X_M)_N}^P=-{f_M}^{QR}{(t_{QR})_N}^P={f_{MN}}^P\quad \textrm{and}\quad {(X_0)_N}^P=-\xi^{QR}{(t_{QR})_N}^P={\xi_N}^P\, .
\end{equation}
We have chosen $SO(5,n)$ generators of the form ${(t_{MN})_P}^Q=\delta^Q_{[M}\eta_{N]P}$ with $\eta_{MN}=\textrm{diag}(-1,-1,-1,-1,-1,1,1,\ldots,1)$ being the $SO(5,n)$ invariant tensor. The gauge covariant derivative reads
\begin{equation}
D_\mu=\nabla_\mu+A_\mu^{M}X_M+A^0_\mu X_0=\nabla_\mu+A^{\mc{M}}X_{\mc{M}}
\end{equation}
with $\nabla_\mu$ being a space-time covariant derivative (possibly) including $SO(5)\times SO(n)$ composite connection.  
\\
\indent The bosonic Lagrangian of a general gauged $N=4$ supergravity can be written as
\begin{eqnarray}
e^{-1}\mc{L}&=&\frac{1}{2}R-\frac{3}{2}\Sigma^{-2}D_\mu \Sigma D^\mu \Sigma +\frac{1}{16} D_\mu M_{MN}D^\mu
M^{MN}-V\nonumber \\
& &-\frac{1}{4}\Sigma^2M_{MN}\mc{H}^M_{\mu\nu}\mc{H}^{N\mu\nu}-\frac{1}{4}\Sigma^{-4}\mc{H}^0_{\mu\nu}\mc{H}^{0\mu\nu}+e^{-1}\mc{L}_{\textrm{top}}
\end{eqnarray}
where $e$ is the vielbein determinant. $\mc{L}_{\textrm{top}}$ is the topological term whose explicit form can be found in \cite{N4_gauged_SUGRA}. 
\\
\indent The covariant gauge field strength tensors read
\begin{equation}
\mc{H}^{\mc{M}}_{\mu\nu}=2\pd_{[\mu}A^{\mc{M}}_{\nu]}+{X_{\mc{N}\mc{P}}}^{\mc{M}}A^{\mc{N}}_\mu A^{\mc{P}}_\nu+Z^{\mc{M}\mc{N}}B_{\mu\nu\mc{N}}\label{covariant_field_strength}
\end{equation}
with 
\begin{equation}
Z^{MN}=\frac{1}{2}\xi^{MN}\qquad \textrm{and} \qquad Z^{0M}=-Z^{M0}=\frac{1}{2}\xi^M=0\, .
\end{equation}
\indent In the embedding tensor formalism, the two-form fields $B_{\mu\nu \mc{M}}$ are introduced off-shell. These fields do not have kinetic terms and couple to vector fields via the topological term. It is useful to note the first-order field equations for these two-form fields 
\begin{equation}
Z^{\mc{M}\mc{N}}\left[\frac{1}{6\sqrt{2}}\epsilon_{\mu\nu\rho\lambda\sigma}\mc{H}^{(3)\rho\lambda\sigma}_{\mc{N}}-\mc{M}_{\mc{N}\mc{P}}
\mc{H}^{\mc{P}}_{\mu\nu}\right]=0\label{2-form_field_eq}
\end{equation}
with $\mc{M}_{00}=\Sigma^{-4}$, $\mc{M}_{0M}=0$ and $\mc{M}_{MN}=\Sigma^2M_{MN}$. The field strength $\mc{H}^{(3)}_{\mc{M}}$ is defined by
\begin{equation}
Z^{\mc{M}\mc{N}}\mc{H}^{(3)}_{\mu\nu\rho\mc{N}}=Z^{\mc{M}\mc{N}}\left[3D_{[\mu}B_{\nu\rho]\mc{N}}
+6d_{\mc{NPQ}}A^{\mc{P}}_{[\mu}\left(\pd_\nu A^{\mc{Q}}_{\rho]}+\frac{1}{3}{X_{\mc{RS}}}^{\mc{Q}}A^{\mc{R}}_\nu A^{\mc{S}}_{\rho]}\right)\right]\label{H3_def}
\end{equation}
for $d_{0MN}=d_{MN0}=d_{M0N}=\eta_{MN}$ and 
\begin{equation}
{X_{MN}}^P={f_{MN}}^P,\qquad {X_{M0}}^0=0,\qquad {X_{0M}}^N={\xi_M}^N\, . 
\end{equation}
\indent The scalar potential is given by
\begin{eqnarray}
V&=&-\frac{1}{4}\left[f_{MNP}f_{QRS}\Sigma^{-2}\left(\frac{1}{12}M^{MQ}M^{NR}M^{PS}-\frac{1}{4}M^{MQ}\eta^{NR}\eta^{PS}\right.\right.\nonumber \\
& &\left.+\frac{1}{6}\eta^{MQ}\eta^{NR}\eta^{PS}\right) +\frac{1}{4}\xi_{MN}\xi_{PQ}\Sigma^4(M^{MP}M^{NQ}-\eta^{MP}\eta^{NQ})\nonumber \\
& &\left.
+\frac{\sqrt{2}}{3}f_{MNP}\xi_{QR}\Sigma M^{MNPQR}\right]
\end{eqnarray}
where $M^{MN}$ is the inverse of $M_{MN}$, and $M^{MNPQR}$ is obtained from
\begin{equation}
M_{MNPQR}=\epsilon_{mnpqr}\mc{V}_{M}^{\phantom{M}m}\mc{V}_{N}^{\phantom{M}n}
\mc{V}_{P}^{\phantom{M}p}\mc{V}_{Q}^{\phantom{M}q}\mc{V}_{R}^{\phantom{M}r}
\end{equation}
by raising the indices with $\eta^{MN}$. 
\\
\indent Supersymmetry transformations of fermionic fields are given by
\begin{eqnarray}
\delta\psi_{\mu i} &=&D_\mu \epsilon_i+\frac{i}{\sqrt{6}}\Omega_{ij}A^{jk}_1\gamma_\mu\epsilon_k\nonumber \\
& &-\frac{i}{6}\left(\Omega_{ij}\Sigma{\mc{V}_M}^{jk}\mc{H}^M_{\nu\rho}-\frac{\sqrt{2}}{4}\delta^k_i\Sigma^{-2}\mc{H}^0_{\nu\rho}\right)({\gamma_\mu}^{\nu\rho}-4\delta^\nu_\mu\gamma^\rho)\epsilon_k,\\
\delta \chi_i &=&-\frac{\sqrt{3}}{2}i\Sigma^{-1} D_\mu
\Sigma\gamma^\mu \epsilon_i+\sqrt{2}\Omega_{ij}A_2^{kj}\epsilon_k\nonumber \\
& &-\frac{1}{2\sqrt{3}}\left(\Sigma \Omega_{ij}{\mc{V}_M}^{jk}\mc{H}^M_{\mu\nu}+\frac{1}{\sqrt{2}}\Sigma^{-2}\delta^k_i\mc{H}^0_{\mu\nu}\right)\gamma^{\mu\nu}\epsilon_k,\\
\delta \lambda^a_i&=&i\Omega^{jk}({\mc{V}_M}^aD_\mu
{\mc{V}_{ij}}^M)\gamma^\mu\epsilon_k+\sqrt{2}\Omega_{ij}A_{2}^{akj}\epsilon_k-\frac{1}{4}\Sigma{\mc{V}_M}^a\mc{H}^M_{\mu\nu}\gamma^{\mu\nu}\epsilon_i
\end{eqnarray}
in which the fermion shift matrices are defined by 
\begin{eqnarray}
A_1^{ij}&=&-\frac{1}{\sqrt{6}}\left(\sqrt{2}\Sigma^2\Omega_{kl}{\mc{V}_M}^{ik}{\mc{V}_N}^{jl}\xi^{MN}+\frac{4}{3}\Sigma^{-1}{\mc{V}^{ik}}_M{\mc{V}^{jl}}_N{\mc{V}^P}_{kl}{f^{MN}}_P\right),\nonumber
\\
A_2^{ij}&=&\frac{1}{\sqrt{6}}\left(\sqrt{2}\Sigma^2\Omega_{kl}{\mc{V}_M}^{ik}{\mc{V}_N}^{jl}\xi^{MN}-\frac{2}{3}\Sigma^{-1}{\mc{V}^{ik}}_M{\mc{V}^{jl}}_N{\mc{V}^P}_{kl}{f^{MN}}_P\right),\nonumber
\\
A_2^{aij}&=&-\frac{1}{2}\left(\Sigma^2{{\mc{V}_M}^{ij}\mc{V}_N}^a\xi^{MN}-\sqrt{2}\Sigma^{-1}\Omega_{kl}{\mc{V}_M}^a{\mc{V}_N}^{ik}{\mc{V}_P}^{jl}f^{MNP}\right).
\end{eqnarray}
\indent $\mc{V}_M^{\phantom{M}ij}$ is defined in terms of ${\mc{V}_M}^m$ and $SO(5)$ gamma matrices ${\Gamma_{mi}}^j$ as
\begin{equation}
{\mc{V}_M}^{ij}=\frac{1}{2}{\mc{V}_M}^{m}\Gamma^{ij}_m
\end{equation}
with $\Gamma^{ij}_m=\Omega^{ik}{\Gamma_{mk}}^j$. Similarly, the inverse ${\mc{V}_{ij}}^M$ can be written as
\begin{equation}
{\mc{V}_{ij}}^M=\frac{1}{2}{\mc{V}_m}^M(\Gamma^{ij}_m)^*=\frac{1}{2}{\mc{V}_m}^M\Gamma_{m}^{kl}\Omega_{ki}\Omega_{lj}\,
.
\end{equation}
We will use the following representation of $SO(5)$ gamma matrices
\begin{eqnarray}
\Gamma_1&=&-\sigma_2\otimes \sigma_2,\qquad \Gamma_2=i\mathbb{I}_2\otimes \sigma_1,\qquad \Gamma_3=\mathbb{I}_2\otimes \sigma_3,\nonumber\\
\Gamma_4&=&\sigma_1\otimes \sigma_2,\qquad \Gamma_5=\sigma_3\otimes \sigma_2
\end{eqnarray}
with $\sigma_i$, $i=1,2,3$, being the Pauli matrices.
\\
\indent The covariant derivative on $\epsilon_i$ is given by
\begin{equation}
D_\mu \epsilon_i=\pd_\mu \epsilon_i+\frac{1}{4}\omega_\mu^{ab}\gamma_{ab}\epsilon_i+{Q_{\mu i}}^j\epsilon_j
\end{equation}
with the composite connection defined by
\begin{equation}
{Q_{\mu i}}^j={\mc{V}_{ik}}^M\pd_\mu {\mc{V}_M}^{kj}-A^0_\mu\xi^{MN}\mc{V}_{Mik}{\mc{V}_N}^{kj}-A^M_\mu{\mc{V}_{ik}}^N\mc{V}^{kjP}f_{MNP}\, .
\end{equation}
\indent In this paper, we will consider $N=4$ gauged supergravity coupled to $n=5$ vector multiplets with $SO(2)_D\times SO(3)\times SO(3)$ gauge group previously studied in \cite{5D_flowII}. The corresponding embedding tensor is given by, see \cite{5D_flowII} for more detail,
\begin{eqnarray}
\xi^{MN}&=&g_1(\delta^M_1\delta^N_2-\delta^M_2\delta^N_1)-g_2(\delta^M_{10}\delta^N_9-\delta^M_9\delta^N_{10}),\label{embedding_tensor}\\ 
f_{\tilde{m}+2,\tilde{n}+2,\tilde{p}+2}&=&h_1\epsilon_{\tilde{m}\tilde{n}\tilde{p}},\qquad \tilde{m},\tilde{n},\tilde{p}=1,2,3,\\
f_{\tilde{a}\tilde{b}\tilde{c}}&=&h_2\epsilon_{\tilde{a}\tilde{b}\tilde{c}},\qquad \tilde{a},\tilde{b},\tilde{c}=1,2,3
\end{eqnarray} 
with the coupling constants $g_1$, $g_2$, $h_1$ and $h_2$. It is useful to note that the first $SO(3)\sim SO(3)_R\subset SO(5)_R$ is a subgroup of the R-symmetry while the second $SO(3)$ factor is a subgroup of $SO(5)$ symmetry of the vector multiplets. $SO(2)_D$ is the diagonal subgroup of $SO(2)'_R\subset SO(2)'_R\times SO(3)_R\subset SO(5)_R$ and $SO(2)'\subset SO(2)'\times SO(3)\subset SO(5)$.   
\\
\indent We end this section by giving an explicit parametrization of the scalar coset $SO(5,5)/SO(5)\times SO(5)$. With $SO(5,5)$ non-compact generators given by
\begin{equation}
Y_{ma}=t_{m,a+5},\qquad m=1,2,\ldots, 5,\qquad a=1,2,\ldots, 5,
\end{equation}
the coset representative can be written as
\begin{equation}
\mc{V}=e^{\phi^{ma}Y_{ma}}\, .
\end{equation}
\section{Supersymmetric $AdS_5$ black strings with $SO(2)_{\textrm{diag}}\times SO(2)$ symmetry}\label{SO2_SO2diag_twist}
We now look for supersymmetric $AdS_5$ black string solutions with a near horizon geometry given by $AdS_3\times \Sigma^2$ for $\Sigma^2=S^2,H^2$. The metric ansatz is taken to be
\begin{equation}
ds^2=e^{2F(r)}dx^2_{1,1}+dr^2+e^{2G(r)}(d\theta^2+f_\kappa(\theta)^2d\phi^2)\label{metric_ansatz}
\end{equation}
with 
\begin{equation}
f_\kappa(\theta)=\begin{cases}
  \sin\theta,  & \kappa=1\quad \textrm{for}\quad \Sigma^2=S^2 \\
  \sinh\theta,  & \kappa=-1\quad \textrm{for}\quad \Sigma^2=H^2
\end{cases}\, .
\end{equation}
$dx_{1,1}^2=\eta_{\alpha\beta}dx^\alpha dx^\beta$, $\alpha,\beta=0,1$, is the flat metric on the two-dimensional Minkowski space. Relevant components of the spin connection are given by
\begin{eqnarray}
& &\omega^{\hat{\alpha}\hat{r}}=F'e^{\hat{\alpha}},\qquad \omega^{\hat{\theta}\hat{r}}=G'e^{\hat{\theta}},\nonumber \\
& &\omega^{\hat{\phi}\hat{r}}=G'e^{\hat{\phi}},\qquad \omega^{\hat{\theta}{\phi}}=e^{-G}\frac{f'_\kappa(\theta)}{f_{\kappa}(\theta)}e^{\hat{\phi}}\, .
\end{eqnarray}
Throughout the paper, $r$-derivatives are denoted by $'$ except for $f'_\kappa(\theta)=\frac{df_\kappa(\theta)}{d\theta}$. It is also useful to point out that the metric ansatz \eqref{metric_ansatz} leads to solutions interpolating between $AdS_5$ and $AdS_3\times\Sigma^2$ geometries. Near the asymptotic $AdS_5$ space, we have
\begin{equation}
F(r)\sim G(r)\sim \frac{r}{L_{AdS_5}}
\end{equation}
with $L_{AdS_5}$ being the corresponding $AdS_5$ radius while the $AdS_3\times \Sigma^2$ near horizon geometry corresponds to
\begin{equation}
F(r)\sim \frac{r}{L_{AdS_3}}\qquad \textrm{and}\qquad G'(r)\sim 0
\end{equation}
with the $AdS_3$ radius $L_{AdS_3}$.
\\
\indent We first consider solutions obtained from a topological twist along $\Sigma^2$ by turning on $SO(2)_{\textrm{diag}}\times SO(2)$ gauge fields of the form
\begin{eqnarray}
A^0=a_0f'_\kappa(\theta)d\phi=\frac{h_1}{g_2}A^3,\qquad A^3=a_3f'_\kappa(\theta)d\phi,\qquad 
A^6=a_6f'_\kappa(\theta)d\phi\label{gauge_ansatz}
\end{eqnarray}
with $a_0$, $a_3$ and $a_6$ being constants. The $SO(2)_{\textrm{diag}}$ is the diagonal subgroup of $SO(2)_D\times SO(2)_R$ generated by the gauge generators $X_0$ and $X_3$ with $SO(2)_R\subset SO(3)_R\subset SO(5)_R$. From the embedding tensor \eqref{embedding_tensor} and the gauge field ansatz \eqref{gauge_ansatz}, we can verify that setting all the two-form fields to zero is a consistent truncation satisfying the two-form field equation \eqref{2-form_field_eq}. The corresponding field strength tensors are then given by
\begin{equation}
\mc{H}^{\mc{M}}=dA^{\mc{M}}=-\kappa a_{\mc{M}}f_\kappa(\theta)d\theta\wedge d\phi,\qquad \textrm{for}\quad \mc{M}=0,3,6 
\end{equation}
in which we have used the relation $f''_\kappa(\theta)=-\kappa f_\kappa(\theta)$. It should also be noted that the relation $a_0=\frac{h_1}{g_2}a_3$ implements the diagonal subgroup of $SO(2)_D\times SO(2)_R$. 
\\
\indent Among the $25$ scalar fields from the vector multiplets, there are $5$ singlets under $SO(2)_{\textrm{diag}}\times SO(2)$ corresponding to the following non-compact generators
\begin{eqnarray}
& &\hat{Y}_1=Y_{31},\phantom{+Y_{55}}\qquad \hat{Y}_3=Y_{44}+Y_{55},\qquad \hat{Y}_5=Y_{45}-Y_{54},\nonumber \\
& &\tilde{Y}_3=Y_{14}+Y_{25},\qquad \tilde{Y}_4=Y_{15}-Y_{24}\, .
\end{eqnarray}
Accordingly, the coset representative can be written as
\begin{equation}
\mc{V}=e^{\phi_1\hat{Y}_1}e^{\phi_3\hat{Y}_3}e^{\phi_5\hat{Y}_5}e^{\varphi_3\tilde{Y}_3}e^{\varphi_4\tilde{Y}_4}\, .
 \end{equation} 
As pointed out in \cite{5D_flowII}, non-vanishing $\phi_3$ and $\phi_5$ break half of the supersymmetry with the corresponding Killing spinors given by $\epsilon_{1,3}$ or $\epsilon_{2,4}$.
 \\ 
 \indent We are now in a position to perform a topological twist by canceling component $\omega^{\hat{\theta}\hat{\phi}}$ of the spin connection. It turns out that this can be achieved only for $g_2=g_1$ or $\varphi_3=\varphi_4=0$. Both of these possibilities lead to equivalent results, so we will choose to set $\varphi_3=\varphi_4=0$ in the following analysis. With this, the composite connection along the $\phi$-direction is given by
 \begin{equation}
 {Q_i}^j=\frac{ih_1}{2g_2}A^3\left[g_2{(\sigma_2\otimes \mathbb{I}_2)_i}^j-g_1{(\sigma_2\otimes \sigma_3)_i}^j\right].
 \end{equation}  
It should be noted that $A^6$ does not appear in the composite connection since this vector field gauges the $SO(2)$ subgroup of the $SO(3)\subset SO(5)$ symmetry of the vector multiplets under which the gravitini are not charged. 
\\
\indent From the supersymmetry transformation $\delta \psi_{\hat{\phi}i}$, the twist requires
\begin{equation}
\gamma_{\hat{\theta}\hat{\phi}}\epsilon_i=\frac{ih_1a_3}{g_2}\left[g_2{(\sigma_2\otimes \mathbb{I}_2)_i}^j-g_1{(\sigma_2\otimes \sigma_3)_i}^j\right]\epsilon_j\, .\label{twist_con1}
\end{equation} 
The identity $(\gamma_{\hat{\theta}\hat{\phi}})^2=-\mathbb{I}_4$ imposes the consistency condition 
\begin{equation}
\epsilon_i=\frac{h_1^2a_3^2}{g_2^2}\left[(g_1^2+g_2^2)\delta_i^j-2g_1g_2{(\mathbb{I}_2\otimes \sigma_3)_i}^j\right]\epsilon_j\, .
\end{equation} 
This condition can be satisfied by setting $g_1g_2=0$ or imposing a projector
\begin{equation}
{(\mathbb{I}_2\otimes \sigma_3)_i}^j\epsilon_j=\pm \epsilon_i\, .\label{extra_proj}
\end{equation}  
The existence of supersymmetric $AdS_5$ vacua requires non-vanishing $g_1$ \cite{AdS5_N4_Jan}. On the other hand, the $SO(2)_{\textrm{diag}}$ twist requires non-vanishing $g_2$, see \eqref{gauge_ansatz}. Therefore, we need to impose the projector \eqref{extra_proj} on the Killing spinors. Explicitly, the two sign choices of this projector give $\epsilon_{2,4}=0$ and $\epsilon_{1,3}=0$, respectively. This is precisely in agreement with the $N=2$ unbroken supersymmetry preserved by non-vanishing scalars $\phi_3$ and $\phi_5$ as noted before. For definiteness, we will choose the plus sign.
\\
\indent Imposing the projector \eqref{extra_proj}, equation \eqref{twist_con1} leads to a twist condition
\begin{equation}
1=h_1a_3\left(1-\frac{g_1}{g_2}\right)\label{twist_con}
\end{equation}
and a projector
\begin{equation}
\gamma_{\hat{\theta}\hat{\phi}}\epsilon_i={(i\sigma_2\otimes \mathbb{I}_2)_i}^j\epsilon_j\, .\label{theta_phi_proj} 
\end{equation}      

\subsection{Supersymmetric $AdS_5$ vacua}         
We first look at supersymmetric $AdS_5$ vacua within the aforementioned truncation to $SO(2)_{\textrm{diag}}\times SO(2)$ residual symmetry. For $\varphi_3=\varphi_4=0$, the scalar potential is given by
\begin{eqnarray}
V&=&\frac{1}{8\Sigma^2}(1+\cosh2\phi_3\cosh2\phi_5)\left[h_1^2[\cosh2\phi_1(\cosh2\phi_3\cosh2\phi_5-1)-2]\phantom{\sqrt{2}} \right.\nonumber \\
& &\left.+4\sqrt{2}g_1h_1\cosh\phi_1\Sigma^3+g_2^2\Sigma^6(\cosh2\phi_3\cosh2\phi_5-1)\right].
\end{eqnarray}
The superpotential is given in terms of the first ($\alpha_1$) or third ($\alpha_3$) eigenvalue of $A^{ij}_1$ tensor as $W=\sqrt{\frac{2}{3}}|\alpha_{1,3}|$ with the explicit form 
\begin{eqnarray}
W&=&\frac{1}{12\Sigma}\left[2h_1\cosh\phi_1(1+\cosh2\phi_3\cosh2\phi_5) \phantom{\sqrt{2}}\right.\nonumber \\
& &\left.+\sqrt{2}\Sigma^3(g_2-2g_1-g_2\cosh2\phi_3\cosh2\phi_5) \right].
\end{eqnarray}
The superpotential admits two supersymmetric $AdS_5$ critical points. The first one is given by the trivial critical point at the origin of the scalar manifold $SO(5,5)/SO(5)\times SO(5)$
\begin{eqnarray}
& &\phi_1=\phi_3=\phi_5=0,\qquad \Sigma=-\left(\frac{h_1}{\sqrt{2}g_1}\right)^{\frac{1}{3}},\nonumber \\  
& &V_0=-3\left(\frac{g_1h_1^2}{2}\right)^{\frac{2}{3}} ,\qquad L_{AdS_5}=\sqrt{2}\left|\frac{2}{g_1h_1^2}\right|^{\frac{1}{3}}\, . 
\end{eqnarray}
$V_0$ is the cosmological constant, and the $AdS_5$ radius is given by $L=\sqrt{-\frac{6}{V_0}}$. We can also choose $g_1=-\frac{h_1}{\sqrt{2}}$ to bring this critical point to $\Sigma=1$. This critical point preserves the full $N=4$ supersymmetry and $SO(2)_D\times SO(3)\times SO(3)$ gauge symmetry.
\\
\indent The second critical point is $N=2$ supersymmetric and given by
\begin{eqnarray}
& &\phi_1=0,\qquad \phi_5=\textrm{constant},\qquad \Sigma=\left(\frac{\sqrt{2}h_1}{g_2}\right)^{\frac{1}{3}},\nonumber \\
& &\phi_3=\frac{1}{2}\ln\left[\frac{2g_2-8g_1+\sqrt{2}\sqrt{32g_1^2-16g_1g_2-7g_2^2-9g_2^2\cosh4\phi_5}}{6g_2\cosh2\phi_5}\right], \nonumber \\
& &V_0=-\frac{2}{3}(g_2-g_1)^2\left(\frac{h_1^2}{\sqrt{2}g_2^2}\right)^{\frac{2}{3}} ,\qquad L_{AdS_5}=\frac{3}{g_2-g_1}\left(\frac{\sqrt{2}g_2^2}{h_1^2}\right)^{\frac{1}{3}}\, . 
\end{eqnarray}
We have taken $g_2>g_1$ for simplicity. For $\phi_5=0$, we recover the results of \cite{5D_N4_flow_Davide} and \cite{5D_flowII} with
\begin{equation}
\phi_3=\frac{1}{2}\ln\left[\frac{g_2-4g_1+2\sqrt{4g_1^2-2g_1g_2-2g_2^2}}{3g_2}\right].
\end{equation}
Similar to the result of \cite{5D_flowII}, the gauge group is broken to $SO(2)_{\textrm{diag}}\times SO(3)$ with $SO(2)_{\textrm{diag}}$ being the diagonal subgroup of $SO(2)_D\times SO(2)_R$. The $SO(2)$ symmetry is enhanced to $SO(3)\subset SO(5)$ due to the vanishing of $\phi_1$. We also note that for $\phi_3=0$, we have 
\begin{equation}
\phi_5=\frac{1}{2}\ln\left[\frac{g_2-4g_1+2\sqrt{4g_1^2-2g_1g_2-2g_2^2}}{3g_2}\right]
\end{equation}
which shows a symmetric role between $\phi_3$ and $\phi_5$. Indeed, all values of $\phi_5$ lead to physically equivalent $N=2$ $AdS_5$ vacua with $SO(2)_{\textrm{diag}}$ generator given by different linear combinations of $X_0$ and $X_3$.

\subsection{Supersymmetric $AdS_3\times \Sigma^2$ fixed points}
With the twist condition \eqref{twist_con} and the projector \eqref{theta_phi_proj}, the supersymmetry transformations $\delta\psi_{\hat{\phi}i}$ and $\delta\psi_{\hat{\theta}i}$ lead to the same BPS equation as usually the case in performing topological twists. Setting $\epsilon_2=\epsilon_4=0$, we arrive at the following BPS equations  
\begin{eqnarray}
F'&=&\frac{1}{48g_2^2\Sigma^2}\left[4g_2^2\Sigma\left\{2h_1\cosh\phi_1(1+\cosh2\phi_3\cosh2\phi_5)\right.\right.\nonumber \\
& &
\left.+\sqrt{2}\Sigma^3(g_2-2g_1-g_2\cosh2\phi_3\cosh2\phi_5)\right\}\nonumber \\
& &\left. -8\kappa e^{-2G}\left\{\sqrt{2}a_3h_1+2g_2\Sigma^3(a_3\cosh\phi_1+a_6\sinh\phi_1)\right\}\right],\\
G'&=&\frac{1}{48g_2^2\Sigma^2}\left[4g_2^2\Sigma\left\{2h_1\cosh\phi_1(1+\cosh2\phi_3\cosh2\phi_5)\right.
\right.\nonumber \\
& &
\left.+\sqrt{2}\Sigma^3(g_2-2g_1-g_2\cosh2\phi_3\cosh2\phi_5)\right\} \nonumber \\
& &\left. +16\kappa e^{-2G}\left\{\sqrt{2}a_3h_1+2g_2\Sigma^3(a_3\cosh\phi_1+a_6\sinh\phi_1)\right\}\right],\\
\Sigma'&=&\frac{1}{6}h_1\cosh\phi_1(1+\cosh2\phi_3\cosh2\phi_5)+\frac{\sqrt{2}}{6}\Sigma^3(2g_1-g_2+g_2\cosh2\phi_3\cosh2\phi_5)\nonumber \\
& &+\frac{\kappa}{6g_2\Sigma}e^{-2G}\left[2\sqrt{2} a_3h_1-2 g_2\Sigma^3(a_3\cosh\phi_1+a_6\sinh\phi_1)\right],\\
\phi'_1&=&-h_1\Sigma^{-1}\sinh\phi_1(1+\cosh2\phi_3\cosh2\phi_5)\nonumber \\
& &-\kappa e^{-2G}\Sigma(a_3\sinh\phi_1+a_6\cosh\phi_1),\label{phi1_eq}\\
\phi'_3&=&\frac{1}{4}\Sigma^{-1}\textrm{sech}2\phi_5\sinh2\phi_3(\sqrt{2}g_2\Sigma^3-2h_1\cosh\phi_1),\\
\phi'_5&=&\frac{1}{4}\Sigma^{-1}\cosh2\phi_3\sinh2\phi_5(\sqrt{2}g_2\Sigma^3-2h_1\cosh\phi_1).
\end{eqnarray}
As in \cite{5D_flowII}, we have also used the $\gamma_{\hat{r}}$-projector of the form
\begin{equation}
\gamma_{\hat{r}}\epsilon_i=-i{I_i}^j\epsilon_j\label{gamma_r_proj}
\end{equation}
for ${I_i}^j=\sqrt{\frac{2}{3}}\frac{\Omega_{ik}A_1^{kj}}{|W|}$. It can be verified that these equations imply the second-order field equations.
\\
\indent An $AdS_3\times \Sigma^2$ fixed point is obtained from the conditions 
\begin{equation}
\phi'_1=\phi_3'=\phi_5'=\Sigma'=G'=0\qquad \textrm{and}\qquad F'=\frac{1}{L_{AdS_3}}\, .
\end{equation}
From the BPS equations, we find the following $AdS_3\times \Sigma^2$ fixed points:
\begin{eqnarray}
1. \qquad & &\phi_3=\phi_6=0,\qquad \phi_1=\frac{1}{2}\ln\left[\frac{(a_3-a_6)(a_3g_1-a_6g_2)}{(a_3+a_6)(a_3g_1+a_6g_2)}\right],\nonumber \\
& &\Sigma=\left(-\frac{\sqrt{2}a_3^2h_1(g_1+g_2)}{g_2\sqrt{(a_3^2-a_6^2)(a_3^2g_1^2-a_6^2g_2^2)}}\right)^{\frac{1}{3}}, \nonumber \\
& &
G=\frac{1}{2}\ln\left[-\frac{2\kappa a_3g_2(a_3^2-a_6^2)^2}{h_1(g_1+g_2)(a_3^2g_1^2-a_6^2g_2^2)}\right],\nonumber \\
& &L_{AdS_3}=\left(\frac{8\sqrt{2}a_3^2g_2^2(g_1+g_2)(a_3^2-a_6^2)(a_6^2g_2^2-a_3^2g_1^2)}{h_1^2(a_6^2g_2^2+g_1a_3^2(g_1+2g_2))^3}\right)^{\frac{1}{3}}\\
2. \qquad & &a_6=0,\quad \phi_1=0,\quad \Sigma=\left(\frac{\sqrt{2}h_1}{g_2}\right)^{\frac{1}{3}},\quad \phi_5=\phi^*_5=\textrm{constant},\nonumber \\
& &\phi_3=\frac{1}{2}\ln\left[\frac{2g_2-8g_1+\sqrt{2}\sqrt{32g_1^2-16g_1g_2-7g_2^2-9g_2^2\cosh4\phi_5}}{6g_2\cosh2\phi_5}\right],\nonumber \\ 
& &G=\frac{1}{2}\ln\left[\frac{3\kappa a_3}{(g_1-g_2)}\left(\frac{2g_2}{h_1}\right)^{\frac{1}{3}}\right],\qquad L_{AdS_3}=\frac{2}{(g_2-g_1)}\left(\frac{\sqrt{2}g_2^2}{h_1^2}\right)^{\frac{1}{3}}\\
3. \qquad & &\phi_1=\frac{1}{2}\ln\left[\frac{a_3+a_6}{a_3-a_6}\right],\quad \Sigma=\left(\frac{\sqrt{2}\kappa a_3h_1}{g_2\sqrt{a_3^2-a_6^2}}\right)^{\frac{1}{3}},\quad \phi_5=\phi^*_5=\textrm{constant},\nonumber \\
& &\phi_3=\frac{1}{2}\ln\left[\sqrt{4e^{4\phi_5}[a_3^2(4g_1-g_2)+a_6^2g_2]^2-g_2^2(3a_3^2+a_6^2)^2(1+e^{4\phi_5})^2}\right.\nonumber \\
& &\left. \phantom{\sqrt{a_3^2}} -2e^{2\phi_5}[a_3^2(4g_1-g_2)+a_6^2g_2]\right]-\frac{1}{2}\ln\left[g_2(3a_3^2+a_6^2)(1+e^{4\phi_5})\right],\nonumber \\
& &G=\frac{1}{6}\ln\left[\frac{2\kappa g_2(3a_3^2+a_6^2)^3}{a_3h_1(a_3^2-a_6^2)(g_1-g_2)^3}\right], \nonumber \\
& & L_{AdS_3}=\left(\frac{8\sqrt{2}g_2^2(a_3^2-a_6^2)}{a_3^2h_1^2(g_2-g_1)^3}\right)^{\frac{1}{3}}\, . 
\end{eqnarray}
It turns out that all of these fixed points only exist for $\kappa=-1$ leading to $AdS_3\times H^2$ geometries. For critical points $2$ and $3$, this is obviously seen by the reality of $G$ and the twist condition \eqref{twist_con}. It should also be pointed out that at critical point $2$, $a_6=0$ is required by consistency of equation \eqref{phi1_eq} for $\phi_1=0$.
\\
\indent Although in this work, we are mainly interested in the cases of $\Sigma^2=S^2$ and $\Sigma^2=H^2$, it could be useful to point out the possibility of obtaining black string solutions with $AdS_3\times \mathbb{R}^2$ near horizon geometry. Since in this case, the $\Sigma^2=\mathbb{R}^2$ is flat, there is no need to perform a twist. A similar analysis leading to the twist condition \eqref{twist_con} would require that $a_3=0$ which also implies $a_0=0$. Therefore, the possible solutions will not be charged under any subgroup of the $SU(4)_R\sim SO(6)_R$ R-symmetry. In this case, only $A^6$ gauge field corresponding to an $SO(2)$ subgroup of the $SO(3)$ flavor symmetry in the dual four-dimensional SCFTs has a non-trivial magnetic flux along $\Sigma^2=\mathbb{R}^2$. Setting $a_3=0$ in the above construction of solutions with $SO(2)_{\textrm{diag}}\times SO(2)$ twist, we have not found any $AdS_3\times \mathbb{R}^2$ vacua from the resulting BPS equations. It is possible that different twists such as turning on $SO(2)_D\times SO(2)\times SO(2)\subset SO(2)_D\times SO(3)\times SO(3)$ gauge fields could lead to $AdS_3\times \mathbb{R}^2$ solutions. However, we will not further consider this type of solutions in the present paper.                 
\\
\indent Before giving explicit solutions interpolating between supersymmetric $AdS_5$ vacua and these $AdS_3\times H^2$ fixed points, we first note the unbroken supersymmetry of the solutions. Due to the projectors \eqref{extra_proj}, \eqref{theta_phi_proj} and \eqref{gamma_r_proj}, the black string solutions preserve $\frac{16}{2^3}=2$ supercharges. Using the relation among $SO(1,4)$ gamma matrices $i=\gamma_{\hat{0}}\gamma_{\hat{1}}\gamma_{\hat{r}}\gamma_{\hat{\theta}}\gamma_{\hat{\phi}}$, we find the chirality matrix on the two-dimensional Minkowski space
\begin{equation}
\gamma_*\epsilon_i=\gamma_{\hat{0}}\gamma_{\hat{1}}\epsilon_i=-i\gamma_{\hat{r}}\gamma_{\hat{\theta}}\gamma_{\hat{\phi}}=\pm (\epsilon_1,0,\epsilon_3,0)
\end{equation}
in which we have used the projector \eqref{theta_phi_proj} and an explicit form of the $\gamma_{\hat{r}}$-projector \eqref{gamma_r_proj}, $\gamma_{\hat{r}}\epsilon_i=\pm {(\sigma_2\otimes \sigma_3)_i}^j\epsilon_j$. We then see that the Killing spinors $\epsilon_1$ and $\epsilon_3$ have definite two-dimensional chiralities, so the flow solutions will preserve $N=(2,0)$ or $N=(0,2)$ Poincare supersymmetry in two dimensions. At the $AdS_3\times H^2$ fixed points, the supersymmetry is enhanced to $4$ supercharges since the $\gamma_{\hat{r}}$-projector is not necessary for constant scalars. This results in $N=(2,0)$ or $N=(0,2)$ superconformal symmetry in two dimensions. Accordingly, the aforementioned $AdS_3\times H^2$ fixed points are dual to two-dimensional $N=(2,0)$ or $N=(0,2)$ SCFTs.

\subsection{Supersymmetric black string solutions}
\begin{figure}
         \centering
         \begin{subfigure}[b]{0.45\textwidth}
                 \includegraphics[width=\textwidth]{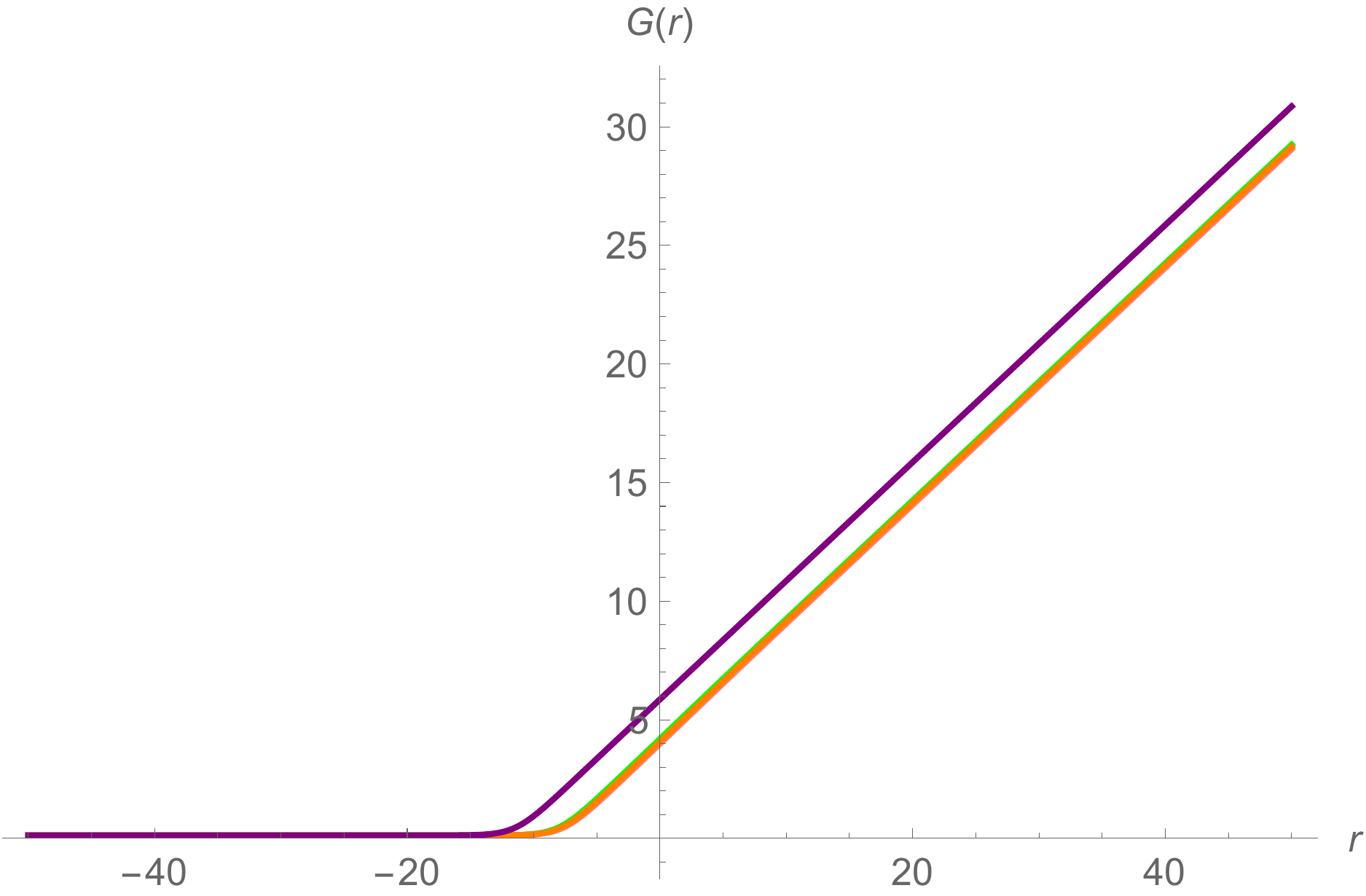}
                 \caption{Solution for $G(r)$}
         \end{subfigure} \qquad
\begin{subfigure}[b]{0.45\textwidth}
                 \includegraphics[width=\textwidth]{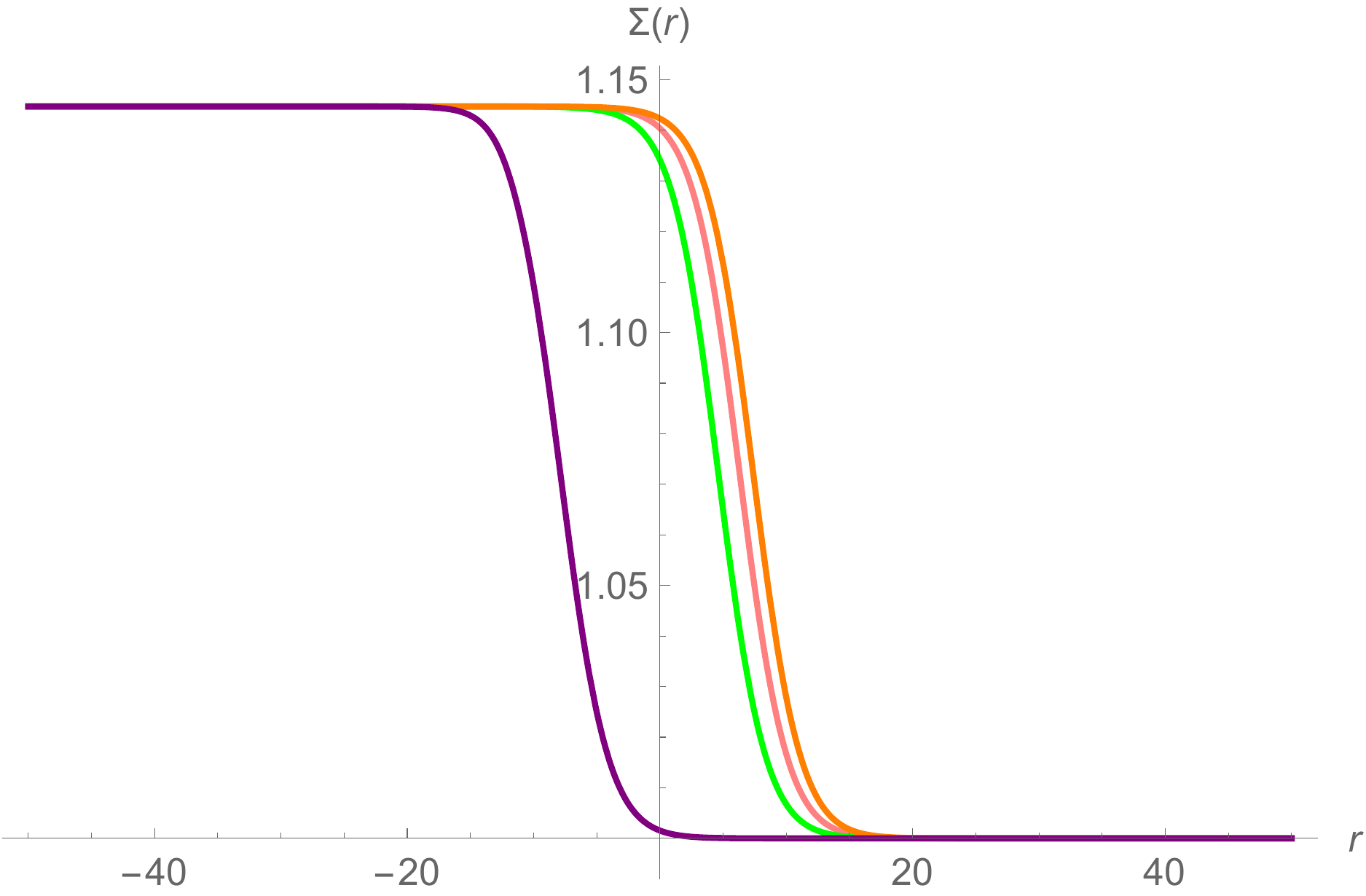}
                 \caption{Solution for $\Sigma(r)$}
         \end{subfigure}\\
         \begin{subfigure}[b]{0.45\textwidth}
                 \includegraphics[width=\textwidth]{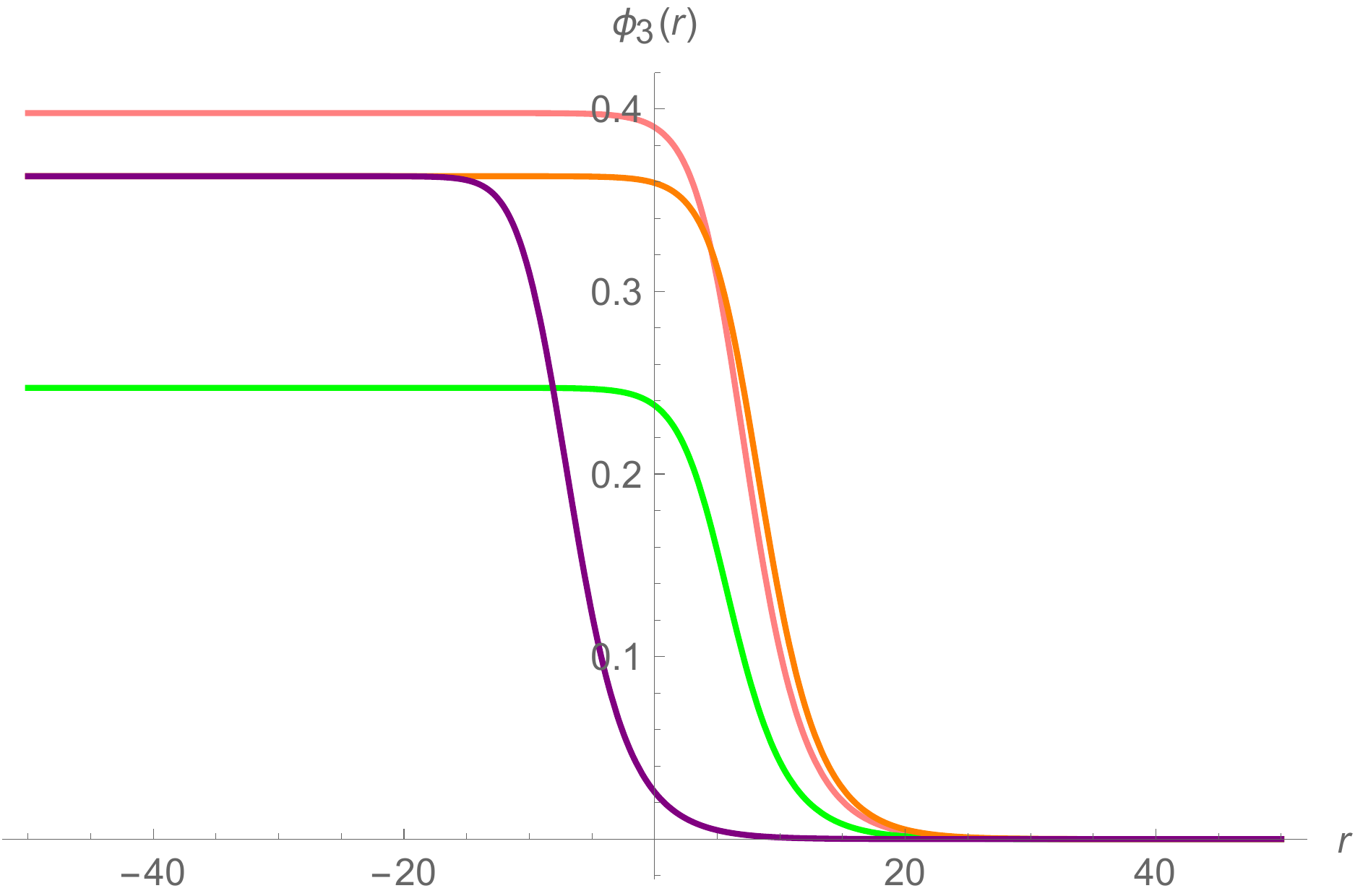}
                 \caption{Solution for $\phi_3(r)$}
         \end{subfigure}\qquad 
         \begin{subfigure}[b]{0.45\textwidth}
                 \includegraphics[width=\textwidth]{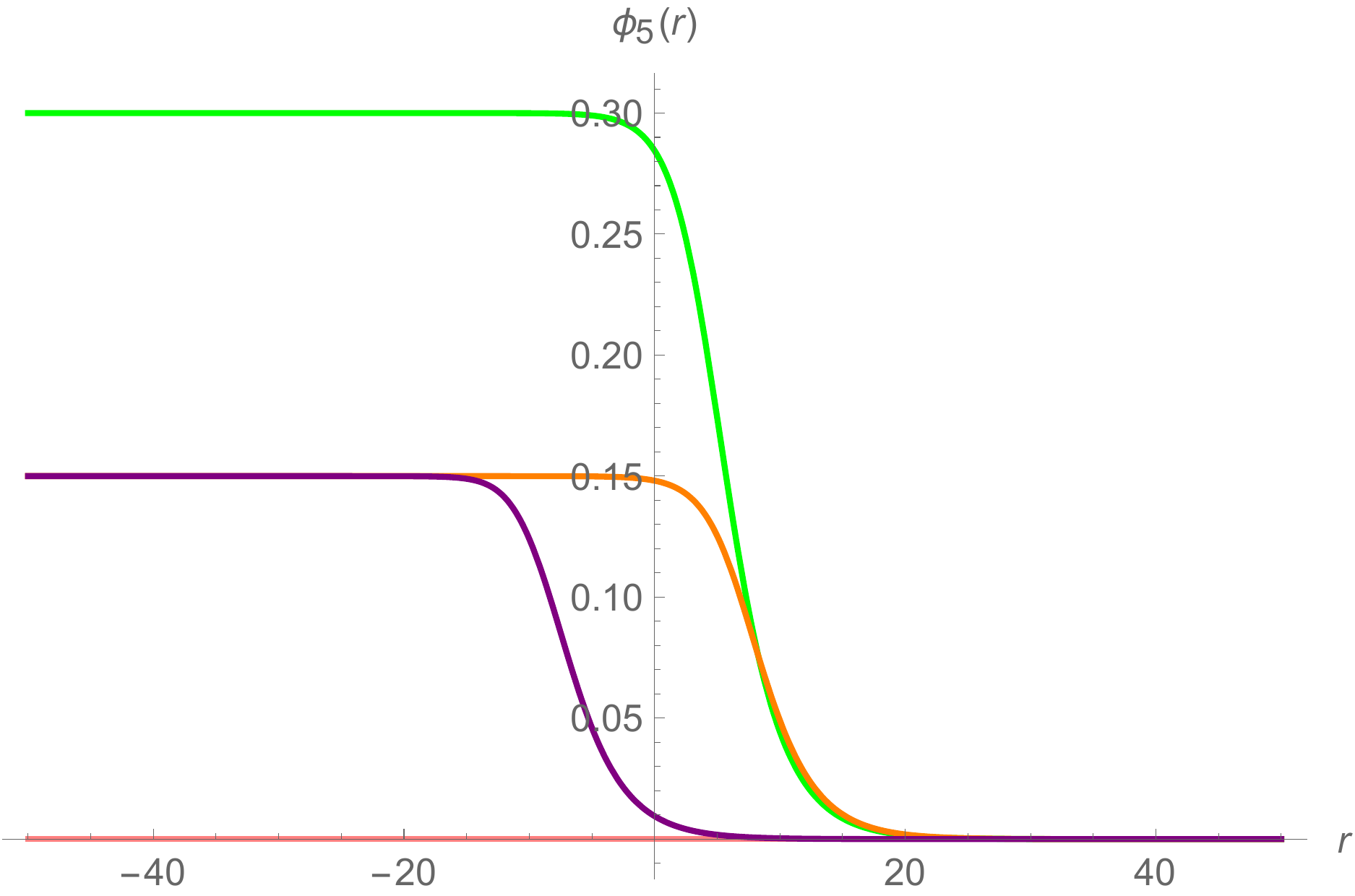}
                 \caption{Solution for $\phi_5(r)$}
         \end{subfigure}\\
          \begin{subfigure}[b]{0.55\textwidth}
                 \includegraphics[width=\textwidth]{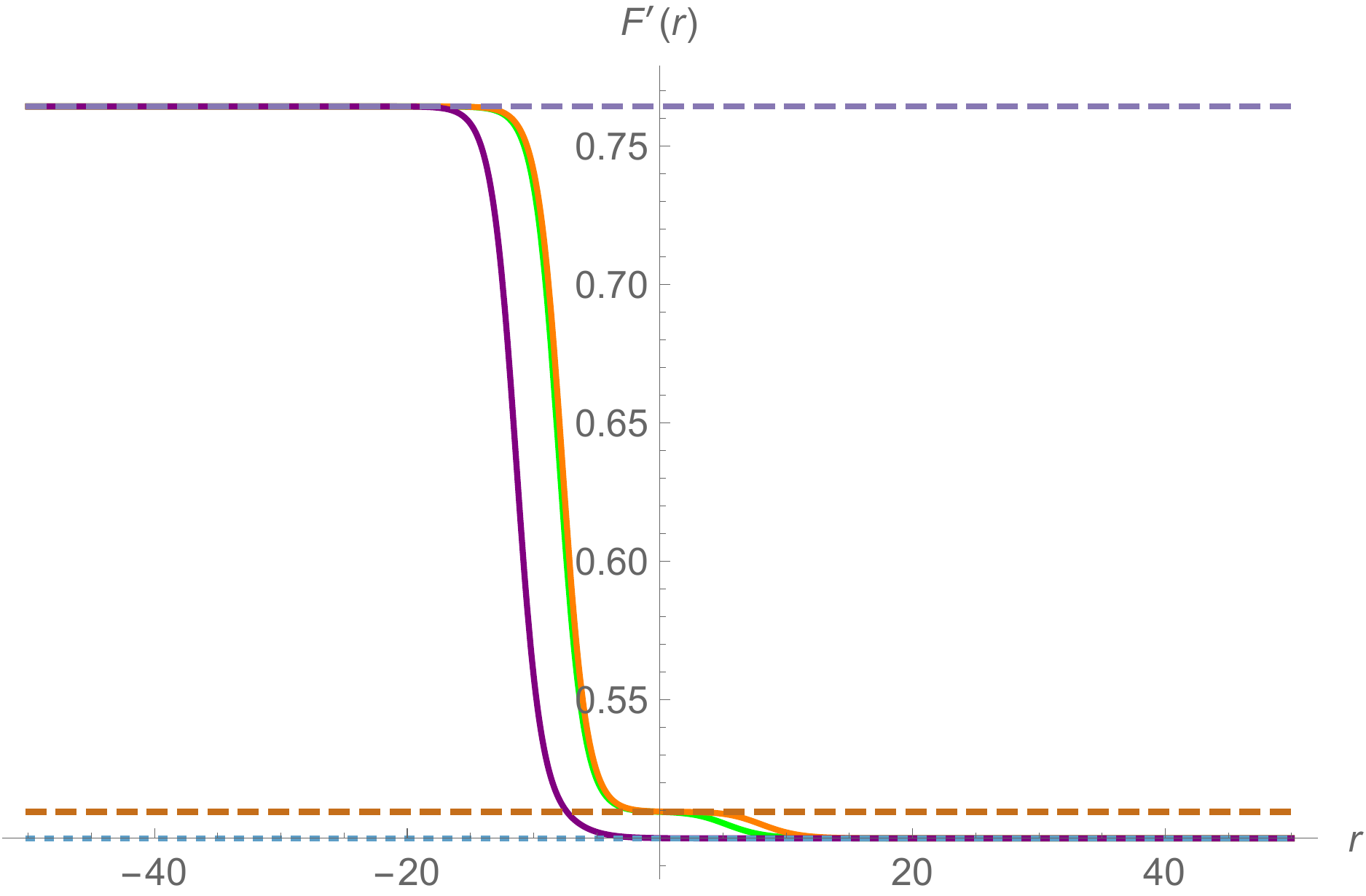}
                 \caption{Solution for $F'(r)$}
         \end{subfigure} 
         \caption{Examples of RG flows from $N=4$ and $N=2$ $AdS_5$ critical points to $N=(2,0)$ $AdS_3\times H^2$ fixed point $2$ with $SO(2)_{\textrm{diag}}\times SO(2)$ symmetry in the IR with $\phi_1=0$, $a_6=0$, $h_1=1$, $\rho=\frac{3}{2}$ and $\zeta=\frac{1}{2}$ for $\phi^*_5=0$ (pink), $\phi^*_5=0.15$ (orange and purple) and $\phi^*_5=0.30$ (green).}\label{fig1}
 \end{figure}

We now find solutions to the BPS equations that interpolate between supersymmetric $AdS_5$ critical points and $AdS_3\times H^2$ fixed points identified previously. We begin with a simple solution interpolating between the $N=2$ $AdS_5$ vacuum to $AdS_3\times H^2$ fixed point $2$. It should be noted that $\Sigma$, $\phi_1$ and $\phi_3$ take the same values at both of these critical points. We can then truncate these scalar fields out by setting them to the values at the critical points together with $a_6=0$. For $\kappa=-1$, the remaining BPS equations are simply given by
\begin{eqnarray}
F'&=&\frac{(g_2-g_1)}{3}\left(\frac{h_1^2}{\sqrt{2}g_2^2}\right)^{\frac{1}{3}}+\frac{1}{2} a_3\left(\frac{\sqrt{2}h_1}{g_2}\right)^{\frac{1}{3}}e^{-2G},\\
G'&=&\frac{(g_2-g_1)}{3}\left(\frac{h_1^2}{\sqrt{2}g_2^2}\right)^{\frac{1}{3}}- a_3\left(\frac{\sqrt{2}h_1}{g_2}\right)^{\frac{1}{3}}e^{-2G}\, .
  \end{eqnarray}  
Accordingly, the flow from $N=2$ $AdS_5$ vacuum to $AdS_3\times H^2$ critical point $2$ is driven only by the metric function $G(r)$. The above equations can be readily solved by
\begin{eqnarray}
G&=&\frac{1}{2}\ln\left[\frac{e^{\frac{2(r-r_0)}{L_5}}+6a_3g_2^{\frac{1}{3}}}{2^{\frac{2}{3}}(g_2-g_1)h_1^{\frac{1}{3}}}\right],\qquad L_5=\frac{3}{(g_2-g_1)}\left(\frac{\sqrt{2}g_2^2}{h_1^2}\right)^{\frac{1}{3}},\nonumber \\
F&=&\frac{3r}{2L_5}-\frac{1}{4}\ln\left[e^{\frac{2(r-r_0)}{L_5}}+6 a_3g_2^{\frac{1}{3}}\right]
\end{eqnarray}
in which we have removed an additive integration constant in $F(r)$. The constant $r_0$ can also be removed by shifting the radial coordinate $r$. From this solution, we immediately see that as $r\rightarrow \infty$, 
\begin{equation}
F\sim G\sim \frac{r}{L_5}
\end{equation}
which gives locally asymptotically $N=2$ $AdS_5$ critical point. On the other hand, for $r\rightarrow -\infty$, we find that the solution becomes $AdS_3\times H^2$ fixed point $2$
\begin{equation}
G\sim \frac{1}{2}\ln\left[\frac{3a_3}{g_2-g_1}\left(\frac{2g_2}{h_1}\right)^{\frac{1}{3}}\right] \qquad  \textrm{and}\qquad F\sim \frac{(g_2-g_1)}{2}\left(\frac{h_1^2}{\sqrt{2}g_2^2}\right)^{\frac{1}{3}}r\, .
 \end{equation} 
\indent For more general solutions with non-vanishing scalars, we need to solve the BPS equations numerically to find the solutions. We follow \cite{5D_flowII} to fix numerical values of various parameters by defining 
\begin{equation}
g_1=-\frac{h_1}{\sqrt{2}},\qquad g_2=\frac{\sqrt{2}h_1}{\rho},\qquad h_2=\frac{h_1}{\zeta}\, .
\end{equation}
Examples of solutions for $\phi_1=0$, $a_6=0$, $h_1=1$, $\zeta=\frac{1}{2}$ and $\rho=\frac{3}{2}$ are given in figure \ref{fig1}. In the figure, we have given two types of solutions. A solution interpolating between the $N=4$ $AdS_5$ critical point to the $AdS_3\times H^2$ fixed point with $\phi^*_5=0.15$ is given by the purple line. There are also solutions that flow from the $N=4$ $AdS_5$ critical point to $N=2$ $AdS_5$ vacuum and then proceed to $AdS_3\times H^2$ fixed point $2$ in the IR. Examples of these solutions for three different values of $\phi^*_5=0.00,0.15,0.30$ at the latter two fixed points are given by pink, orange and green lines, respectively. The dashed lines in the solutions for $F'(r)$ indicate the values of $\frac{1}{L_{AdS_5}}$ and $\frac{1}{L_{AdS_3}}$ at different critical points. 
\\
\indent To obtain these solutions, we have performed numerical integrations from the $AdS_3\times H^2$ vacua in the IR to the $AdS_5$ critical points in the UV. Generic choices of boundary conditions lead to the first type of solutions interpolating between the $N=4$ $AdS_5$ and $AdS_3\times H^2$ fixed points. On the other hand, by tuning the boundary conditions very close to the $AdS_3\times H^2$ fixed points, we find the second type of solutions interpolating between $N=4$ and $N=2$ $AdS_5$ vacua and $AdS_3\times H^2$ geometries in the IR.    
 
\begin{figure}
         \centering
         \begin{subfigure}[b]{0.45\textwidth}
                 \includegraphics[width=\textwidth]{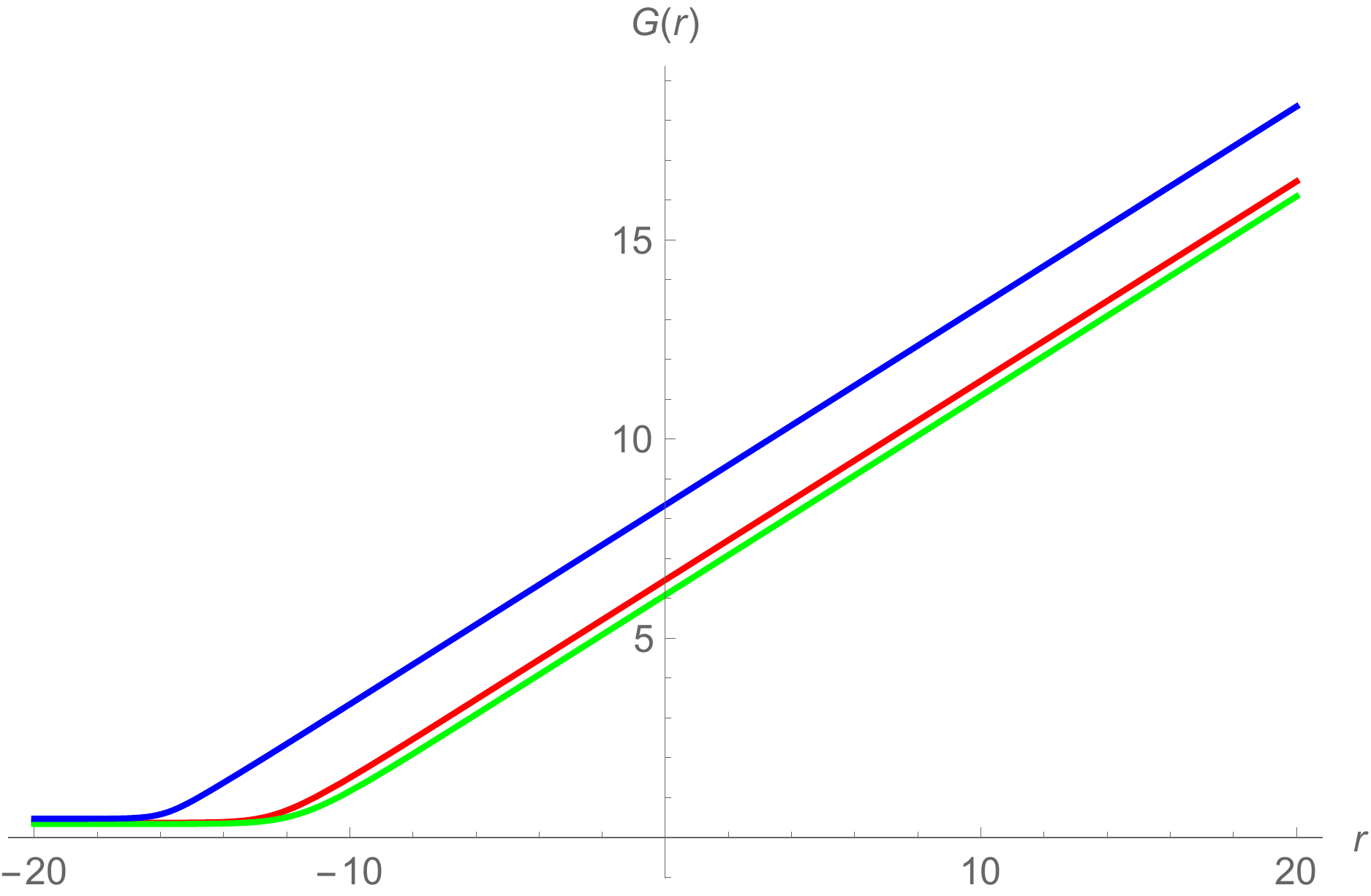}
                 \caption{Solution for $G(r)$}
         \end{subfigure} \qquad
\begin{subfigure}[b]{0.45\textwidth}
                 \includegraphics[width=\textwidth]{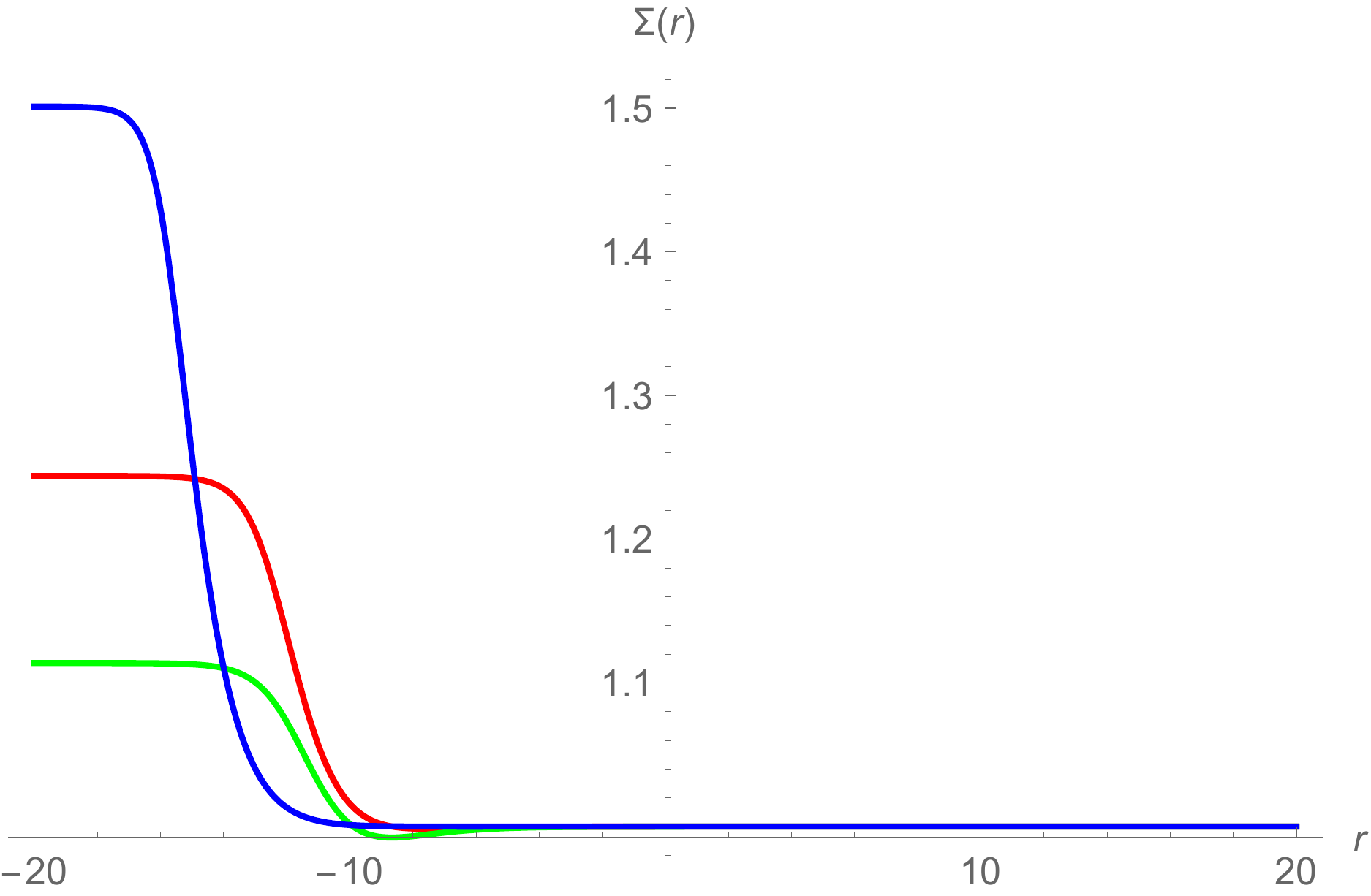}
                 \caption{Solution for $\Sigma(r)$}
         \end{subfigure}\\
         \begin{subfigure}[b]{0.45\textwidth}
                 \includegraphics[width=\textwidth]{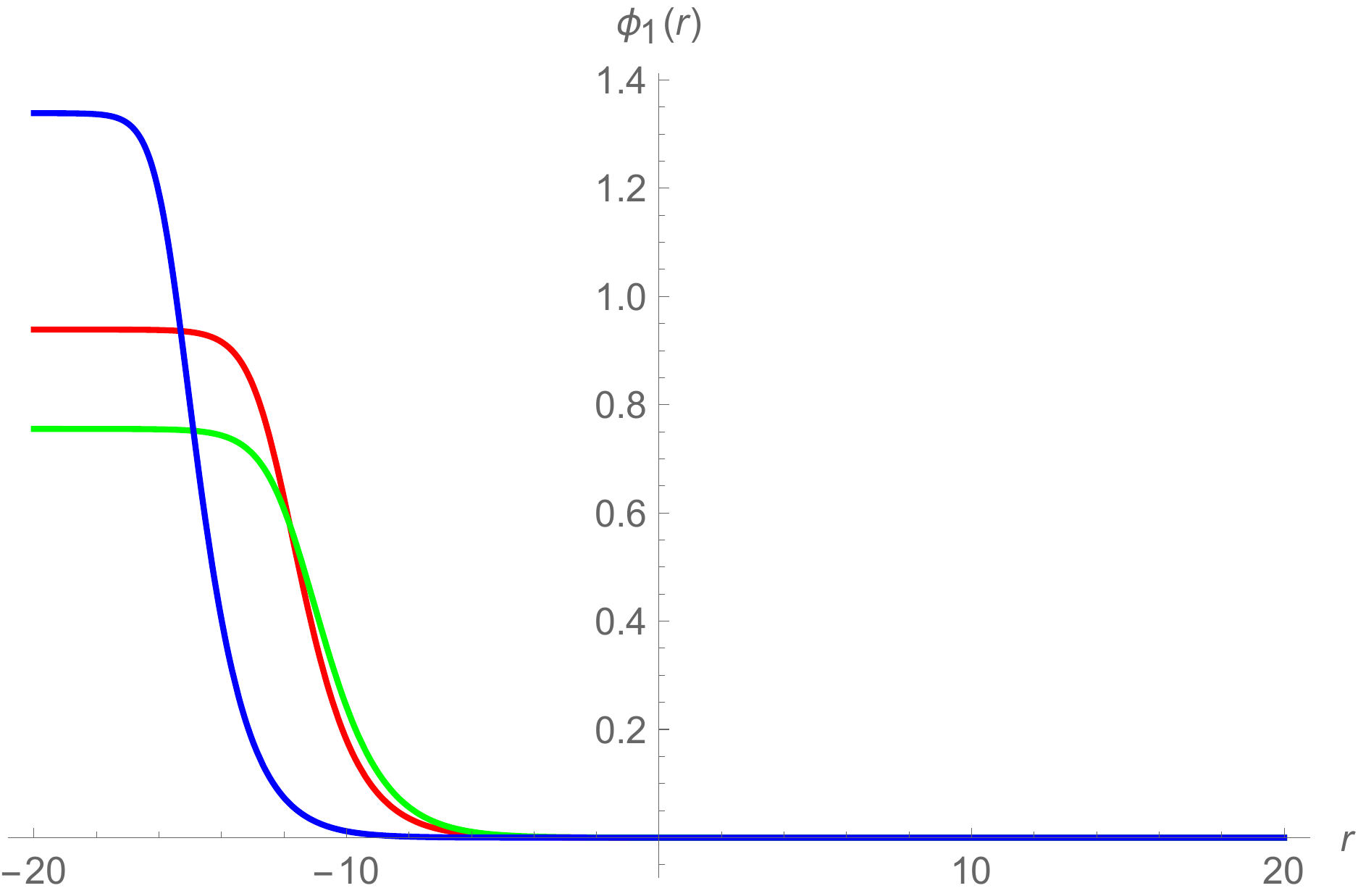}
                 \caption{Solution for $\phi_1(r)$}
         \end{subfigure}\qquad 
          \begin{subfigure}[b]{0.45\textwidth}
                 \includegraphics[width=\textwidth]{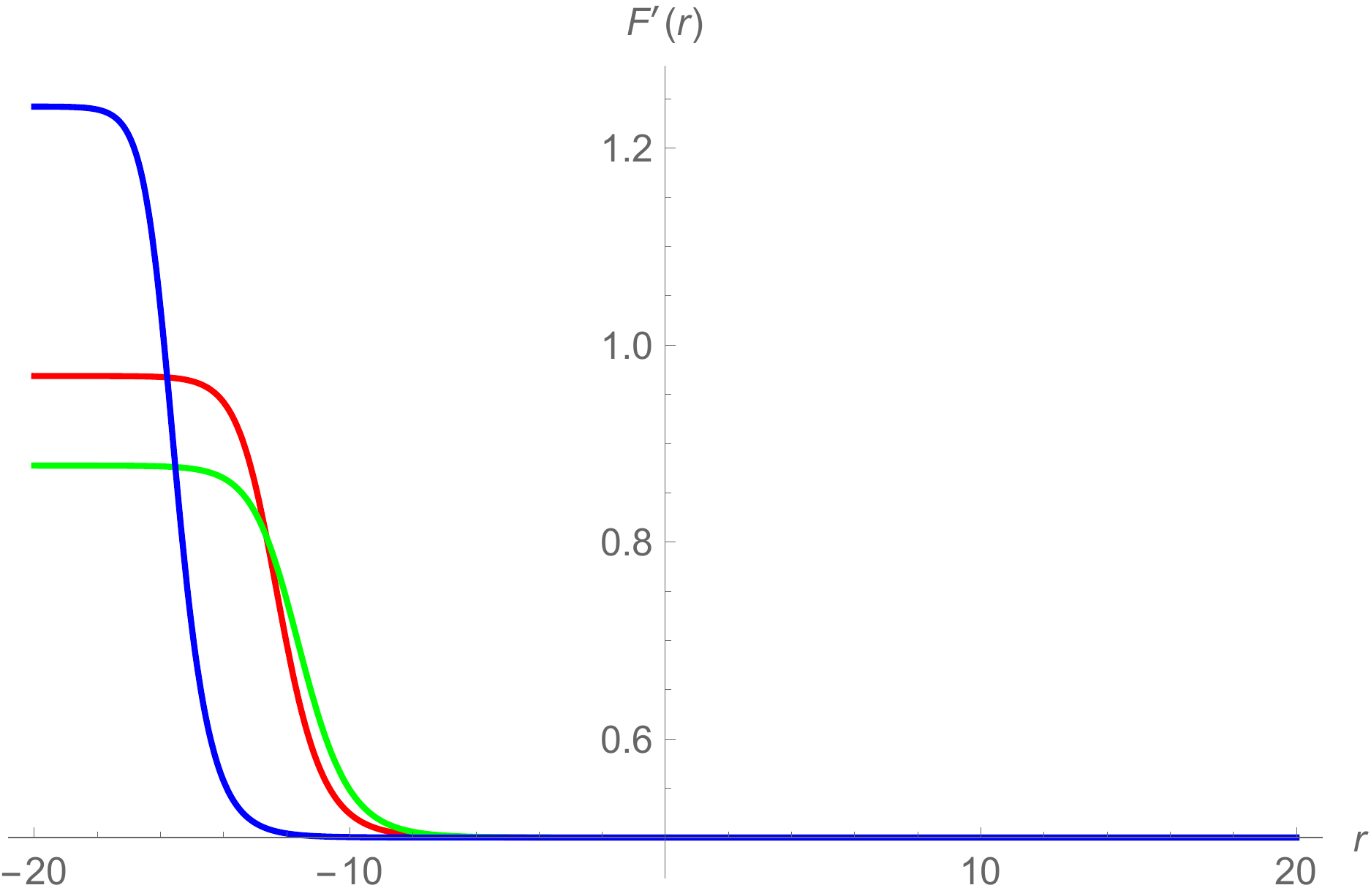}
                 \caption{Solution for $F'(r)$}
         \end{subfigure} 
         \caption{Examples of RG flows from $N=4$ $AdS_5$ critical point to $N=(2,0)$ $AdS_3\times H^2$ fixed point $1$ in the IR with $\phi_3=\phi_5=0$, $h_1=1$, $\rho=4$ and $\zeta=\frac{1}{2}$ for $a_6=0.70$ (blue), $a_6=0.75$ (red) and $a_6=0.80$ (green).}\label{fig2}
 \end{figure}

We now consider solutions flowing to $AdS_3\times H^2$ critical point $1$ with only $\phi_1$ non-vanishing. For $\phi_3=\phi_5=0$, the truncated BPS equations only admit the $N=4$ $AdS_5$ critical point as an asymptotic geometry. Therefore, in this case, there are only solutions interpolating between this $AdS_5$ critical point and the $AdS_3\times H^2$ geometry in the IR. Examples of these solutions for different values of $a_6=0.70,0.75,0.80$ are given in figure \ref{fig2}.    

\begin{figure}
         \centering
         \begin{subfigure}[b]{0.45\textwidth}
                 \includegraphics[width=\textwidth]{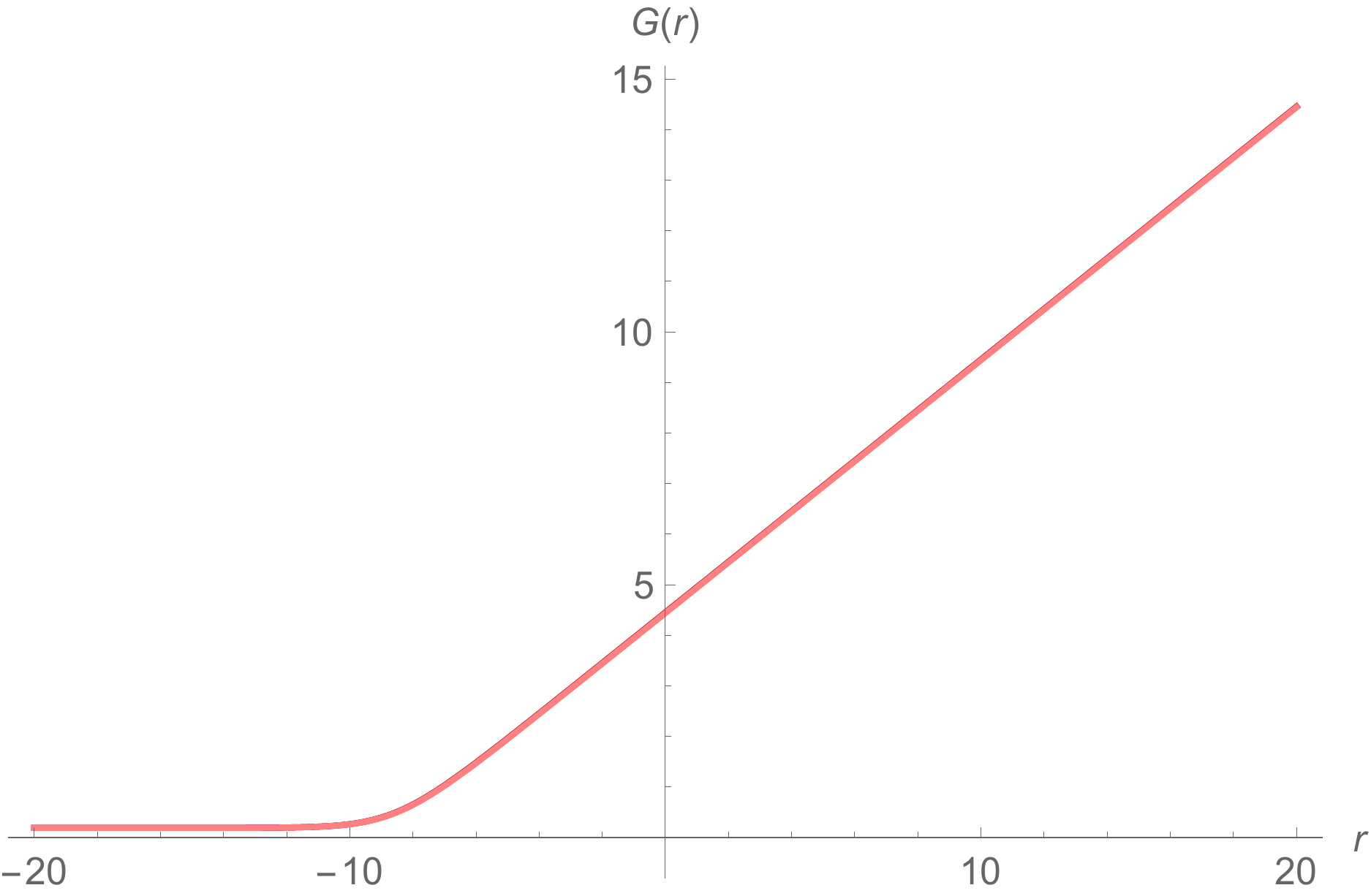}
                 \caption{Solution for $G(r)$}
         \end{subfigure} \qquad
\begin{subfigure}[b]{0.45\textwidth}
                 \includegraphics[width=\textwidth]{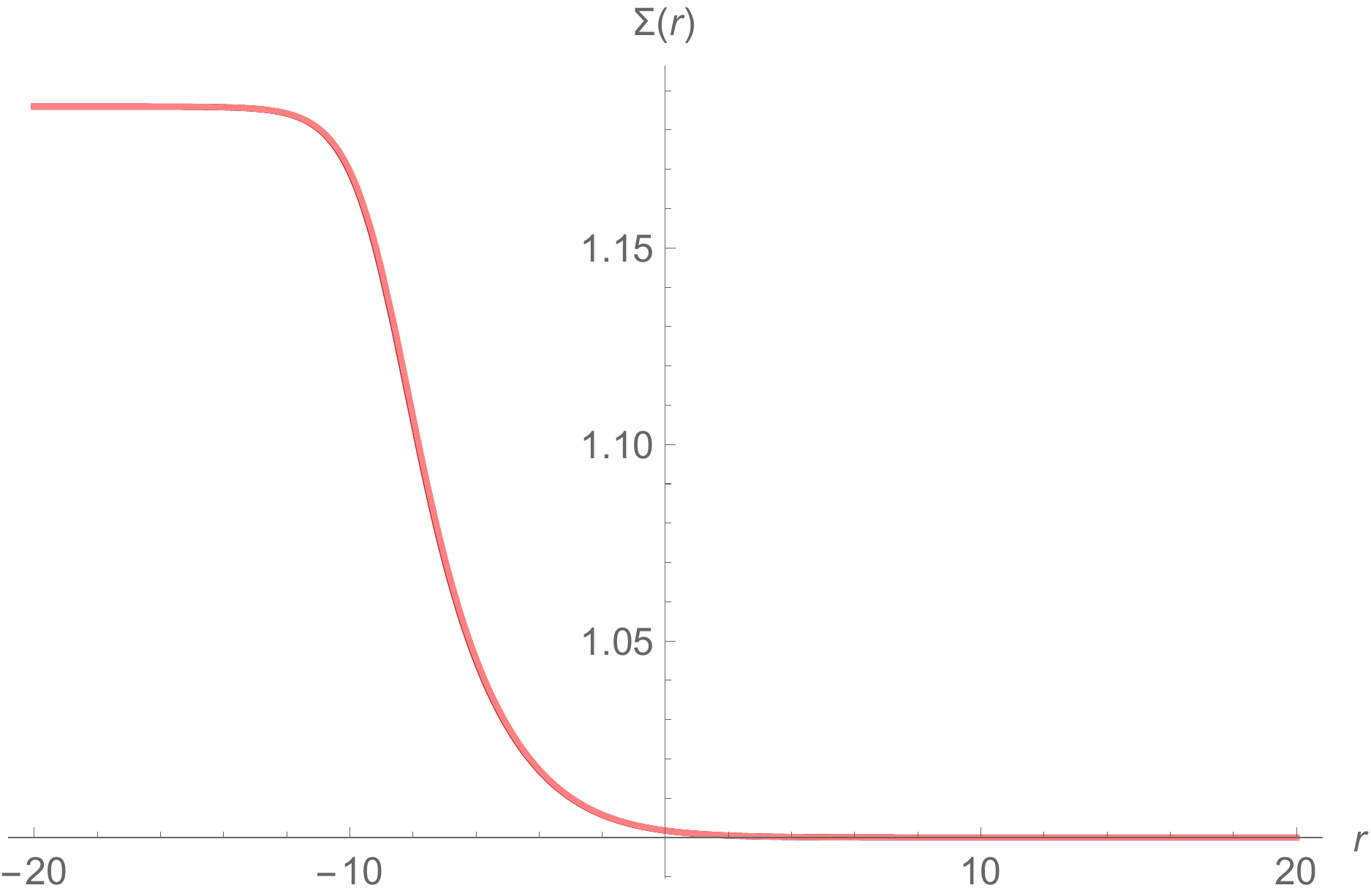}
                 \caption{Solution for $\Sigma(r)$}
         \end{subfigure}\\
         \begin{subfigure}[b]{0.45\textwidth}
                 \includegraphics[width=\textwidth]{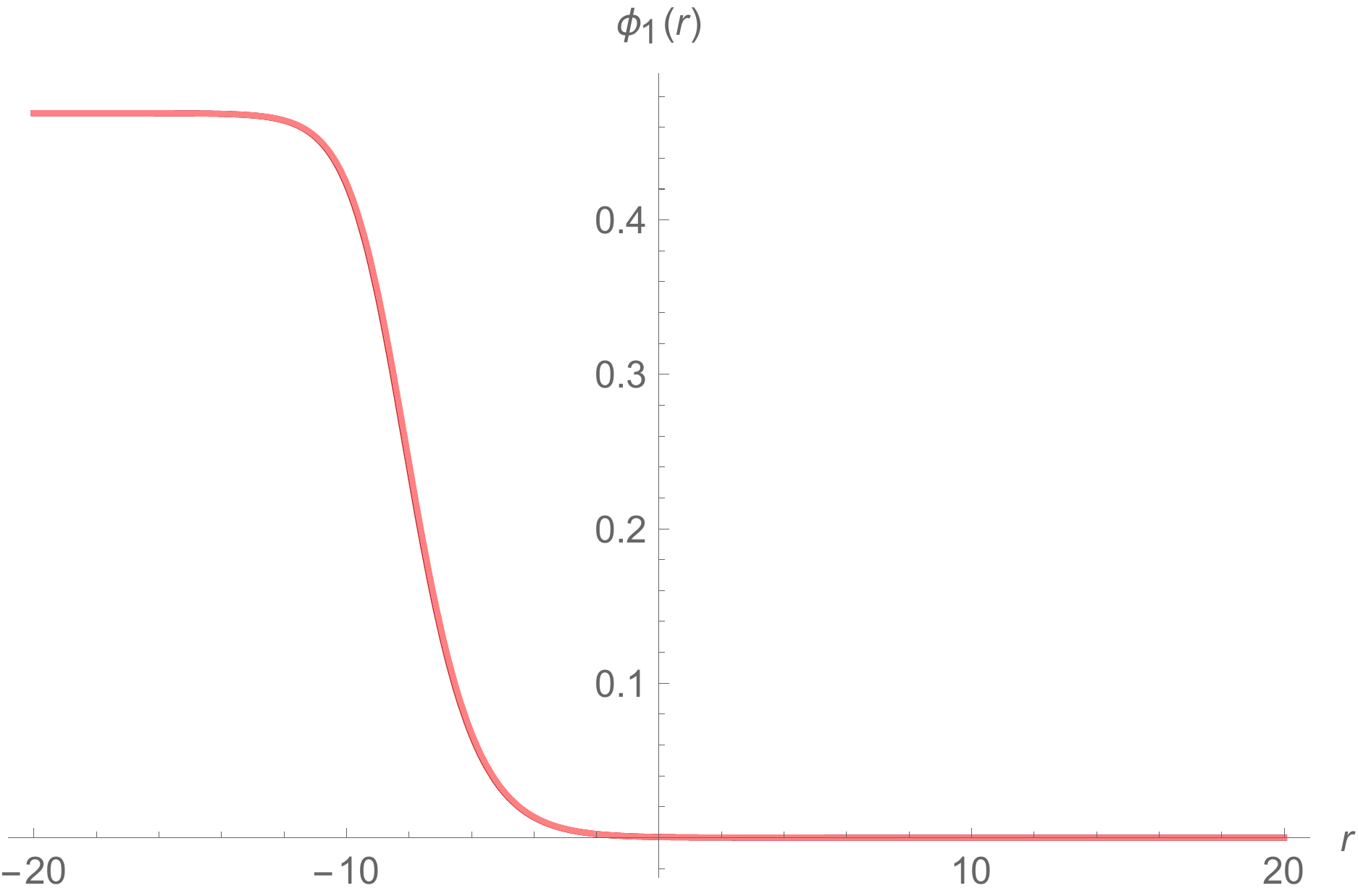}
                 \caption{Solution for $\phi_1(r)$}
         \end{subfigure}\qquad 
         \begin{subfigure}[b]{0.45\textwidth}
                 \includegraphics[width=\textwidth]{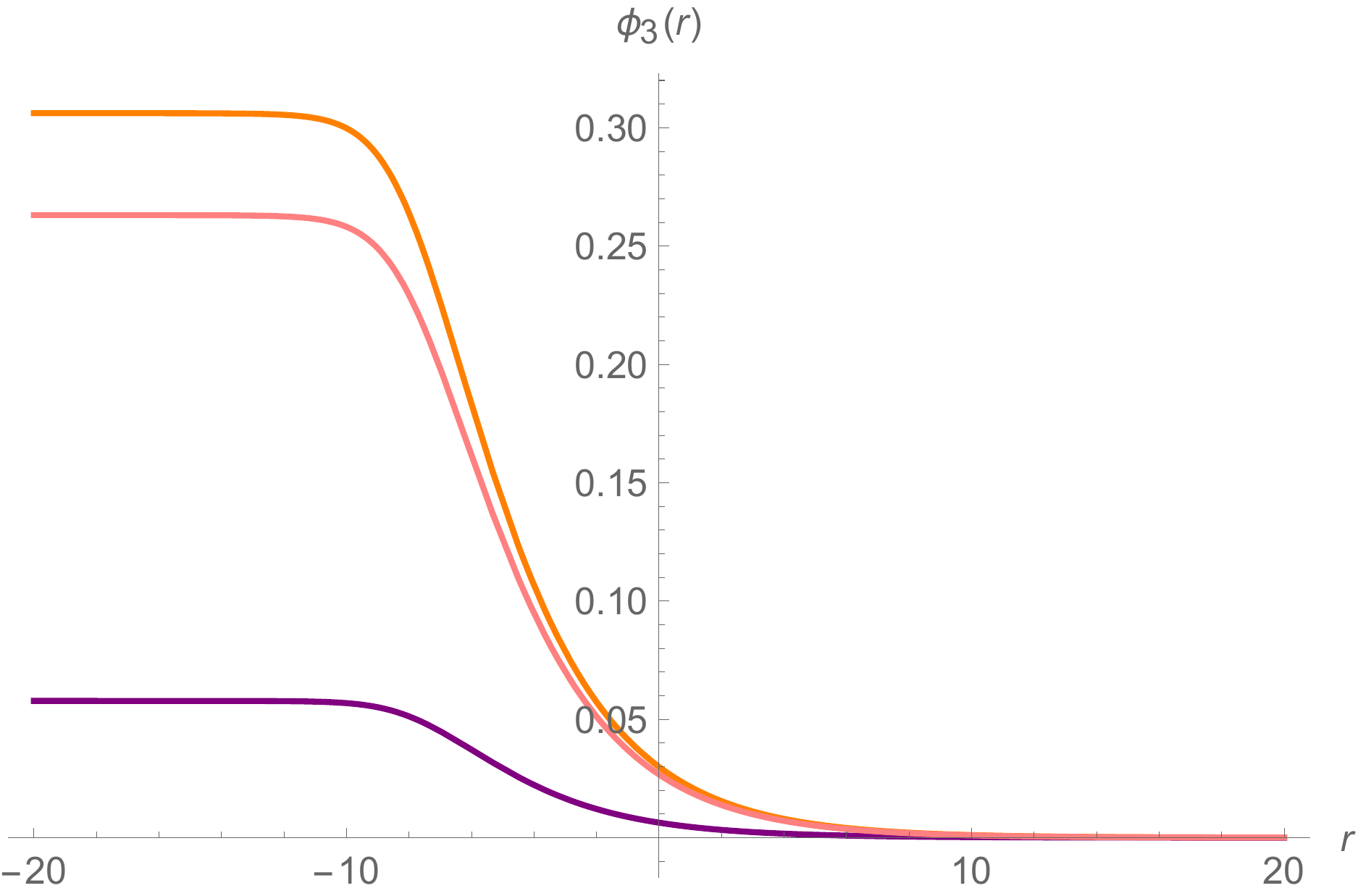}
                 \caption{Solution for $\phi_3(r)$}
         \end{subfigure}\\
         \begin{subfigure}[b]{0.45\textwidth}
                 \includegraphics[width=\textwidth]{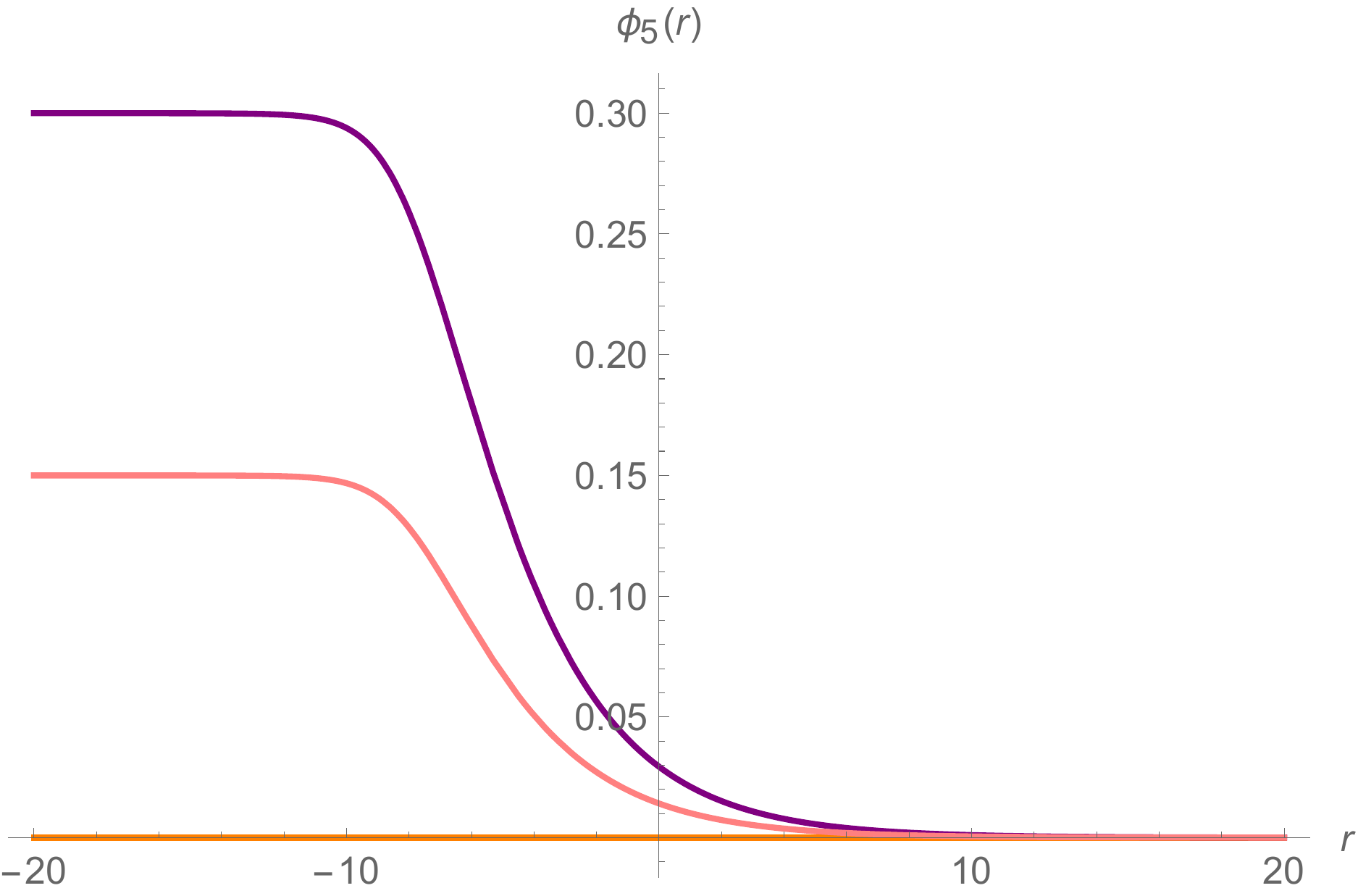}
                 \caption{Solution for $\phi_5(r)$}
         \end{subfigure}\qquad
                    \begin{subfigure}[b]{0.45\textwidth}
                 \includegraphics[width=\textwidth]{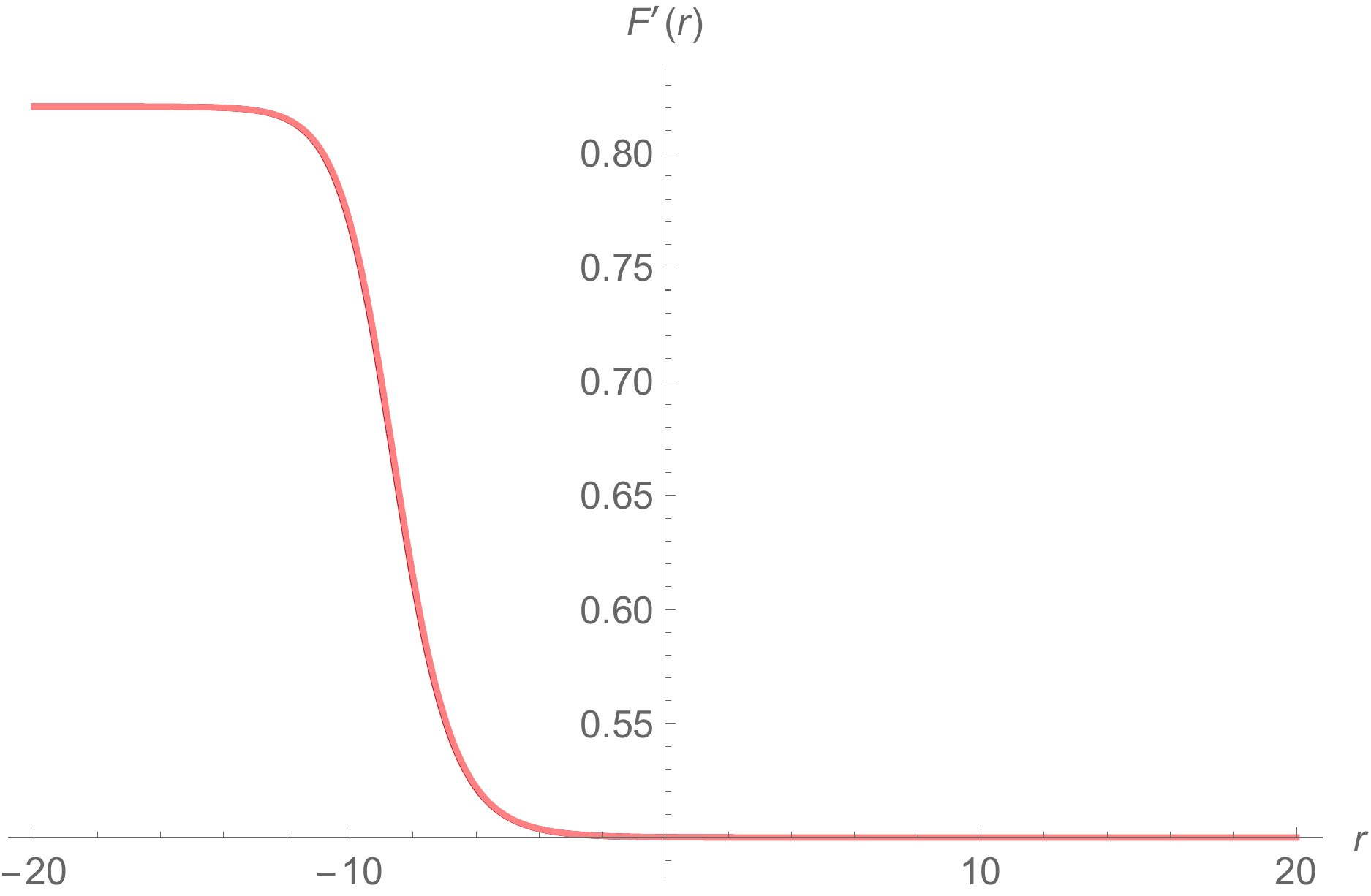}
                 \caption{Solution for $F'(r)$}
         \end{subfigure} 
         \caption{Examples of RG flows from $N=4$ $AdS_5$ critical point to $N=(2,0)$ $AdS_3\times H^2$ fixed point $3$ in the IR with $h_1=1$, $\rho=\frac{3}{2}$, $\zeta=\frac{1}{2}$ and $a_6=\frac{1}{4}$ for $\phi^*_5=0$ (orange), $\phi^*_5=0.15$ (pink) and $\phi^*_5=0.30$ (purple).}\label{fig3}
 \end{figure}

Finally, we look for more complicated solutions flowing to $AdS_3\times H^2$ fixed point $3$ with all scalars non-vanishing. Examples of solutions interpolating between $N=4$ $AdS_5$ critical point and this $AdS_3\times H^2$ fixed point are given in figure \ref{fig3}. It should be noted that at the $AdS_3\times H^2$ fixed point, only the value of $\phi_3$ is affected by the value of $\phi^*_5$. In addition, the solutions in the figure indicate that apart from the solutions for $\phi_3$ and $\phi_5$, the entire flow solutions for other fields are not affected by different values of $\phi^*_5$ at the $AdS_3\times H^2$ fixed point.  

All of these solutions describe black strings with a near horizon geometry of the form $AdS_3\times H^2$ in asymptotically locally $AdS_5$ spaces. Holographically, the solutions can also be considered as holographic RG flows from the four-dimensional $N=2$ and $N=1$ SCFTs to two-dimensional $N=(2,0)$ SCFTs in the IR.
 
\section{Supersymmetric $AdS_5$ black strings with $SO(2)_{\textrm{diag}}$ symmetry}\label{SO2diag_twist}
In this section, we repeat the same analysis for a smaller residual symmetry $SO(2)_{\textrm{diag}}$ that is the diagonal subgroup of $SO(2)_D\times SO(2)_R\times SO(2)$ generated by $X_0$, $X_3$ and $X_6$. As pointed out in \cite{5D_flowII}, there are nine singlet scalars under $SO(2)_{\textrm{diag}}$ corresponding to the non-compact generators
\begin{eqnarray}
& &\hat{Y}_1=Y_{31},\phantom{-Y_{31}}\qquad \hat{Y}_2=Y_{42}+Y_{53},\qquad \hat{Y}_3=Y_{44}+Y_{55},\nonumber \\
& &\hat{Y}_4=Y_{43}-Y_{52},\qquad \hat{Y}_5=Y_{45}-Y_{54},\qquad \tilde{Y}_1=Y_{12}+Y_{23},\nonumber \\
& &\tilde{Y}_2=Y_{13}-Y_{22},\qquad \tilde{Y}_3=Y_{14}+Y_{25},\qquad \tilde{Y}_4=Y_{15}-Y_{24}\, . 
\end{eqnarray}
The coset representative is then given by
\begin{equation}
\mc{V}=e^{\phi_1\hat{Y}_1}e^{\phi_2\hat{Y}_2}e^{\phi_3\hat{Y}_3}e^{\phi_4\hat{Y}_4}e^{\phi_5\hat{Y}_5}
e^{\varphi_1\tilde{Y}_1}e^{\varphi_2\tilde{Y}_2}e^{\varphi_3\tilde{Y}_3}e^{\varphi_4\tilde{Y}_4}\, .\label{coset_rep2}
\end{equation}
It turns out that the analysis in this case is much more complicated than the previous case of $SO(2)_{\textrm{diag}}\times SO(2)$ symmetry. To proceed further, we will consider a subtruncation of this sector with $\varphi_1=\varphi_2=\varphi_3=\varphi_4=0$. However, consistency among the BPS equations and compatibility between the BPS equations and the field equations further require that $\phi_4=\phi_5=0$. The resulting truncation is accordingly the same as that studied in \cite{5D_flowII} with only $\phi_1$, $\phi_2$ and $\phi_3$ non-vanishing.  
\\
\indent The ansatz for the metric and $(A^0, A^3,A^6)$ gauge fields are the same as in the previous section. To implement the $SO(2)_{\textrm{diag}}$ symmetry, in this case, we impose the following conditions
\begin{equation}
A^0=\frac{h_1}{g_2}A^3\qquad \textrm{and}\qquad A^6=\frac{h_1}{h_2}A^3\, . 
\end{equation}
The twist can be performed in the same way as in the previous section with the twist condition \eqref{twist_con} and projectors \eqref{extra_proj} and \eqref{theta_phi_proj}.

\subsection{Supersymmetric $AdS_5$ vacua}
Since the scalar sector considered here is the same as in \cite{5D_flowII}, the $AdS_5$ vacua are given by those identified in \cite{5D_flowII}. We will only give the superpotential and supersymmetric $AdS_5$ critical points but refer to \cite{5D_flowII} for the explicit form of the scalar potential.
\\
\indent The superpotential is given by
\begin{eqnarray}
W&=&\frac{1}{3}\Sigma^{-1}\cosh^2\phi_3\left(h_1\cosh\phi_1\cosh^2\phi_2+h_2\sinh\phi_1\sinh^2\phi_2\right)\nonumber \\
& &+\frac{\sqrt{2}}{12}(g_2- 2g_1-g_2\cosh2\phi_3)\Sigma^2
\end{eqnarray}
which admits the following $AdS_5$ critical points:
\begin{itemize}
\item I. The trivial critical point, at the origin of $SO(5,5)/SO(5)\times SO(5)$, is given by
\begin{equation}
\phi_1=\phi_2=\phi_3=0,\qquad \Sigma=-\left(\frac{h_1}{\sqrt{2}g_1}\right)^{\frac{1}{3}},\qquad V_0=-3\left(\frac{g_1h_1^2}{2}\right)^{\frac{2}{3}}\, .
\end{equation}
As in the previous section, this critical point preserves the full $SO(2)_D\times SO(3)\times SO(3)$ gauge symmetry and $N=4$ supersymmetry. 

\item II. Unlike the previous case of $SO(2)_{\textrm{diag}}\times SO(2)$ symmetry, there is another $N=4$ $AdS_5$ critical point given by
\begin{eqnarray}
& &\phi_1=\phi_2=\frac{1}{2}\ln\left[\frac{h_2-h_1}{h_2+h_1}\right],\qquad \Sigma=-\left(\frac{h_1h_2}{\sqrt{2}g_1\sqrt{h_2^2-h_1^2}}\right)^{\frac{1}{3}},\nonumber \\
& &\phi_3=0,\qquad V_0=-\frac{3}{2}\left(\frac{\sqrt{2}g_1h_1^2h_2^2}{h_2^2-h_1^2}\right)^{\frac{2}{3}}\, .
\end{eqnarray}
This critical point preserves $SO(2)_D\times SO(3)_{\textrm{diag}}$ symmetry. 

\item III. The next critical point is given by
\begin{eqnarray}
& &\phi_1=\phi_2=0,\qquad \phi_3=\frac{1}{2}\ln\left[\frac{g_2-4g_1+ 2\sqrt{4g_1^2-2g_1g_2-2g_2^2}}{3g_2}\right],\nonumber \\
& & \Sigma=-\left(\frac{\sqrt{2}h_1}{g_2}\right)^{\frac{1}{3}},\qquad V_0=-\frac{1}{3}(g_1-g_2)^2\left(\frac{\sqrt{2}h_1}{g_2}\right)^{\frac{4}{3}}\, .\label{PointIII}
\end{eqnarray}
This critical point preserves $N=2$ supersymmetry and $SO(2)_{\textrm{diag}}\times SO(3)$ symmetry.

\item IV. There is another $N=2$ supersymmetric critical point given by
\begin{eqnarray}
& &\phi_1=\phi_2=\frac{1}{2}\ln\left[\frac{h_2-h_1}{h_2+h_1}\right],\qquad \Sigma=\left(\frac{\sqrt{2}h_1h_2}{g_2\sqrt{h_2^2-h_1^2}}\right)^{\frac{1}{3}},\nonumber \\
& &\phi_3=\frac{1}{2}\ln\left[\frac{g_2-4g_1+ 2\sqrt{4g_1^2-2g_1g_2-2g_2^2}}{3g_2}\right],\nonumber \\
& & V_0=-\frac{2}{3}(g_1-g_2)^2\left(\frac{h_1^2h_2^2}{\sqrt{2}g_2^2(h_2^2-h_1^2)}\right)^{\frac{2}{3}}
\end{eqnarray}
which is invariant under $SO(2)_{\textrm{diag}}$.
 \end{itemize}  
\indent As pointed out in \cite{5D_flowII}, some of these critical points appear in pairs with some sign differences. Since critical points related by these sign changes are physically equivalent, in the above equations, we have chosen a particular sign choice for definiteness.

\subsection{Supersymmetric $AdS_3\times \Sigma^2$ fixed points}
With the same anlysis of supersymmetry transformations of fermionic fields, we obtain the BPS equations
\begin{eqnarray}
F'&=&\frac{1}{12g_2h_2\Sigma^2}\left[\sqrt{2}g_2h_2\Sigma^4(g_2-2g_1-g_2\cosh2\phi_3) +4g_2h_2\cosh^2\phi_3\times\right.\nonumber \\
& &(h_1\cosh\phi_1\cosh^2\phi_2+h_2\sinh\phi_1\sinh^2\phi_2)\Sigma\nonumber \\
& &\left. -2\kappa a_3e^{-2G}\left\{2g_2\Sigma^3(h_2\cosh\phi_1+h_1\sinh\phi_1)+\sqrt{2}h_1h_2 \right\}\right],\\
G'&=&\frac{1}{12g_2h_2\Sigma^2}\left[\sqrt{2}g_2h_2\Sigma^4(g_2-2g_1-g_2\cosh2\phi_3) +4g_2h_2\cosh^2\phi_3\times\right.\nonumber \\
& &(h_1\cosh\phi_1\cosh^2\phi_2+h_2\sinh\phi_1\sinh^2\phi_2)\Sigma\nonumber \\
& &\left. +4\kappa a_3e^{-2G}\left\{2g_2\Sigma^3(h_2\cosh\phi_1+h_1\sinh\phi_1)+\sqrt{2}h_1h_2 \right\}\right],\\
\Sigma'&=&\frac{1}{3}\cosh^2\phi_3\left(h_1\cosh\phi_1\cosh^2\phi_2+h_2\sinh\phi_1\sinh^2\phi_2\right)+\frac{\sqrt{2}}{6}\Sigma^3(g_2\cosh2\phi_3+2g_1\nonumber \\
& &-g_2)+\frac{1}{3g_2h_2\Sigma}\kappa a_3e^{-2G}\left[\sqrt{2}h_1h_2-2g_2\Sigma^3(h_2\cosh\phi_1+h_1\sinh\phi_1)\right],\\
\phi_1'&=&-\Sigma^{-1}\cosh^2\phi_3\left(h_1\cosh^2\phi_2\sinh\phi_1+h_2\cosh\phi_1\sinh^2\phi_2\right)\nonumber \\
& &-\frac{\kappa a_3}{h_2}e^{-2G}\Sigma(h_1\cosh\phi_1+h_2\sinh\phi_1),\\
\phi_2'&=&-\Sigma^{-1}\cosh\phi_2\sinh\phi_2(h_1\cosh\phi_1+h_2\sinh\phi_1),\\
\phi_3'&=&\frac{1}{4}\Sigma^{-1}\sinh2\phi_3\left[\sqrt{2}g_2\Sigma^3-2h_1\cosh\phi_1\cosh^2\phi_2-2h_2\sinh\phi_1\sinh^2\phi_2\right].
\end{eqnarray}
From these equations, we find the following $AdS_3\times \Sigma^2$ fixed points:
\begin{eqnarray}
i&:&\quad \phi_2=\phi_3=0,\quad \phi_1=\frac{1}{2}\ln\left[\frac{(h_2-h_1)(g_1h_2-g_2h_1)}{(h_2+h_1)(g_2h_1+g_1h_2)}\right],\qquad\qquad\qquad\nonumber \\
& &\quad \Sigma=\left(-\frac{\sqrt{2}(g_1+g_2)h_1h_2^2}{g_2\sqrt{(h_2^2-h_1^2)(g_1^2h_2^2-g_2^2h_1^2)}}\right)^{\frac{1}{3}}, \nonumber \\
& &\quad G=\frac{1}{2}\ln \left[\frac{2^{\frac{1}{3}}\kappa a_3 (g_1+g_2)h_1h_2^2}{g_2(g_2^2h_1^2-g_1^2h_2^2)}\right],\nonumber \\
& &\quad L_{AdS_3}=\left(\frac{8\sqrt{2}g_2^2h_2^2(g_1+g_2)(h_2^2-h_1^2)(g^2_2h_1^2-g_1^2h_2^2)}{h_1^2[g_2^2h_1^2+g_1^2h_2^2(g_1+2g_2)]^3}\right)^{\frac{1}{3}},
\end{eqnarray}
\begin{eqnarray}  
ii&:&\quad \phi_3=0,\quad \phi_1=\phi_2=\frac{1}{2}\ln\left[\frac{h_2-h_1}{h_2+h_1}\right],\quad \Sigma=\left(-\frac{\sqrt{2}h_1h_2(g_1+g_2)}{g_1g_2\sqrt{h_2^2-h_1^2}}\right)^{\frac{1}{3}},\nonumber \\
& &\quad G=\frac{1}{6}\ln\left[-\frac{2g_2\kappa a_3^3(h_2^2-h_1^2)^2}{g_1^2h_1h_2^4(g_1+g_2)}\right],\nonumber \\
& &\quad L_{AdS_3}=\left(-\frac{8\sqrt{2}g_2^2(g_1+g_2)(h_2^2-h_1^2)}{g_1h_1^2h_2^2(g_1+2g_2)^3}\right)^{\frac{1}{3}},
\end{eqnarray}
\begin{eqnarray}  
iii&:&\quad \phi_2=0,\quad \phi_1=\frac{1}{2}\ln\left[\frac{h_2+h_1}{h_2-h_1}\right],\quad \Sigma=\left(\frac{\sqrt{2}h_1h_2}{g_2\sqrt{h_2^2-h_1^2}}\right)^{\frac{1}{3}},\nonumber \\
& &\quad \phi_3=\frac{1}{2}\ln\left[\frac{g_2(h_2^2-h_1^2)-4g_1h_2^2+2h_2\sqrt{2(g_1-g_2)[g_2h_1^2+h_2^2(2g_1+g_2)]}}{g_2(h_1^2+3h_2^2)}\right], \nonumber \\
& &\quad G=\frac{1}{6}\ln\left[\frac{2\kappa a_3^3g_2(h_1^2+3h_2^2)^3}{h_1h_2^4(g_1-g_2)^3(h_2^2-h_1^2)}\right],\nonumber \\
& &\quad L_{AdS_3}=\frac{6}{g_2-g_1}\left(1+\sqrt{\frac{4h_2^2}{h_2^2-h_1^2}}\right)^{-1}\left(\frac{\sqrt{2}g_2^2(h_2^2-h_1^2)}{h_1^2h_2^2}\right)^{\frac{1}{3}},
\end{eqnarray}
\begin{eqnarray}  
iv&:&\quad \phi_1=\phi_2=\frac{1}{2}\ln\left[\frac{h_2-h_1}{h_2+h_1}\right],\quad \Sigma=\left(\frac{\sqrt{2}h_1h_2}{g_2\sqrt{h_2^2-h_1^2}}\right)^{\frac{1}{3}},\quad\qquad\qquad\qquad\nonumber \\
& &\quad \phi_3=\frac{1}{2}\ln\left[\frac{g_2-4g_1+2\sqrt{4g_1^2-2g_1g_2-2g_2^2}}{3g_2}\right],\nonumber \\
& &\quad G=\frac{1}{6}\ln\left[\frac{54\kappa a_3^3g_2(h_2^2-h_1^2)^2}{(g_1-g_2)^3h_1h_2^4}\right],\nonumber \\
& &\quad L_{AdS_3}=\frac{2}{g_2-g_1}\left(\frac{\sqrt{2}g_2^2(h_2^2-h_1^2)}{h_1^2h_2^2}\right)^{\frac{1}{3}}\, .
\end{eqnarray}      
Among these fixed points, only $ii$ leads to $AdS_3\times S^2$ geometry. All the remaining fixed points correspond to $AdS_3\times H^2$ solutions.               
            
\subsection{Supersymmetric black string solutions}      
\begin{figure}
         \centering
               \begin{subfigure}[b]{0.45\textwidth}
                 \includegraphics[width=\textwidth]{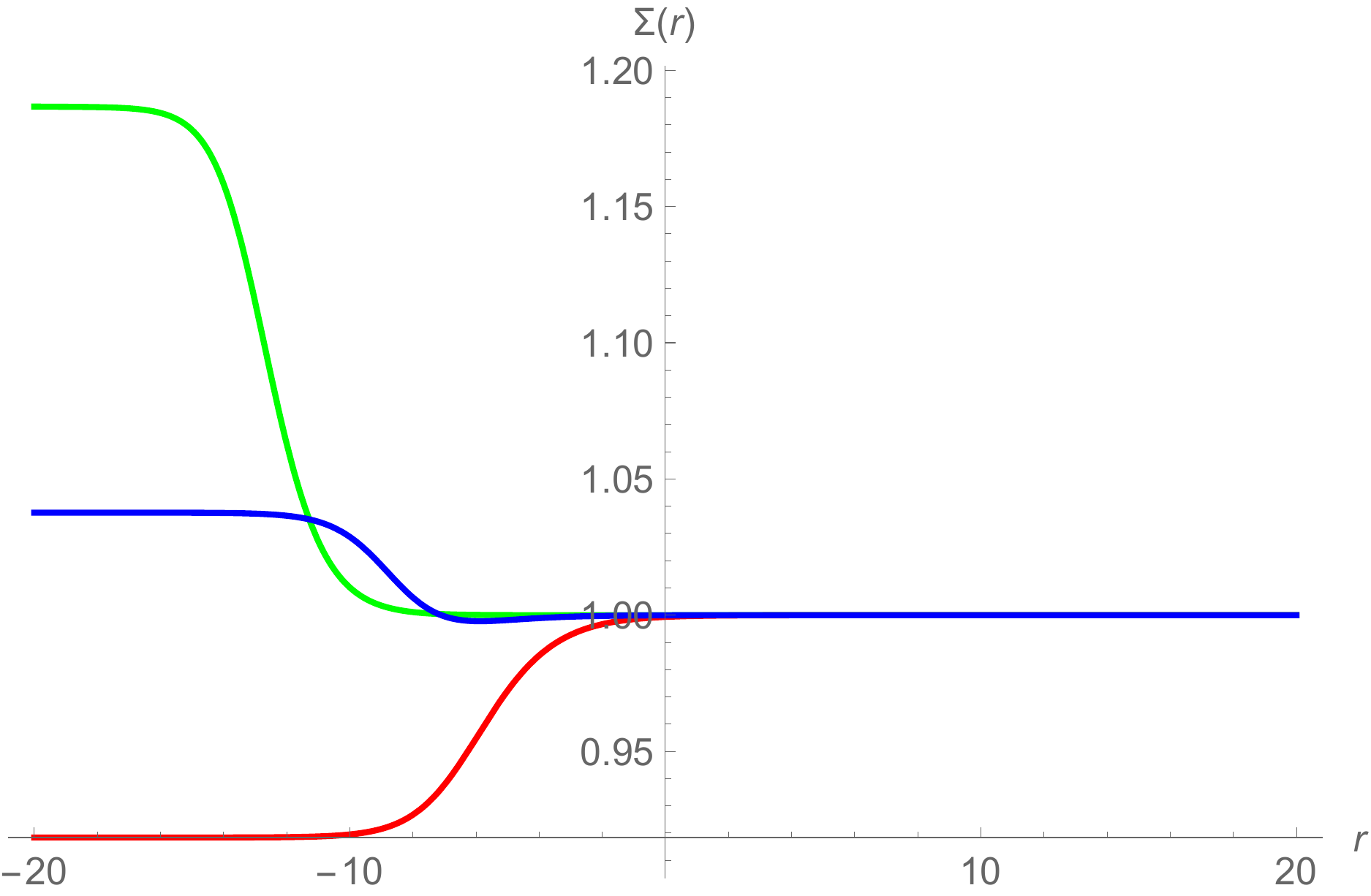}
                 \caption{Solution for $\Sigma(r)$}
         \end{subfigure}
         \begin{subfigure}[b]{0.45\textwidth}
                 \includegraphics[width=\textwidth]{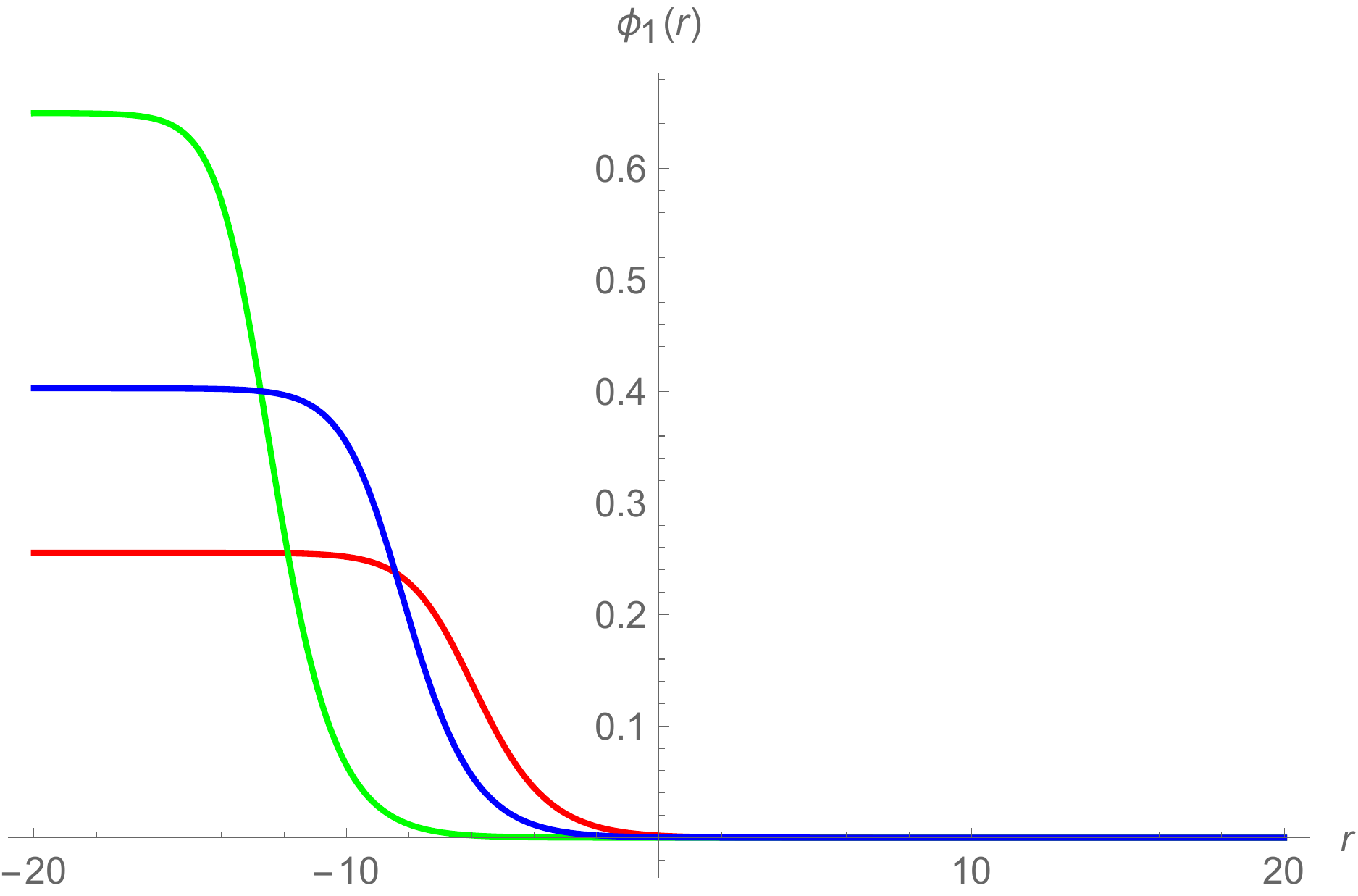}
                 \caption{Solution for $\phi_1(r)$}
         \end{subfigure}\\
          \begin{subfigure}[b]{0.45\textwidth}
                 \includegraphics[width=\textwidth]{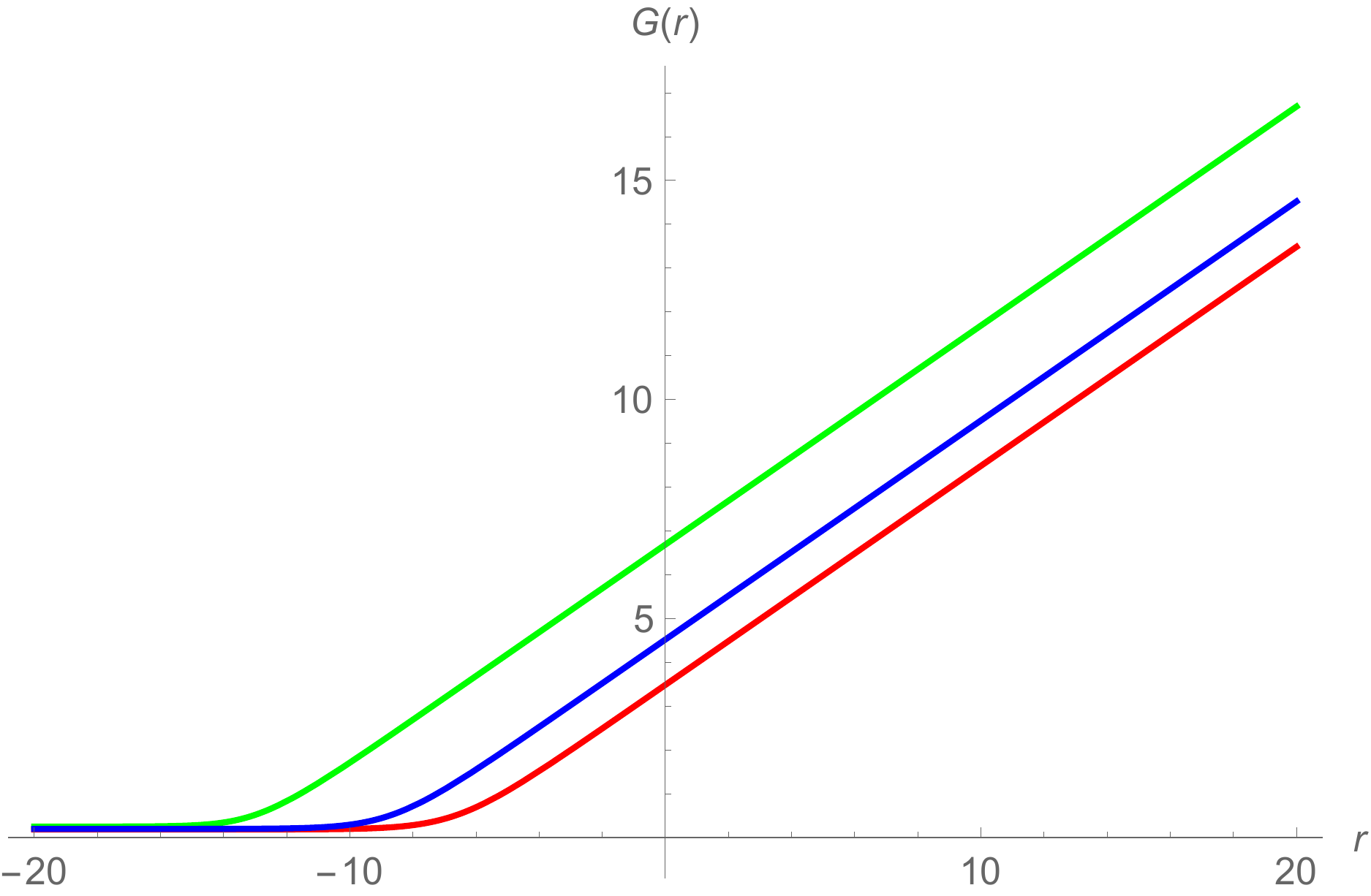}
                 \caption{Solution for $G(r)$}
         \end{subfigure}
          \begin{subfigure}[b]{0.45\textwidth}
                 \includegraphics[width=\textwidth]{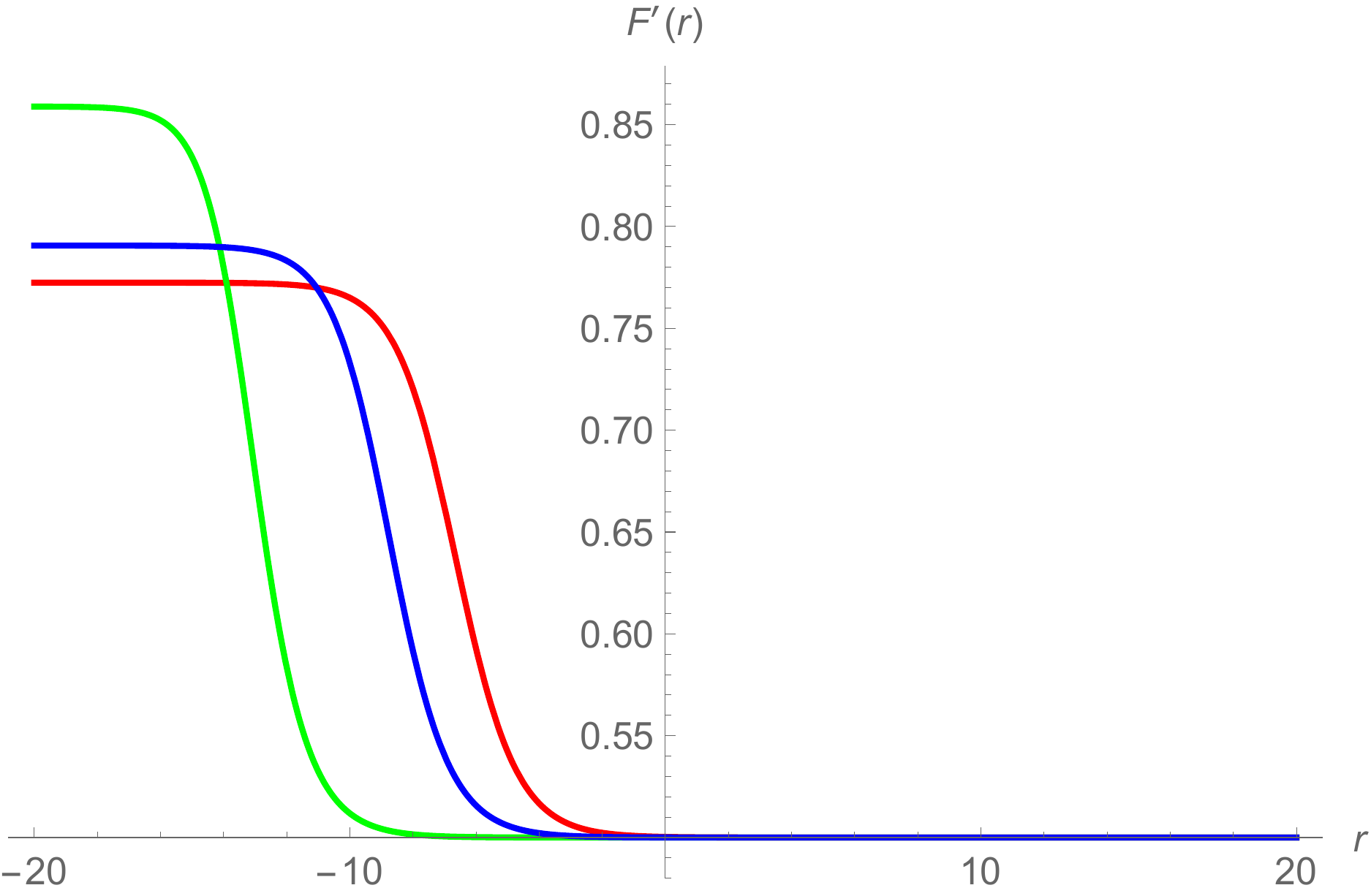}
                 \caption{Solution for $F'(r)$}
         \end{subfigure}
         \caption{RG flows from $N=4$ $AdS_5$ critical point I to $N=(2,0)$ $AdS_3\times H^2$ fixed point $i$ with $\phi_2=\phi_3=0$, $h_1=1$, $\zeta=\frac{1}{2}$ for $\rho=1.20$ (green), $\rho=1.35$ (blue), $\rho=1.50$ (red).}\label{fig4}
 \end{figure}

We now look for supersymmetric black string solutions interpolating between $AdS_5$ and $AdS_3\times \Sigma^2$ geometries. We begin with solutions flowing to $AdS_3\times H^2$ critical point $i$. With $\phi_2=\phi_3=0$ and $\kappa=-1$, examples of solutions interpolating between $N=4$ $AdS_5$ critical point I and $AdS_3\times H^2$ fixed point $i$ are shown in figure \ref{fig4}. 
                                                    
\begin{figure}
         \centering
               \begin{subfigure}[b]{0.45\textwidth}
                 \includegraphics[width=\textwidth]{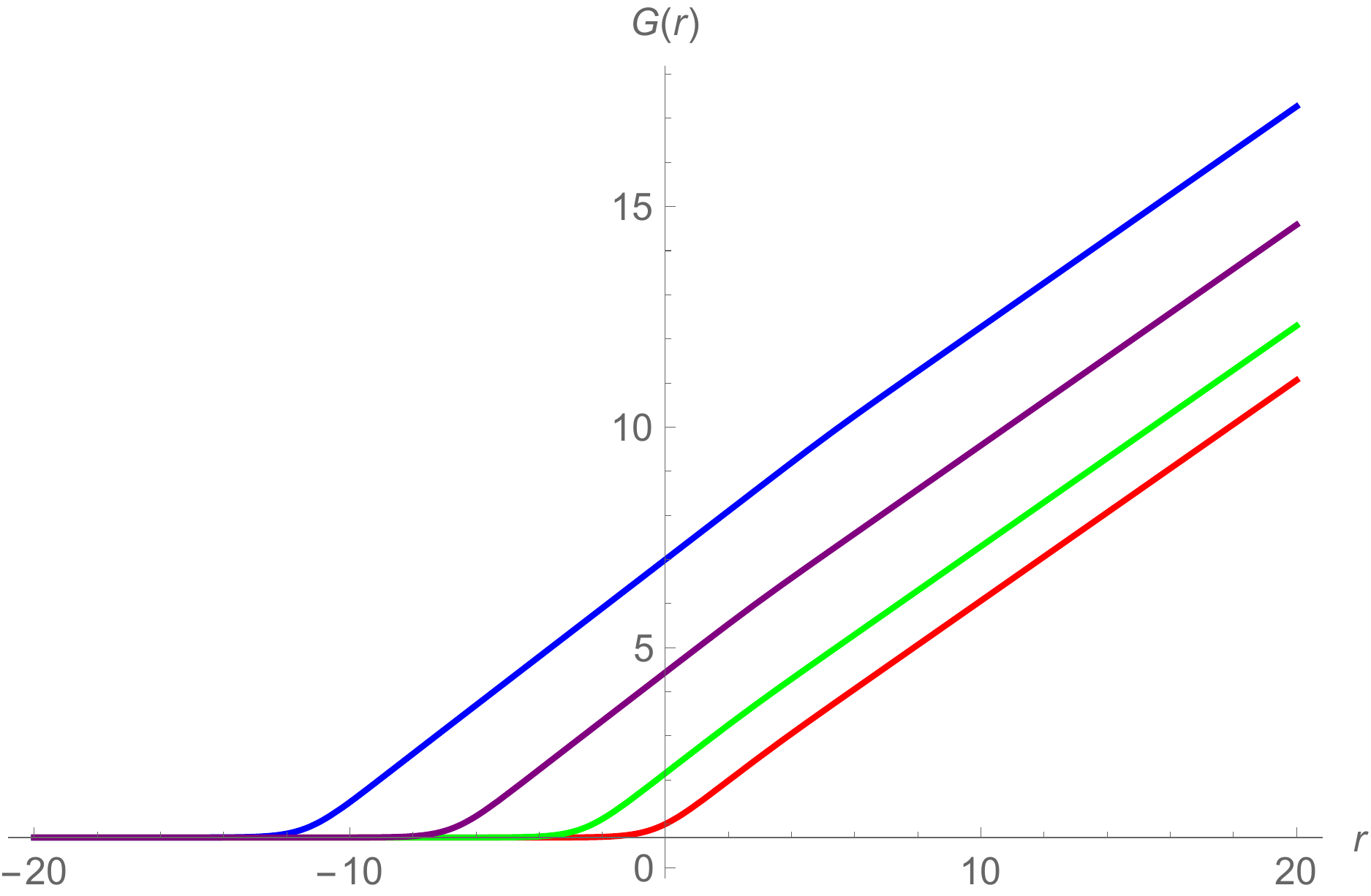}
                 \caption{Solution for $G(r)$}
         \end{subfigure}
         \begin{subfigure}[b]{0.45\textwidth}
                 \includegraphics[width=\textwidth]{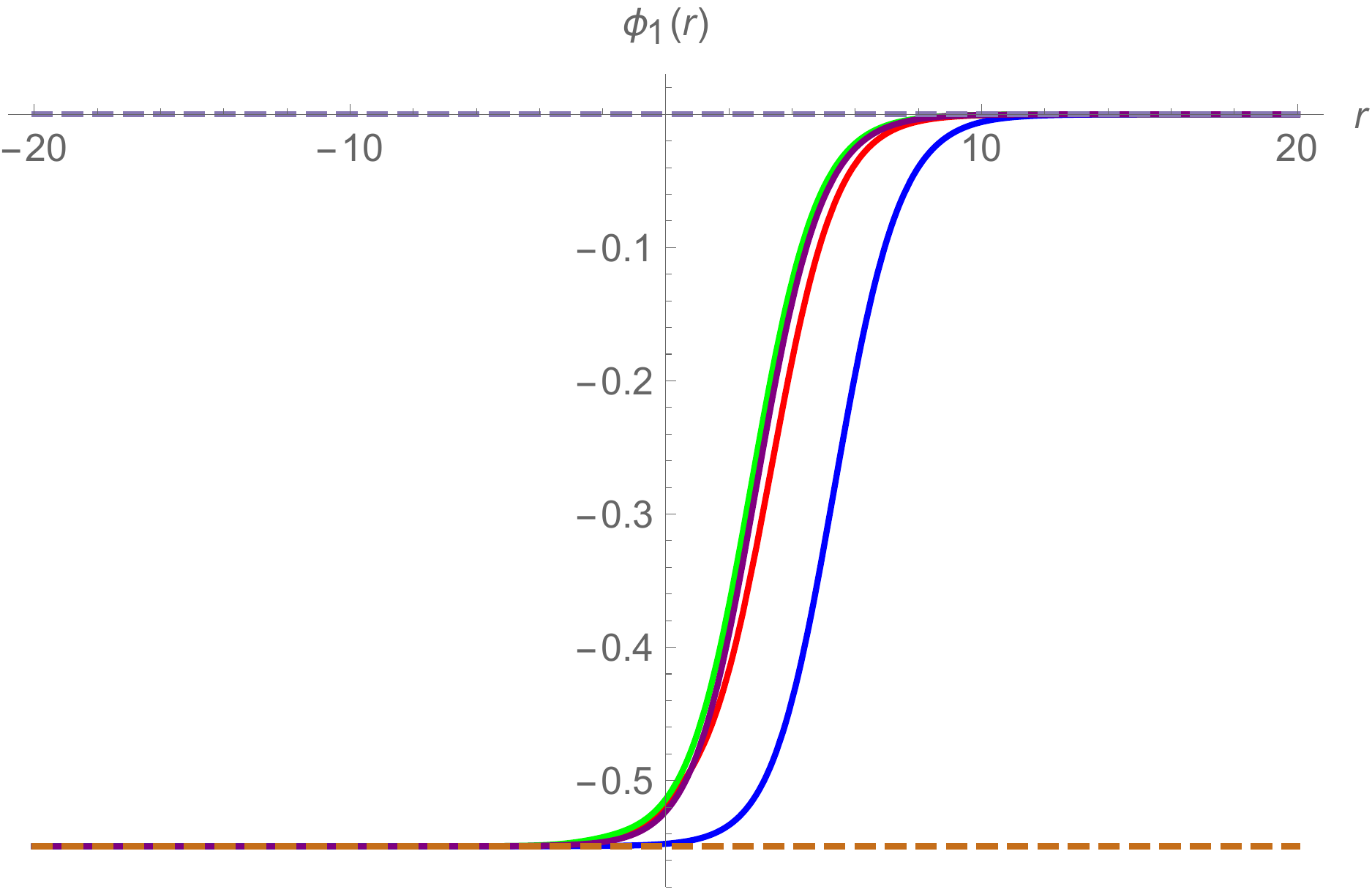}
                 \caption{Solution for $\phi_1(r)$}
         \end{subfigure}\\
         \begin{subfigure}[b]{0.45\textwidth}
                 \includegraphics[width=\textwidth]{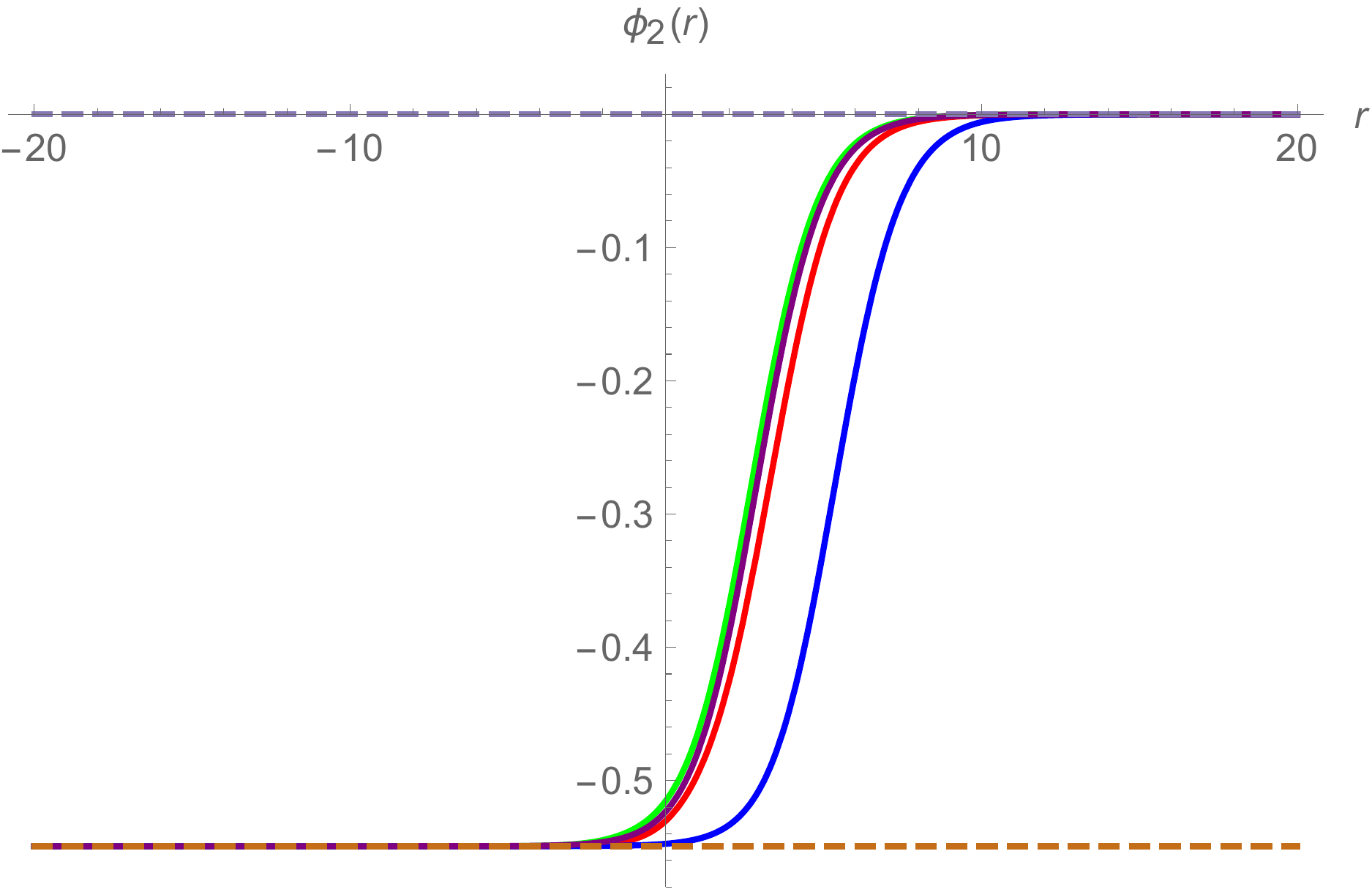}
                 \caption{Solution for $\phi_2(r)$}
         \end{subfigure}
         \begin{subfigure}[b]{0.45\textwidth}
                 \includegraphics[width=\textwidth]{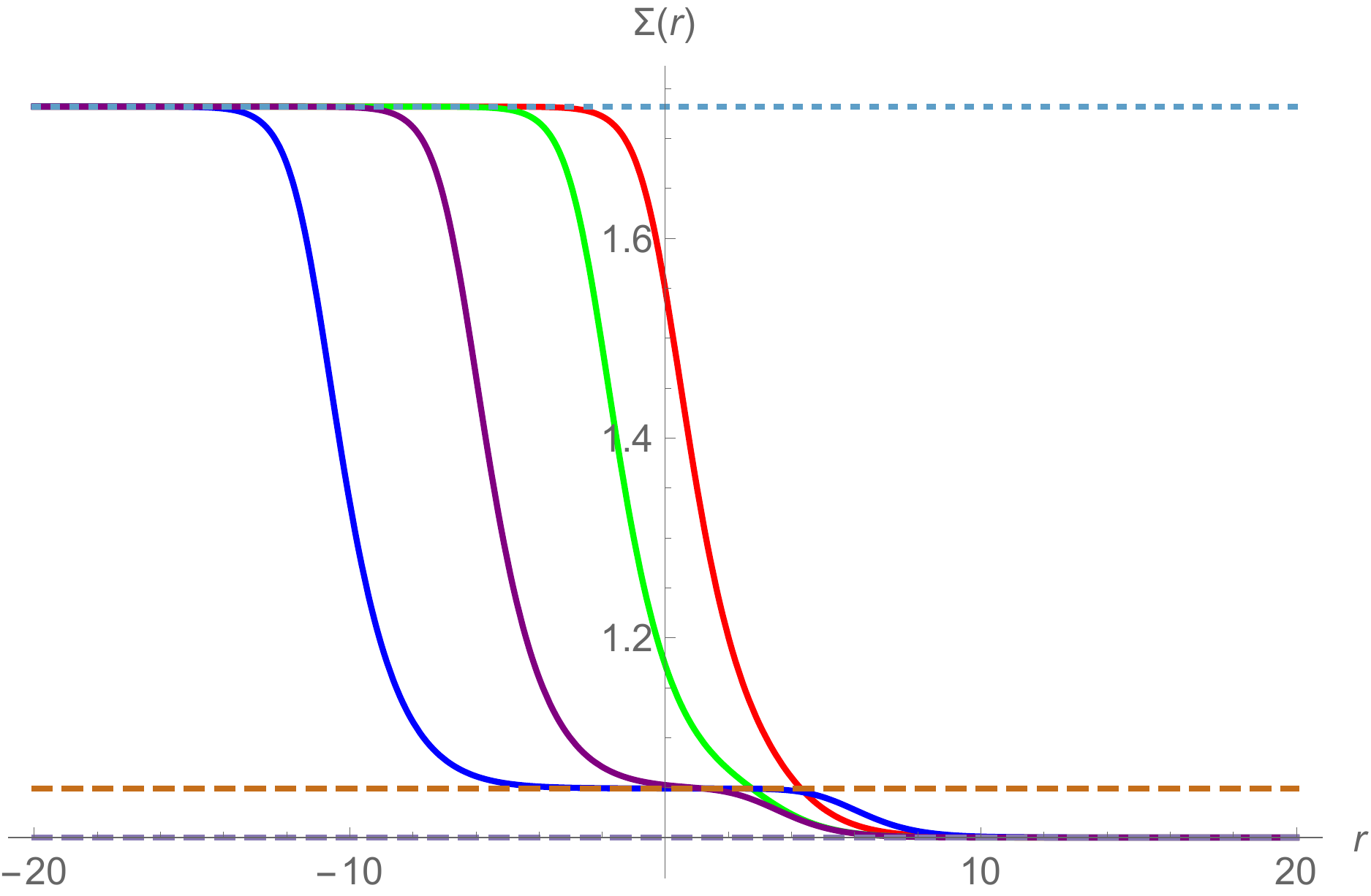}
                 \caption{Solution for $\Sigma(r)$}
         \end{subfigure}\\
          \begin{subfigure}[b]{0.45\textwidth}
                 \includegraphics[width=\textwidth]{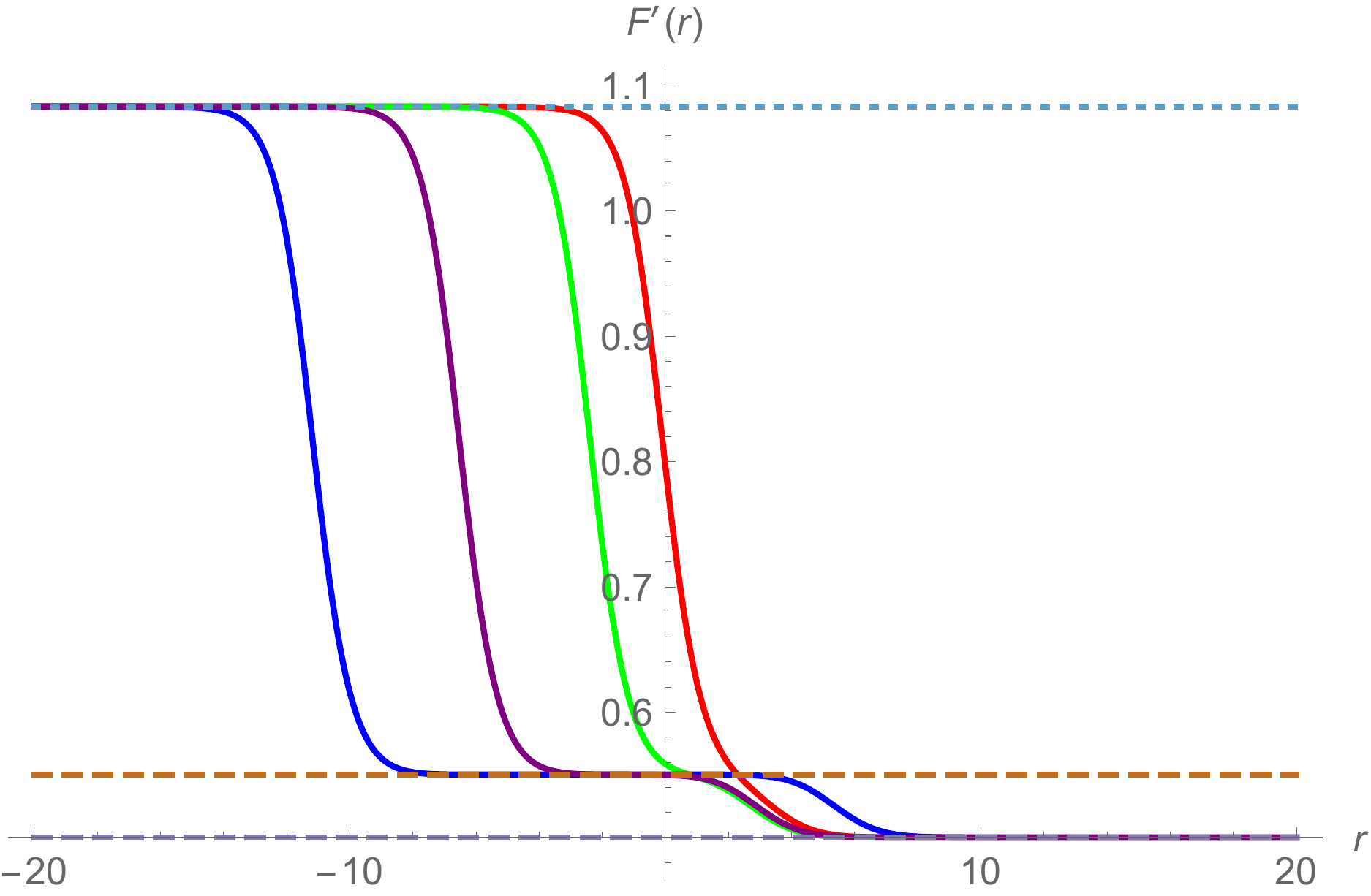}
                 \caption{Solution for $F'(r)$}
         \end{subfigure}
         \caption{RG flows from $N=4$ $AdS_5$ critical points I and II to $N=(2,0)$ $AdS_3\times S^2$ fixed point $ii$ in the IR with $\phi_3=0$, $h_1=1$, $\rho=-\frac{5}{2}$ and $\zeta=\frac{1}{2}$.}\label{fig5}
 \end{figure}

\begin{figure}
         \centering
               \begin{subfigure}[b]{0.45\textwidth}
                 \includegraphics[width=\textwidth]{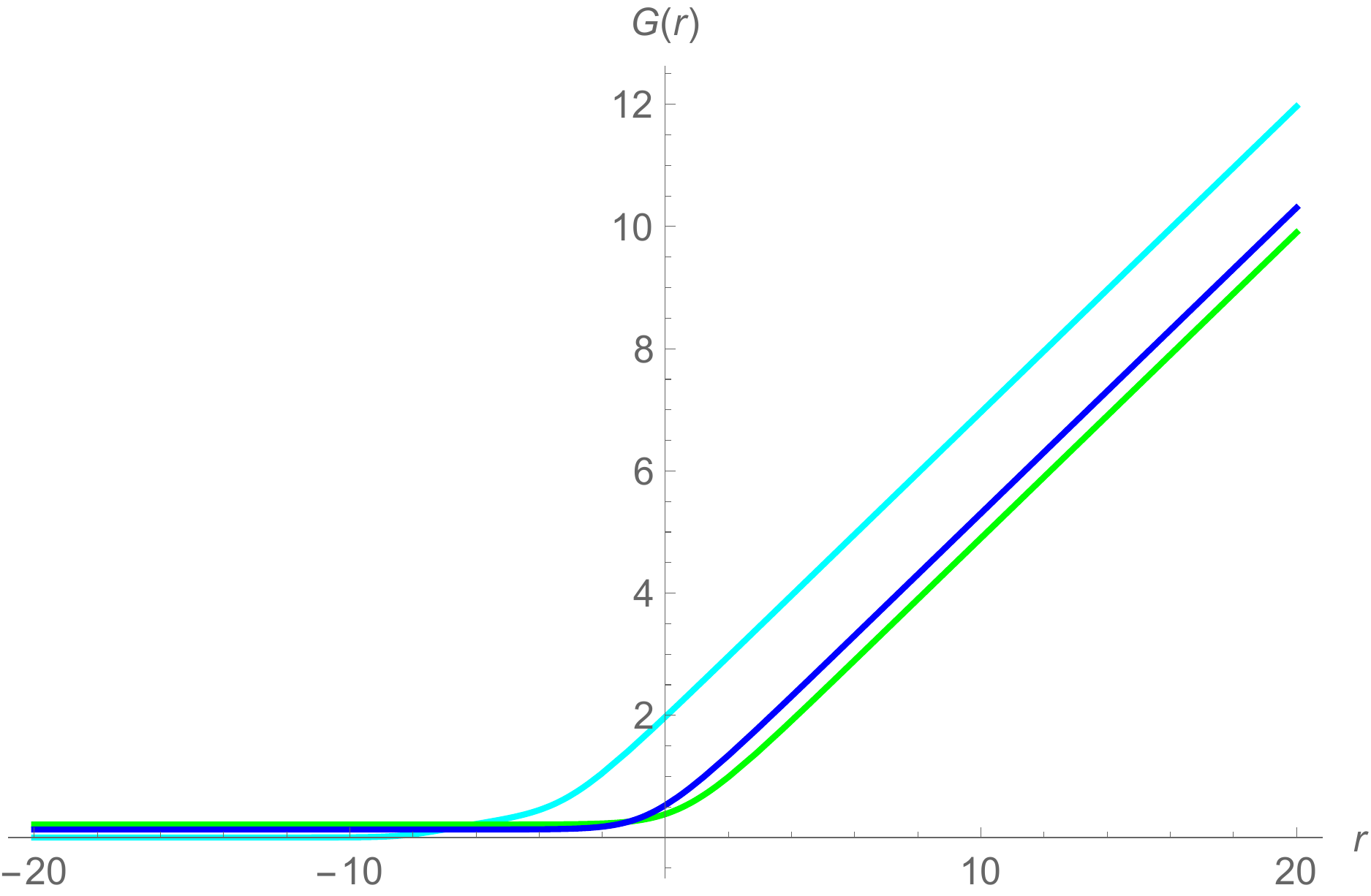}
                 \caption{Solution for $G(r)$}
         \end{subfigure}
         \begin{subfigure}[b]{0.45\textwidth}
                 \includegraphics[width=\textwidth]{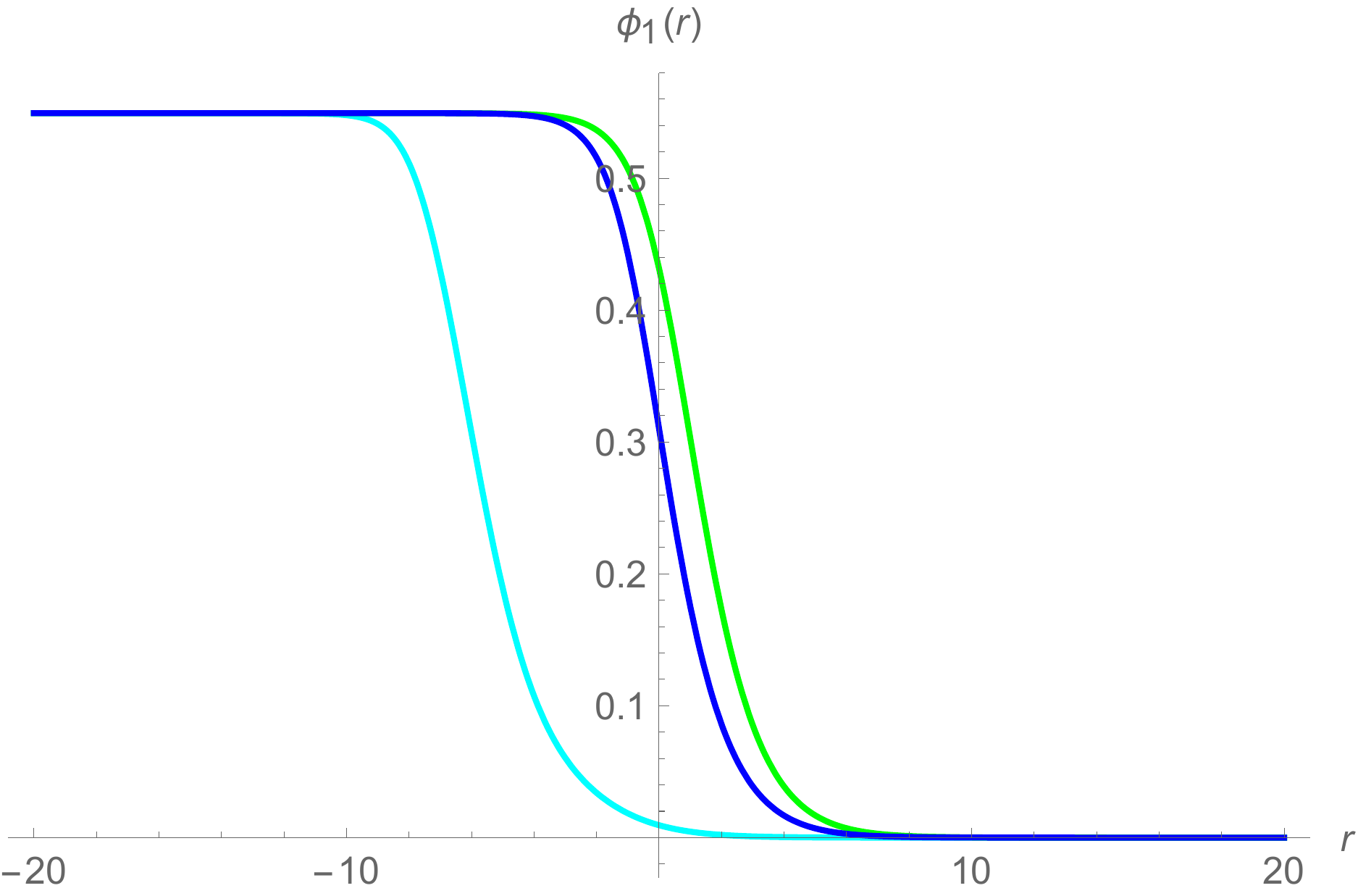}
                 \caption{Solution for $\phi_1(r)$}
         \end{subfigure}\\
         \begin{subfigure}[b]{0.45\textwidth}
                 \includegraphics[width=\textwidth]{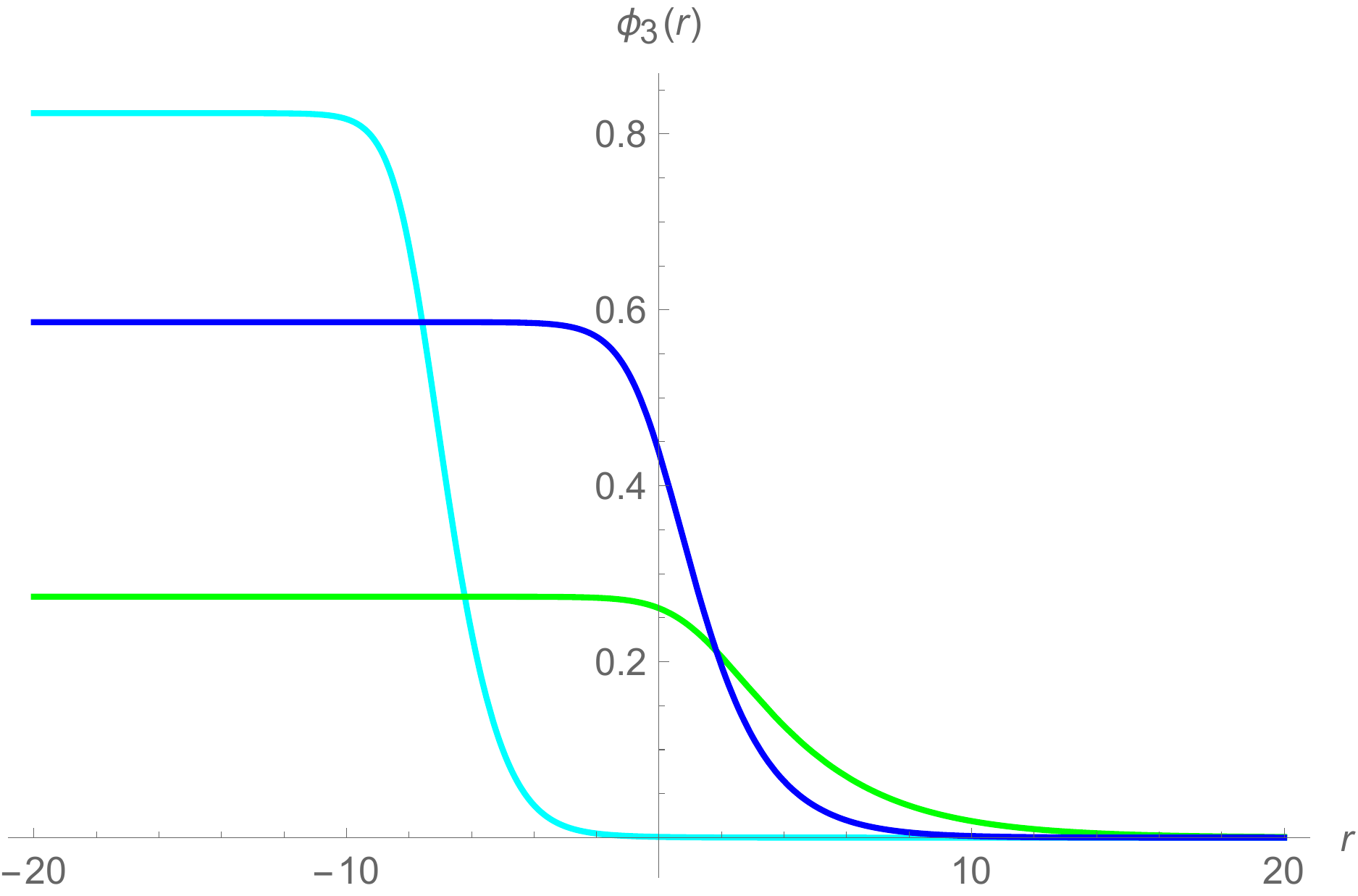}
                 \caption{Solution for $\phi_3(r)$}
         \end{subfigure}
         \begin{subfigure}[b]{0.45\textwidth}
                 \includegraphics[width=\textwidth]{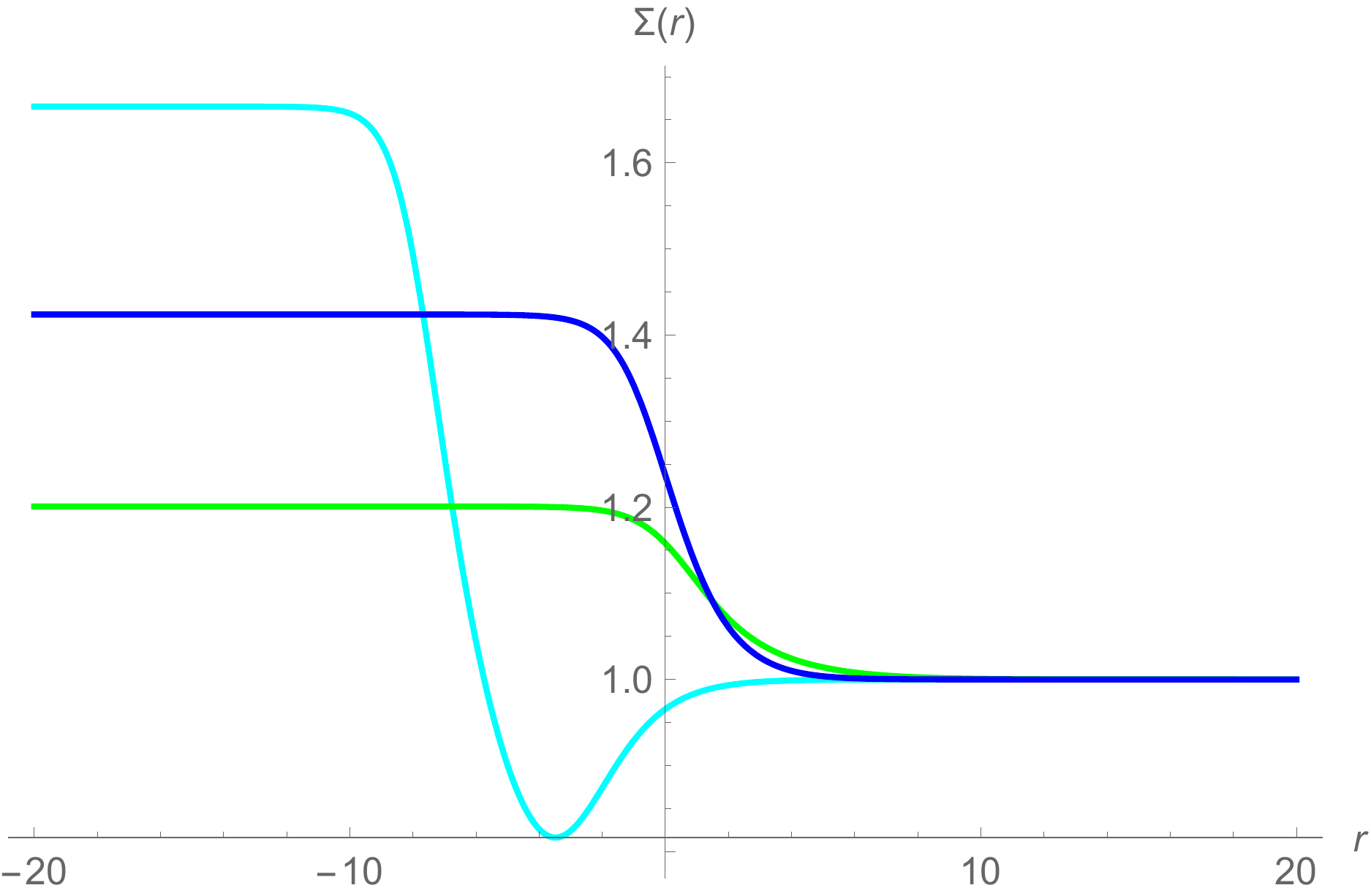}
                 \caption{Solution for $\Sigma(r)$}
         \end{subfigure}\\
          \begin{subfigure}[b]{0.45\textwidth}
                 \includegraphics[width=\textwidth]{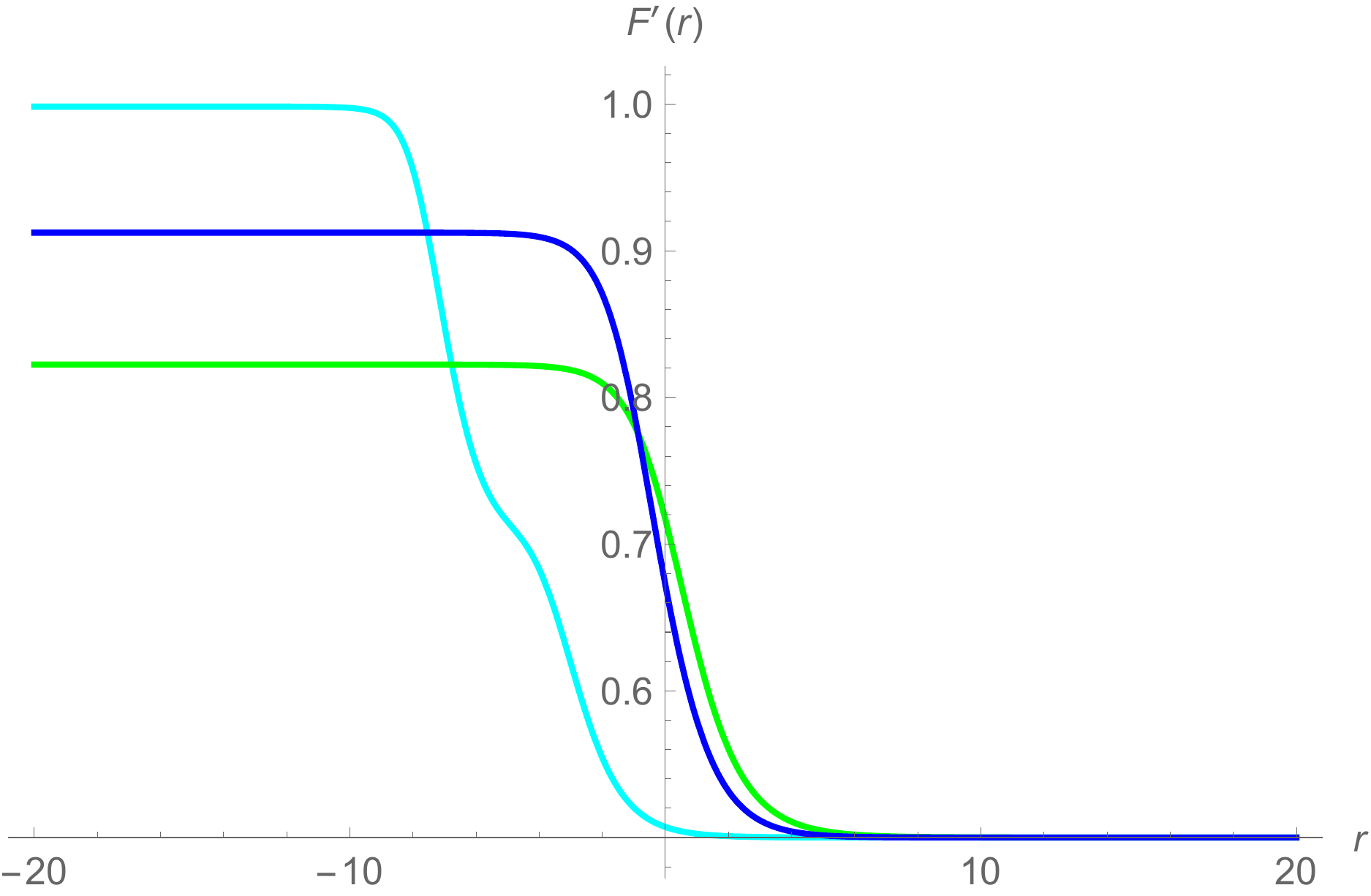}
                 \caption{Solution for $F'(r)$}
         \end{subfigure}
         \caption{RG flows from $N=4$ $AdS_5$ critical point I to $N=(2,0)$ $AdS_3\times H^2$ fixed point $iii$ in the IR with $\phi_2=0$, $h_1=1$, $\zeta=\frac{1}{2}$ and $\rho=\frac{3}{2}$ (green), $\rho=\frac{5}{2}$ (blue), $\rho=4$ (cyan).}\label{fig6}
 \end{figure}

We now move to solutions involving $AdS_3\times \Sigma^2$ fixed point $ii$. Unlike other cases, the solutions in this case only exist for $\kappa=1$. Some examples of solutions for $\phi_3=0$ are given in figure \ref{fig5}. There is a solution flowing directly from $N=4$ $AdS_5$ vacuum I to $AdS_3\times S^2$ fixed point $ii$ (red line). There are also solutions that flow to $AdS_3\times S^2$ fixed point $ii$ and approach arbitrarily close to $N=4$ $AdS_5$ critical point II shown by green, purple and blue lines. Similarly, setting $\phi_2=0$, we find solutions interpolating between $AdS_5$ critical point I and $AdS_3\times H^2$ fixed point $iii$ with some examples of these solutions shown in figure \ref{fig6}. 

\begin{figure}
         \centering
               \begin{subfigure}[b]{0.45\textwidth}
                 \includegraphics[width=\textwidth]{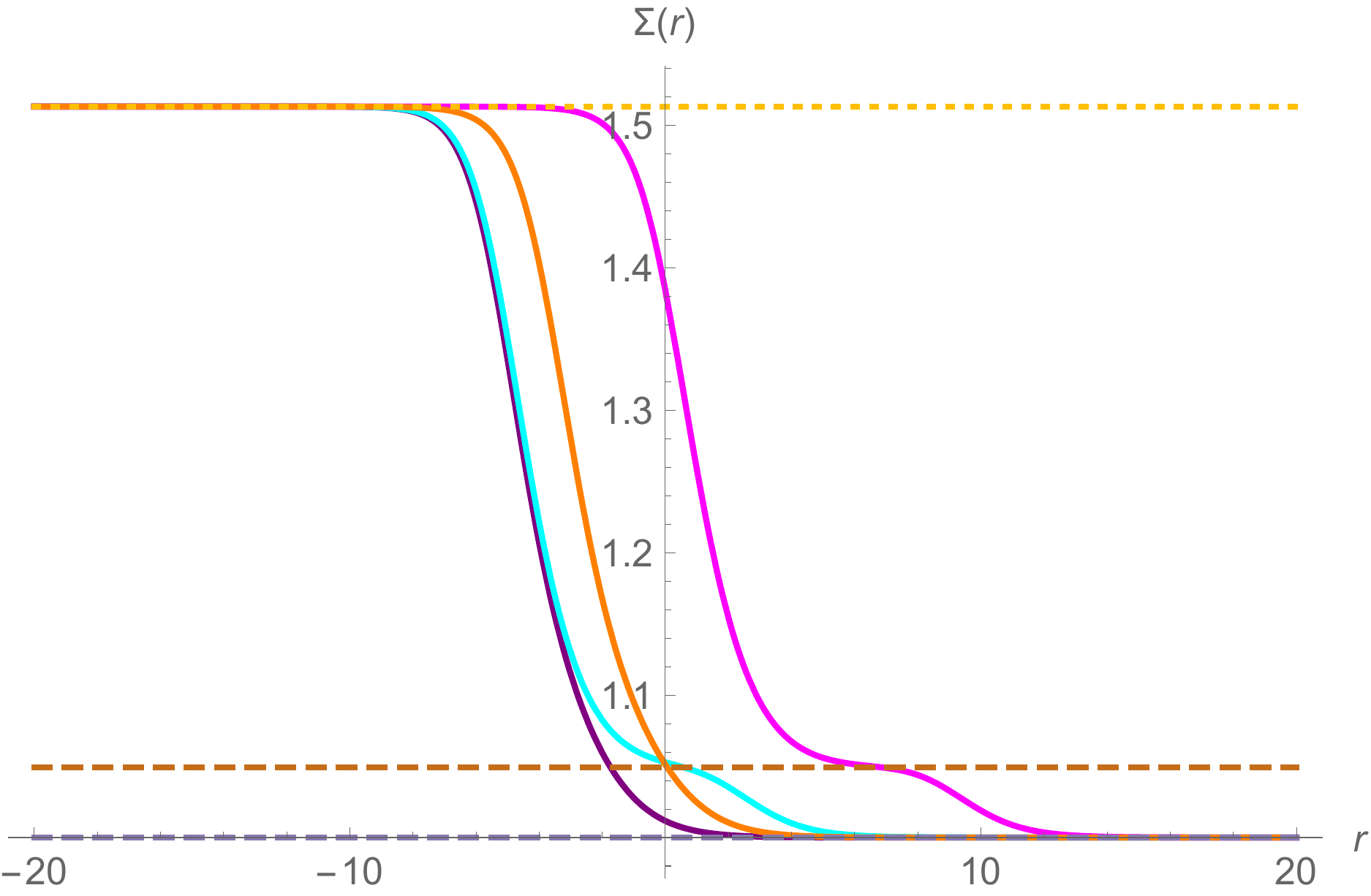}
                 \caption{Solution for $\Sigma(r)$}
         \end{subfigure}
         \begin{subfigure}[b]{0.45\textwidth}
                 \includegraphics[width=\textwidth]{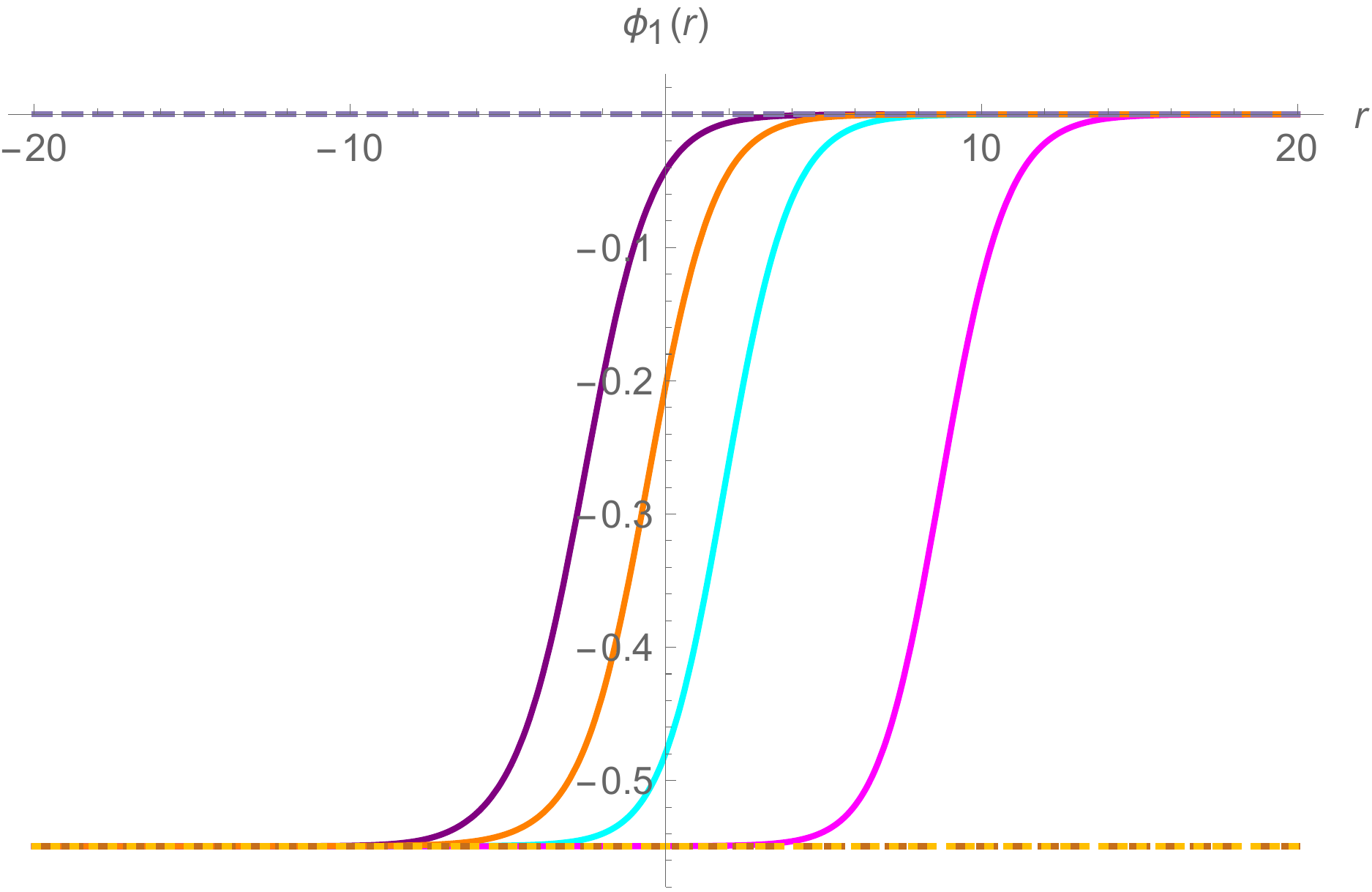}
                 \caption{Solution for $\phi_1(r)$}
         \end{subfigure}\\
         \begin{subfigure}[b]{0.45\textwidth}
                 \includegraphics[width=\textwidth]{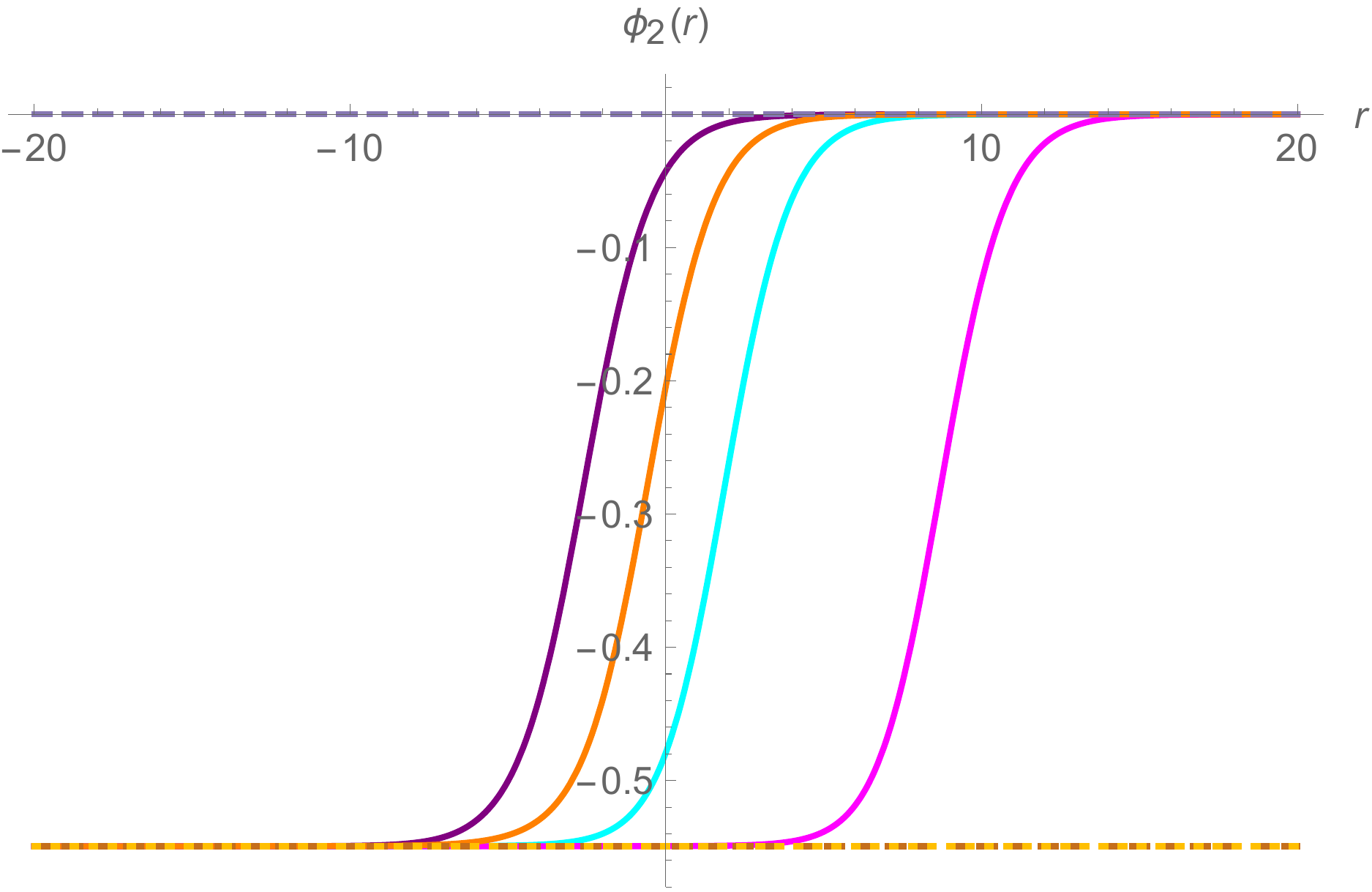}
                 \caption{Solution for $\phi_2(r)$}
         \end{subfigure}
         \begin{subfigure}[b]{0.45\textwidth}
                 \includegraphics[width=\textwidth]{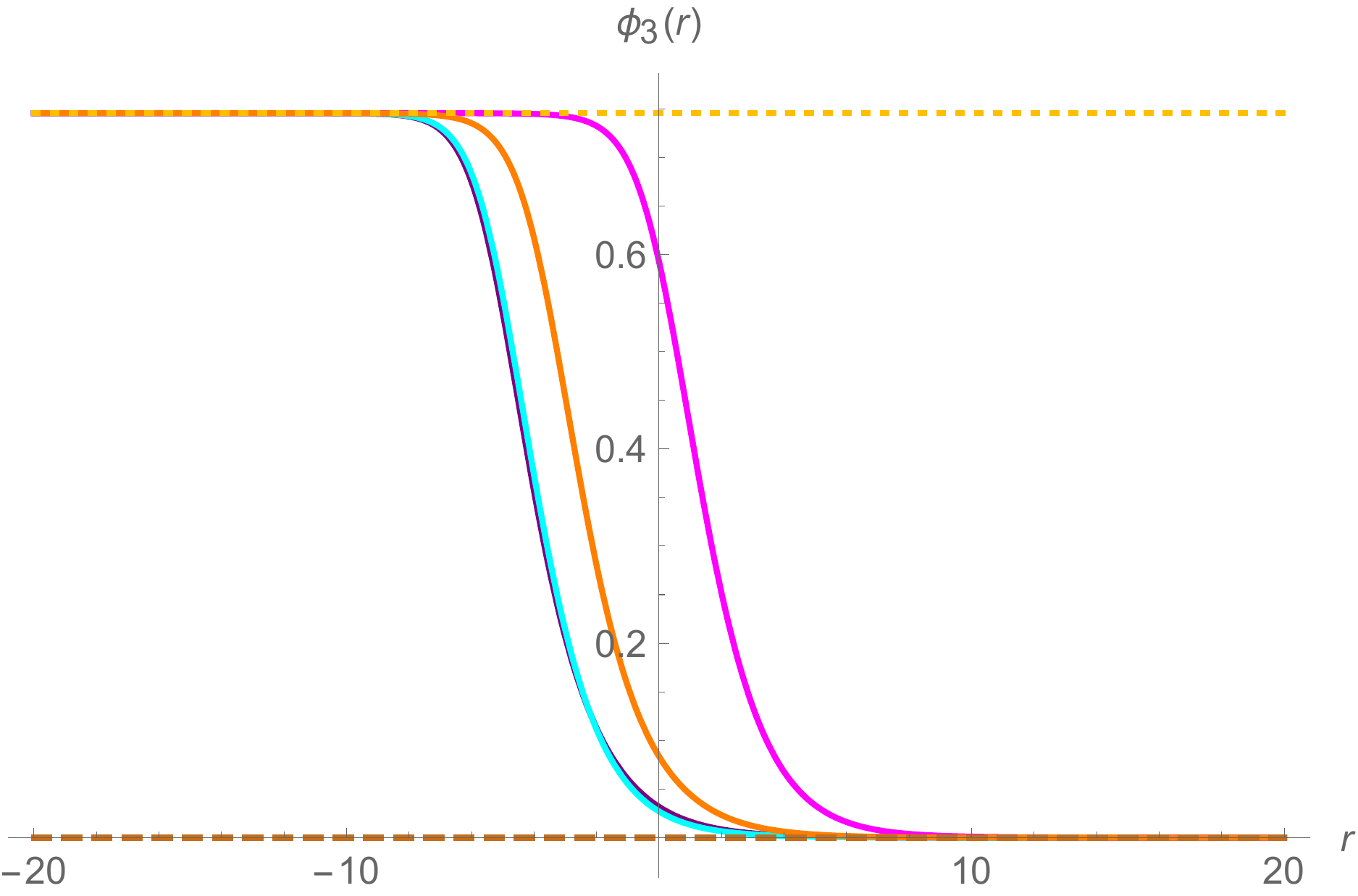}
                 \caption{Solution for $\phi_3(r)$}
         \end{subfigure}\\
          \begin{subfigure}[b]{0.45\textwidth}
                 \includegraphics[width=\textwidth]{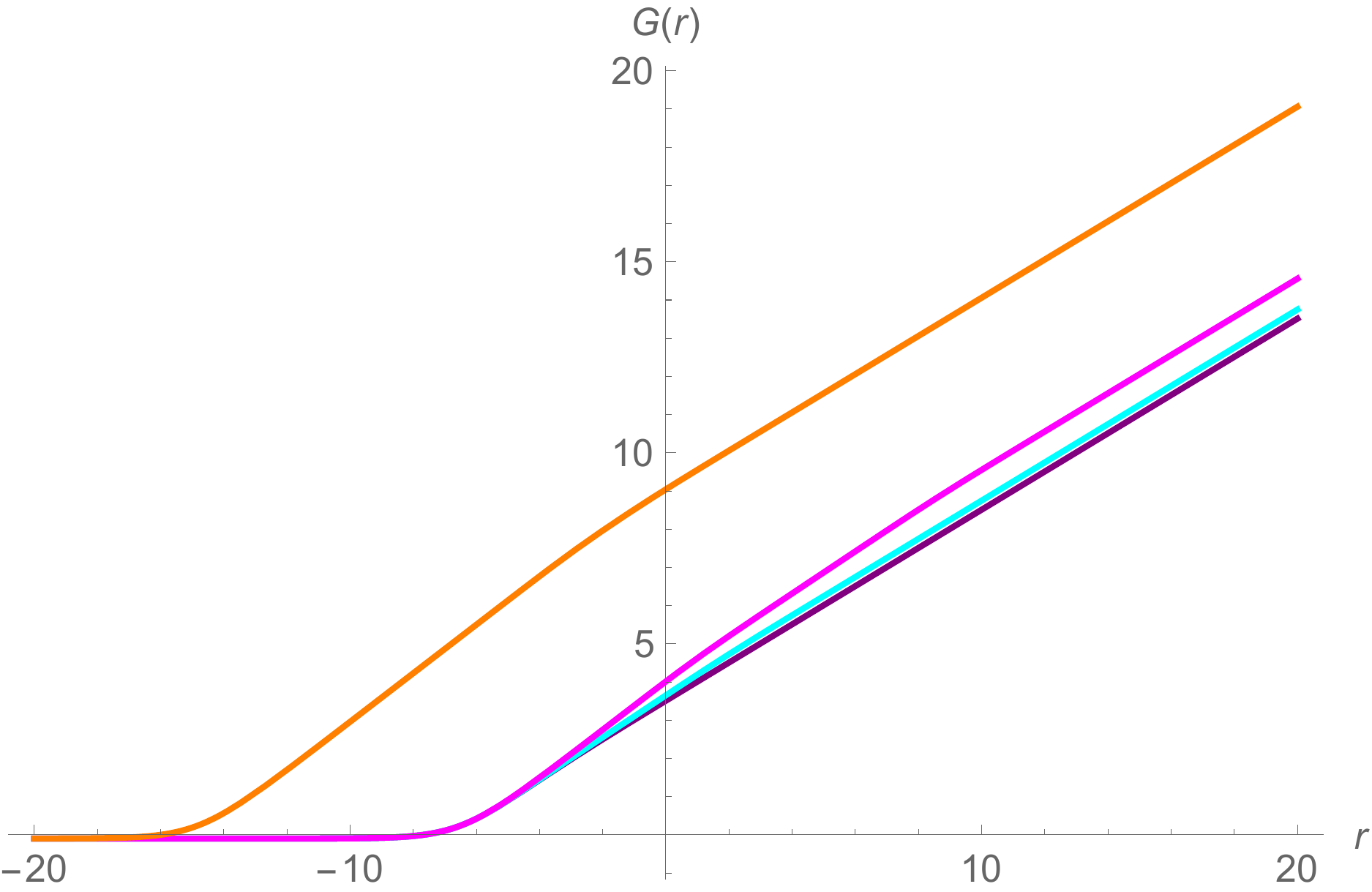}
                 \caption{Solution for $G(r)$}
         \end{subfigure}
          \begin{subfigure}[b]{0.45\textwidth}
                 \includegraphics[width=\textwidth]{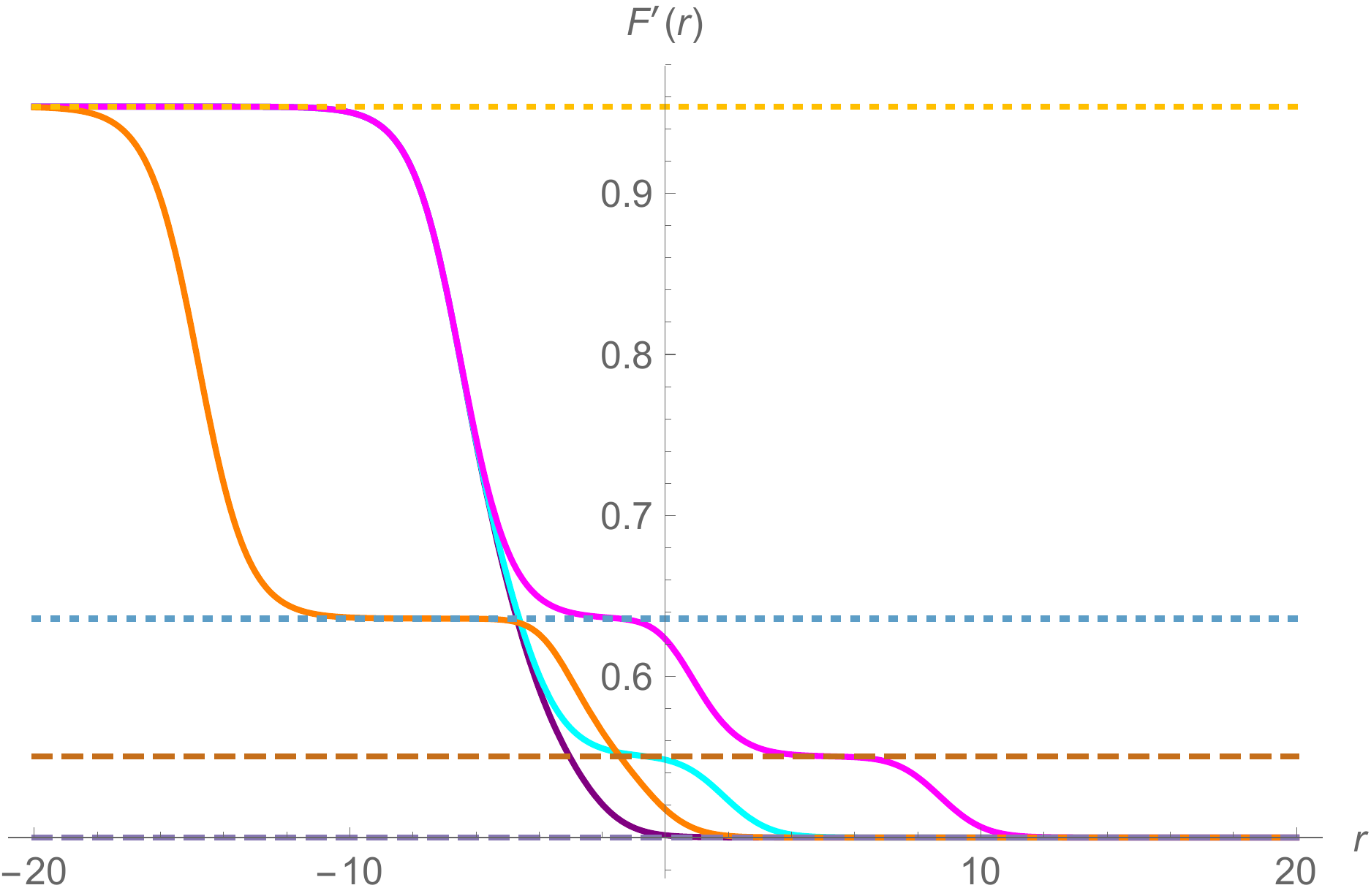}
                 \caption{Solution for $F'(r)$}
         \end{subfigure}
         \caption{RG flows from $AdS_5$ critical points I, II and IV to $N=(2,0)$ $AdS_3\times H^2$ fixed point $iv$ with $h_1=1$, $\zeta=\frac{1}{2}$ and $\rho=3$.}\label{fig7}
 \end{figure}

Finally, we consider solutions that flow to $AdS_3\times H^2$ critical point $iv$. We first note that the values of scalar fields are the same as those for $N=2$ $AdS_5$ critical point IV. By setting all scalar fields to the values at these critical points, we find a solution involving only $F(r)$ and $G(r)$ given by
\begin{eqnarray}
G&=&\frac{1}{2}\ln\left[\frac{e^{\frac{2(r-r_0)}{L_5}}+6 a_3g_2^{\frac{1}{3}}(h_2^2-h_1^2)}{h_2(g_2-g_1)[4h_1h_2(h_2^2-h_1^2)]^{\frac{1}{3}}}\right],\quad L_5=\frac{3}{(g_2-g_1)}\left(\frac{g_2^2(h_2^2-h_1^2)}{4\sqrt{2}h_1^2h_2^2}\right)^{\frac{1}{3}},\nonumber \\
F&=&\frac{3r}{2L_5}-\frac{1}{4}\ln \left[e^{\frac{2(r-r_0)}{L_5}}+6 a_3g_2^{\frac{1}{3}}(h_2^2-h_1^2)\right].
\end{eqnarray}
For more general solutions with scalars depending on $r$, we can find only numerical solutions. Examples of these solutions can be found in figure \ref{fig7}. In this case, the solutions are more interesting than those of the previous cases. There is a solution interpolating between $AdS_5$ critical point I to $AdS_3\times H^2$ fixed point $iv$ shown by the purple line. On the other hand, there are solutions interpolating among $AdS_5$ critical points I, II and IV and $AdS_3\times H^2$ fixed point $iv$. Some of these solutions connects all these critical points within a single flow (pink line). There are also solutions interpolating between two $AdS_5$ vacua and $AdS_3\times H^2$ fixed point $iv$ as shown by the orange and cyan lines. The former begins at critical point I and flows to critical points IV and $iv$ while the latter flows from critical point I to critical points II and $iv$. 
\\
\indent As in the previous section, these solutions should also holographically describe various possible RG flows from $N=2$ and $N=1$ SCFTs in four dimensions to $N=(2,0)$ two-dimensional conformal fixed points in the IR. 

\section{Conclusions and discussions}\label{conclusion}
In this paper, we have constructed a large number of supersymmetric $AdS_5$ black string solutions from five-dimensional $N=4$ gauged supergravity with $SO(2)_D\times SO(3)\times SO(3)$ gauge group. The solutions interpolate between a number of different $AdS_5$ and $AdS_3\times \Sigma^2$ critical points and preserve $SO(2)_{\textrm{diag}}\times SO(2)$ or $SO(2)_{\textrm{diag}}$ symmetry and two supercharges. On the other hand, the $AdS_3\times \Sigma^2$ fixed points preserve $4$ supercharges corresponding to $N=(2,0)$ or $N=(0,2)$ superconformal symmetry in two dimensions. Some of the solutions can even be analytically obtained. Although most of the solutions describe black strings with $AdS_3\times H^2$ near horizon geometry, we have also found one $AdS_3\times S^2$ solution with $SO(2)_{\textrm{diag}}$ symmetry. The solutions could be of interest in microscopic counting of black string entropy along the line of \cite{microstate_black_string1,microstate_black_string2,microstate_black_string3,microstate_black_string4}. Holographically, the solutions also describe various RG flows from four-dimensional $N=1$ and $N=2$ SCFTs to two-dimensional $N=(2,0)$ SCFTs in the IR via twisted compactifications on $\Sigma^2$. The solutions given here are expected to be useful in holographic study of strongly coupled $N=1,2$ SCFTs in four dimensions with topological twists as well.     
\\
\indent It should be pointed out that the $N=2$ $AdS_5$ vacuum with $SO(2)_{\textrm{diag}}\times SO(3)$ symmetry, obtained from $SO(2)_\textrm{diag}\times SO(2)$ sector, together with $AdS_3\times H^2$ fixed point $2$ and related flow solutions can be embedded in $N=4$ gauged supergravity with $n=2,3$ vector multiplets. The latter can be obtained from the gauged supergravity considered here by truncating out $\phi_1$ and the gauge field $A^6_\mu$ resulting in $N=4$ gauged supergravity with $SO(2)_D\times SO(3)_R$ gauge group. It has been shown in \cite{Malek_AdS5_N4_embed} that this gauged supergravity can be embedded in eleven dimensions.Therefore, within the above truncation, the solutions given here can be uplifted to M-theory. It is then of particular interest to construct the truncation ansatz from the result of \cite{Malek_AdS5_N4_embed} and uplift the black string solutions found here to M-theory. This would lead to a new holographic dual of $N=(2,0)$ SCFTs in two dimensions within string/M-theory context.    
\\
\indent Moreover, it could be interesting to identify the $N=1$ and $N=2$ SCFTs dual to the $AdS_5$ vacua and the two-dimensional conformal fixed points in the IR as well as the associated RG flows in the field theory context. Partial results along this direction have been given in \cite{5D_N4_flow_Davide} in which the possible dual $N=1$ and $N=2$ SCFTs have been identified. It would be useful to extend these results to the case with topological twists. Furthermore, constructing similar solutions in the form of $AdS_3\times \mathbb{\Sigma}$ with $\mathbb{\Sigma}$ being a spindle or a half-spindle along the line of recent results in \cite{string_spindle1,string_spindle2,string_spindle_microstate,string_spindle3} is also worth considering. Finally, finding similar solutions within $N=4$ gauged supergravity with gauge groups identified in \cite{Malek_AdS5_N4_embed} as embeddable in eleven dimensions would lead to new holographic solutions in string/M-theory framework. We leave all these and related issues for future work.                 
\vspace{0.5cm}\\
{\large{\textbf{Acknowledgement}}} \\
This work is funded by National Research Council of Thailand (NRCT) and Chulalongkorn University under grant N42A650263.

\end{document}